\pdfoutput=1
\documentclass[12pt,english]{iopart}
\usepackage[T1]{fontenc}
\usepackage[latin9]{inputenc}
\usepackage{babel}
\usepackage{graphicx}
\usepackage{hyperref}
\usepackage{cite}
\usepackage{float}
\usepackage{subfig}
\usepackage{caption}

\begin{document}

\title[Tuning superfluid phases of spin-1 bosons in cubic optical lattice with linear Zeeman effect]{Tuning superfluid phases of spin-1 bosons in cubic optical lattice with linear Zeeman effect}

\author{Mohamed Mobarak}
\address{Insitute f\"ur Theoretische Physik, Freie Universit\"at Berlin, Arnimallee 14, 14195 Berlin, Germany}
\author{Axel Pelster}

\address{Fachbereich Physik und Forschungszentrum OPTIMAS, Technische Universit\"at Kaiserslautern, 67663 Kaiserslautern,Germany}
\address{Hanse-Wissenschaftskolleg, Lehmkuhlenbusch 4, D-27733 Delmenhorst, Germany}
\ead{axel.pelster@physik.uni-kl.de}

\begin{abstract}
We analyze theoretically a spinor Bose gas loaded into a three-dimensional cubic optical lattice. In
order to account for different superfluid phases of spin-1 bosons in the presence of an external magnetic
field, we work out a Ginzburg-Landau theory for the underlying spin-1 Bose-Hubbard model. In particular
at zero temperature, we determine both the Mott and the superfluid phases for the competition between
the anti-ferromagnetic interaction and the linear Zeeman effect within the validity range of the Ginzburg-Landau
theory. Moreover, we find that the phase transition between the superfluid and Mott insulator phases
is of second order and that the transitions between the respective superfluid phases for anti-ferromagnetic
interaction can be both of first and second order.
\end{abstract}

\section{Introduction}

Experimental and theoretical studies on Bose-Einstein condensates (BECs) emerged to be one of the most
interesting topics in physics since their realization in a series of experiments in dilute atomic gases
of alkali atoms such as $^{87}$Rb \cite{key-1}, $^{23}$Na \cite{key-3}, and $^{7}$Li \cite{key-4-1,key-100}.
In these systems the atoms are confined in a magneto-optical trap, cooled to nano-Kelvin temperatures,
and then BEC occurs at a critical phase space density. The main advantage of these quantum-many body
systems is the high degree of tunability of both the shape of the confining trap and the strength of
the two-particle interaction. Thus, they serve as an ideal model for a quantum simulator in the sense
of Richard Feynman to realize various phenomena in the realm of condensed matter physics \cite{key-50}.

An optical lattice represents a periodic potential, which is generated by the interference of counter-propagating
laser beams. The experimental realization of bosons in optical lattices triggered the research on strongly
correlated quantum many-body systems \cite{key-34,key-35}. Most prominently, the quantum phase transition
between a superfl{}uid (SF) and a Mott-insulating (MI) phase of a spinless Bose gas loaded in a periodic
optical potential was experimentally observed by increasing the lattice depth. All properties of this
quantum phase transition are captured by the underlying Bose-Hubbard Hamiltonian \cite{key-33,key-50-1,key-36}
for which different analytical solution methods have been worked out \cite{key-22,key-18,key-23-1,key-3-3,key-24,key-30}
and high-precision Monte Carlo studies have been performed \cite{key-27,key-28}. Furthermore, various
extensions of the Bose-Hubbard model have been investigated, which cover for instance, superlattices
\cite{key-31}, Bose-Fermi mixtures \cite{key-58,key-59,key-60,key-61}, quantum simulations like entanglement
of atoms or quantum teleportation \cite{key-57} and disorder \cite{key-37,key-51,key-52,key-53-1}.

Bosons with an internal spin degrees  of freedom yield a quantum gas with magnetization. The first theoretical
discussion of a BEC with spin degrees of freedom in an optical trap was performed in Refs.~\cite{key-5,key-148}.
There the Hamiltonian of a spinor Bose-Einstein condensate and the mean-field condensate wave function
were determined. This ansatz was verified experimentally by the Ketterle group by studying the ground
state of the spin-1 system consisting of $^{23}$Na atoms \cite{key-145}. Furthermore, the MIT group
succeeded to transfer a spin-polarized $^{23}$Na condensate, which was produced in a traditional magneto-optical
trap, into a dipole trap formed by the focus of a far-off-resonant laser \cite{key-143}. With this,
spinor condensates opened a new area to study various aspects of the quantum magnetism such as spin dynamics
\cite{key-71,key-72,key-73}, spin waves \cite{key-74,key-75}, or spin mixing \cite{key-59-1,key-77}.
These examples result from coherent collisional processes between two atoms where the total magnetization
is constant but the spins of the individual particles can change.

The experimental realization of an optically trapped BEC motivated both theoretical and experimental
studies on spinor Bose gases loaded in an optical lattice. Rich physics with various phenomena in both
MI and SF phases were expected due to the additional spin degree of freedom. For instance, the coherent
collisional spin dynamics in an optical lattice was measured in Ref.~\cite{key-42-2} and the $^{87}$Rb
scattering lengths for $F=1$ and $F=2$ were determined in Ref.~\cite{key-42-3}. Furthermore,  $^{87}$Rb atoms  were loaded   in a frustrated triangular lattice \cite{key-53-2}.
Despite these initial promising investigations, spinor Bose gases in optical lattice seem experimentally
to be so challenging that no further detailed experiments have so far been performed. On the other hand,
the properties of spin-1 Bose gases in an optical lattice were investigated in detail some time ago theoretically
in Refs.~\cite{key-29,key-4}. Several unique MI and SF phases for spin-1 bosons were determined without
external magnetic field at zero temperature in case of an anti-ferromagnetic interaction in an optical
lattice \cite{key-29}. For instance, the MI phase with an even number of atoms is more strongly stabilized
than that with an odd number because of the formation of singlet pairs \cite{key-4}. Moreover, the SF
phase represents a polar state with zero spin expectation value \cite{key-29,key-4}. On the other side,
the influence of the linear Zeeman effect with a non-vanishing external magnetic field upon the MI-SF
phase boundary was determined within a mean-field approximation in Refs.~\cite{key-60-1,key-15}. In
addition, it was also shown in Ref.~\cite{key-60-1} that the superfluid transition occurs into either
a polar spin-1 or a polar spin-(-1) state, but it was not investigated, which magnetic phases may emerge
deeper in the superfluid. 

In this paper, we follow Ref.~\cite{key-120} and study the effect of an external magnetic field on
the emergence of superfluid phases for anti-ferromagnetic spin-1 bosons in a three-dimensional cubic
optical lattice at zero temperature. To this end, we extend the Ginzburg-Landau theory developed in Ref.~\cite{key-23-1,key-3-3}
from the spin-0 to the spin-1 Bose-Hubbard model. Thus, we calculate the effective action which allows
us to obtain the different superfluid phases and to determine the respective order of the transitions
between them.

In detail we proceed as follows. In Sec.~II, we derive the Bose-Hubbard model for spin-1 atoms in a
cubic optical lattice. Afterwards, Sec.~III shows that, already in the atomic limit, when the hopping
of bosons is neglected, a quite complicated phase diagram of different Mott phases emerges. Then, we
add in Sec.~IV site- and spin-dependent sources to the Hamiltonian in order to deal with the system
inherent spontaneous symmetry breaking and determine the grand-canonical free energy in first order of
the hopping parameter and in fourth order of the symmetry-breaking currents. In Sec.~V, we then introduce
the corresponding spin-dependent order parameters via a Legendre transformation with respect to the currents
and calculate the resulting hopping expansion of the effective action up to first order. With this we
study in Sec.~VI the quantum phase transition between the superfluid phase and the Mott insulator. In
Sec.~VII we determine the range of validity of the Ginzburg-Landau theory, which turns out to be limited
due to a sharp increase of the condensate density in the superfluid phase. Subsequently, we distinguish
in Sec.~VIII between various ferromagnetic and anti-ferromagnetic superfluid phases in view of a competition
between an anti-ferromagnetic interaction with a non-vanishing external magnetic field within the determined
range of validity of the Ginzburg-Landau theory. In Sec.~IX we finally find that the superfluid-Mott
insulator phase transition is of second order and that the transitions between different superfluid phases
with anti-ferromagnetic interaction can be both of first and second order in the presence of the Zeeman
effect.

\section{Spinor BOSE-HUBBARD MODEL}

In order to derive the underlying spinor Bose-Hubbard model model, we start from the second quantized
Hamiltonian for a spin-1 Bose gas in the grand-canonical ensemble \cite{key-5,key-29,key-4,key-60-1,key-15}:
\begin{eqnarray}
\hspace{-2.5cm}\hat{H}_{\rm{BH}}=\sum_{\alpha}\int d^{3}x\hat{\Psi}_{\alpha}^{\dagger}(\mathbf{x})\biggl[-\frac{\hbar^{2}}{2M}\nabla^{2}+V(\mathbf{x})
-\mu\biggr]\hat{\Psi}_{\alpha}(\mathbf{x})-\eta\sum_{\alpha,\beta}\int d^{3}x\hat{\Psi}_{\alpha}^{\dagger}(\mathbf{x})F_{\alpha\beta}^{z}\hat{\Psi}_{\beta}
(\mathbf{x}) \nonumber \\
\hspace{-1cm}+\frac{c_{0}}{2}\sum_{\alpha,\beta}\int d^{3}x\hat{\Psi}_{\alpha}^{\dagger}(\mathbf{x})\Psi_{\beta}^{\dagger}(\mathbf{x})\hat{\Psi}_{\beta}(\mathbf{x})
\hat{\Psi}_{\alpha}(\mathbf{x})\nonumber\\
\hspace{-1cm}+\frac{c_{2}}{2}\sum_{\alpha,\beta,\gamma,\delta}\int d^{3}x\hat{\Psi}_{\alpha}^{\dagger}(\mathbf{x})\Psi_{\gamma}^{\dagger}(\mathbf{x})\mathbf{F_{\alpha\beta}
\cdot F_{\gamma\delta}}\hat{\Psi}_{\delta}(\mathbf{x})\hat{\Psi}_{\beta}(\mathbf{x}).\label{eq:1}
\end{eqnarray}
Here $\mu$ is the chemical potential, $\eta$ is an additional parameter which can be interpreted for
the time being as an external magnetic field, and $M$ is the mass of the atom. Furthermore, $V(\mathbf{x})=V_{0}\sum_{\nu=1}^{3}\sin^{2}(k_{L}x_{\nu})$
is a periodic potential of a $3$-dimensional cubic optical lattice with a lattice period $a=\pi/k_{L}$
where $k_{L}=2\pi/\lambda$ is the wave vector of the laser beam and the lattice depth is described by
$V_{0}$ which is measured in units of the recoil energy $E_{R}=\hbar^{2}k_{L}^{2}/2M$. Because of the
bosonic nature of the particles, the field operators fullfill the standard commutator relations:
\begin{eqnarray}
\Bigl[\hat{\Psi}_{\alpha}(\mathbf{x}),\hat{\Psi}_{\beta}(\mathbf{x}^{\prime})\Bigr]=0,\;\Bigl[\hat{\Psi}_{\alpha}^{\dagger}(\mathbf{x}),\hat{\Psi}_{\beta}^{\dagger}(\mathbf{x}^{\prime})\Bigr]=0,
\nonumber \\
\Bigl[\hat{\Psi}_{\alpha}(\mathbf{x}),\hat{\Psi}_{\beta}^{\dagger}(\mathbf{x}^{\prime})\Bigr]=\delta_{\alpha,\beta}\delta(\mathbf{x}-\mathbf{x}^{\prime}).\label{eq:2-1}
\end{eqnarray}
Moreover, $\mathbf{F_{\alpha\beta}}$ are the following spin-1 matrices

\begin{eqnarray}
F_{x}=\frac{1}{\sqrt{2}}
\left( \begin{array}{ccc}
0 & 1 & 0\\
1 & 0 & 1\\
0 & 1 & 0 
\end{array} \right)\qquad,\qquad
F_{y}=\frac{\mathit{i}}{\sqrt{2}}\left(\begin{array}{ccc}
0 & -1 & 0\\
1 & 0 & -1\\
0 & 1 & 0
\end{array}\right)\qquad,\nonumber \\
F_{z}=\left(\begin{array}{ccc}
0 & 1 & 0\\
1 & 0 & 1\\
0 & 1 & 0
\end{array}\right).
\end{eqnarray}\\

The first term in (\ref{eq:1}) results from the one-particle Hamiltonian without a magnetic field, the second one represents
the linear Zeeman effect, the third one the spin-independent interaction and the last one describes the
spin-dependent interaction. The interaction strengths $c_{0}$ and $c_{2}$ can be defined as $c_{0}= 4\pi\hbar^{2}(a_{0}+2a_{2})/3M,\,\,
c_{2}=4\pi\hbar^{2}(a_{2}-a_{0})/3M$,
where $a_{0}$ and $a_{2}$ are the $s$-wave scattering lengths with total angular momenta 0 and 2 since
the total spin $F=1$ is forbidden due to the bosonic symmetry \cite{key-42-1}. The spin-dependent interaction
is ferromagnetic (anti-ferromagnetic) when $c_{2}<0$, i.e., $a_{2}<a_{0}$ ($c_{2}>0$, i.e., $a_{2}>a_{0}$).
In the case of $^{23}$Na atoms the interaction is anti-ferromagnetic as its scattering lengths are $a_{0}=(46\pm5)a{}_{B}$
and $a{}_{2}=(52\pm5)a{}_{B}$, where $a_{B}$ is the Bohr radius \cite{key-3-1}. For $^{87}$Rb, we
have instead $a_{0}=(110\pm4)a_{B}$ and $a_{2}=(107\pm4)a_{B}$, so the interaction is ferromagnetic
\cite{key-5}. In a periodic potential Bloch wave functions are the energy eigenstates of a single atom
with fixed wave vectors. Via a Fourier transformation these states can be converted to Wannier functions
which are localized on the respective lattice sites through the tight-binding limit \cite{key-119}.
We can expand a field operator with respect to the Wannier functions of the lowest energy band for low
enough temperatures as then the energy gap $E_{\rm{gap}}$ between the first and the second band
is much larger than $k_{\rm{B}}T$: 

\begin{eqnarray}
\hat{\Psi}_{\alpha}(\mathbf{x})  =\sum_{i}\hat{a}_{i\alpha}w(\mathbf{x}-\mathbf{x}_{i})\,\,\,,\,\,\,
\hat{\Psi}_{\alpha}^{\dagger}(\mathbf{x})  =\sum_{i}\hat{a}_{i\alpha}^{\dagger}w^{*}(\mathbf{x}-\mathbf{x}_{i}),\label{eq:2}
\end{eqnarray}
where $\hat{a}_{i\alpha}^{\dagger}$ $\left(\hat{a}_{i\alpha}\right)$ is the creation (annihilation)
operator for an atom at site $i$ with hyperfine spin $\alpha$. Using the orthonormality conditions
of the Wannier functions, we obtain from (\ref{eq:2-1}) the commutation relations for the lattice operator
\begin{eqnarray}
\Bigl[\hat{a}_{i\alpha},\hat{a}_{j\beta}\Bigr]=0,\quad \Bigl[\hat{a}_{i\alpha}^{\dagger},\hat{a}_{j\beta}^{\dagger}\Bigr]=0,
\quad \Bigl[\hat{a}_{i\alpha},\hat{a}_{j\beta}^{\dagger}\Bigr]=\delta_{\alpha,\beta}\delta_{i,j}.\label{eq:5}
\end{eqnarray}
Inserting Eq.~(\ref{eq:2}) into (\ref{eq:1}), and using the approximation that the overlap of Wannier
functions at different sites can be neglected for a deep enough lattice potential, the Bose-Hubbard model
for spin-1 bosons in a cubic optical lattices becomes

\begin{eqnarray}
\hat{H}_{\rm{BH}}=\sum_{i}\Biggl[\frac{U_{0}}{2}\sum_{\alpha,\beta}\hat{a}_{i\alpha}^{\dagger}\hat{a}_{i\beta}^{\dagger}\hat{a}_{i\alpha}\hat{a}_{i\beta}+\frac{U_{2}}{2}
\sum_{\alpha,\beta,\gamma,\delta}
\hat{a}_{i\alpha}^{\dagger}\hat{a}_{i\gamma}^{\dagger}\mathbf{F_{\alpha\beta}\cdot F_{\gamma\delta}}\hat{a}_{i\delta}\hat{a}_{i\beta}\nonumber \\
-\mu\sum_{\alpha}\hat{a}_{i\alpha}^{\dagger}\hat{a}_{i\alpha}-\eta\sum_{\alpha,\beta}\hat{a}_{i\alpha}^{\dagger}
F_{\alpha\beta}^{z}\hat{a}_{i\beta}\Biggr]-J\sum_{<i,j>}\sum_{\alpha}\hat{a}_{i\alpha}^{\dagger}\hat{a}_{j\alpha}.\label{eq:9-1}
\end{eqnarray}
 Here \emph{$<i,j>$} describes a summation over all sets of  nearest neighbor sites. The corresponding
hopping matrix element is given by 
\begin{eqnarray}
J=-\int d^{3}x\; w^{*}(\mathbf{x}-\mathbf{x}_{i})\left[-\frac{\hbar^{2}\nabla^{2}}{2M}+V(\mathbf{x})\right]w(\mathbf{x}-\mathbf{x}_{j})\label{eq:1-2-3}
\end{eqnarray}
and turns out to be independent of the spatial dimension. Note that we can drop the site indices due
to translational invariance. Furthermore, \emph{$U_{0}$} and \emph{$U_{2}$} are the on-site spin-independent
and the on-site spin-dependent interaction, respectively: 
\begin{eqnarray}
U_{0,2} & = & c_{0,2}\int d^{3}x\bigl|w(\mathbf{x}-\mathbf{x}_{i})\bigr|^{4}.\label{eq:14}
\end{eqnarray}
Therefore, we have a ferromagnetic (anti-ferromagnetic) interaction when $U_{2}<0$ ($U_{2}>0$). Note
that we have neglected in (\ref{eq:9-1}) a physically irrelevant energy shift which is of the form of
the right-hand side of Eq.~(\ref{eq:1-2-3}) with $i=j$. 

We define the spin operator $\hat{\mathbf{\mathbf{S}}}_{i}=\sum_{\alpha,\beta}\hat{a}_{i\alpha}^{\dagger}\mathbf{\,\mathbf{F}}_{\alpha\beta}\,\hat{a}_{i\beta}$,
the number operator for each spin component $\hat{n}_{i\alpha}=\hat{a}_{i\alpha}^{\dagger}\hat{a}_{i\alpha},\,\,\,\rm{and}$
the total atom number operator $\hat{n}_{i}=\sum_{\alpha}\hat{n}_{i\alpha}$. With this Eq.~(\ref{eq:9-1})
decomposes according to 
\begin{eqnarray}
\hat{H}_{\rm{BH}} & = & \hat{H}^{\left(0\right)}+\hat{H}^{\left(1\right)},\label{eq:8}
\end{eqnarray}
where $\hat{H}^{\left(0\right)}=\sum_{i}\hat{H}_{i}^{\left(0\right)}$ denotes the local part with 
\begin{eqnarray}
\hat{H}_{i}^{\left(0\right)}=  \sum_{i}\biggl[\frac{U_{0}}{2}\hat{n}_{i}(\hat{n}_{i}-1)+\frac{U_{2}}{2}(\hat{\mathbf{S}}_{i}^{2}-2\hat{n}_{i})
  -\mu\hat{n}_{i}-\eta\hat{S}_{iz}\biggr],\label{eq:13-1}
\end{eqnarray}
whereas the hopping represents the bilocal part
\begin{eqnarray}
\hat{H}^{\left(1\right)} & =-J\sum_{<i,j>}\sum_{\alpha}\hat{a}_{i\alpha}^{\dagger}\hat{a}_{j\alpha}.\label{eq:10}
\end{eqnarray}
In order to show that the operator $\hat{\mathbf{\mathbf{S}}}$ behaves like an angular momentum or spin
operator, we write down explicitly each component of the spin operator
\begin{eqnarray}
\hat{S}_{ix}= & \frac{1}{\sqrt{2}}(\hat{a}_{i1}^{\dagger}\hat{a}_{i0}+\hat{a}_{i0}^{\dagger}\hat{a}_{i1}+\hat{a}_{i0}^{\dagger}\hat{a}_{i-1}+\hat{a}_{i-1}^{\dagger}\hat{a}_{i0}),\nonumber \\
\hat{S}_{iy}= & \frac{i}{\sqrt{2}}(-\hat{a}_{i1}^{\dagger}\hat{a}_{i0}+\hat{a}_{i0}^{\dagger}\hat{a}_{i1}-\hat{a}_{i0}^{\dagger}\hat{a}_{i-1}+\hat{a}_{i-1}^{\dagger}\hat{a}_{i0}),\nonumber \\
\hat{S}_{\mathit{iz}}= & \hat{n}_{i1}-\hat{n}_{i-1}.\label{eq:12}
\end{eqnarray}
 With this and (\ref{eq:5}) one can determine that the operators $\hat{S}_{i\sigma}$with $\sigma=x,y,z$
obey the usual angular momentum commutation relation $\left[\hat{S}_{i},\hat{S}_{j}\right]=i\sum_{k}\epsilon_{ijk}\hat{S}_{k}$.
Using Eq.~(\ref{eq:12}) we get furthermore 

\begin{eqnarray}
\hspace{-2.5cm}\hat{\mathbf{S}}_{i}^{2}=2\hat{n}_{i1}\hat{n}_{i0}+2\hat{n}_{i0}\hat{n}_{i-1}+\hat{n}_{i-1}^{2}+2\hat{n}_{i0}+\hat{n}_{i-1}+\hat{n}_{i1}^{2}
-2\hat{n}_{i1}\hat{n}_{i-1}+\hat{n}_{i1}+2\hat{a}_{i1}^{\dagger}\hat{a}_{i-1}^{\dagger}\hat{a}_{i0}^{2}
\nonumber \\
+2\hat{a}_{i0}^{\dagger}\hat{a}_{i0}^{\dagger}\hat{a}_{i1}\hat{a}_{i-1}.
\end{eqnarray}
All these relations turn out to be useful in the subsequent section for studying the system properties
in the atomic limit, i.e. $J=0$, at zero temperature. 

\begin{figure}[t!]
\centering{}\includegraphics[width=7.5cm]{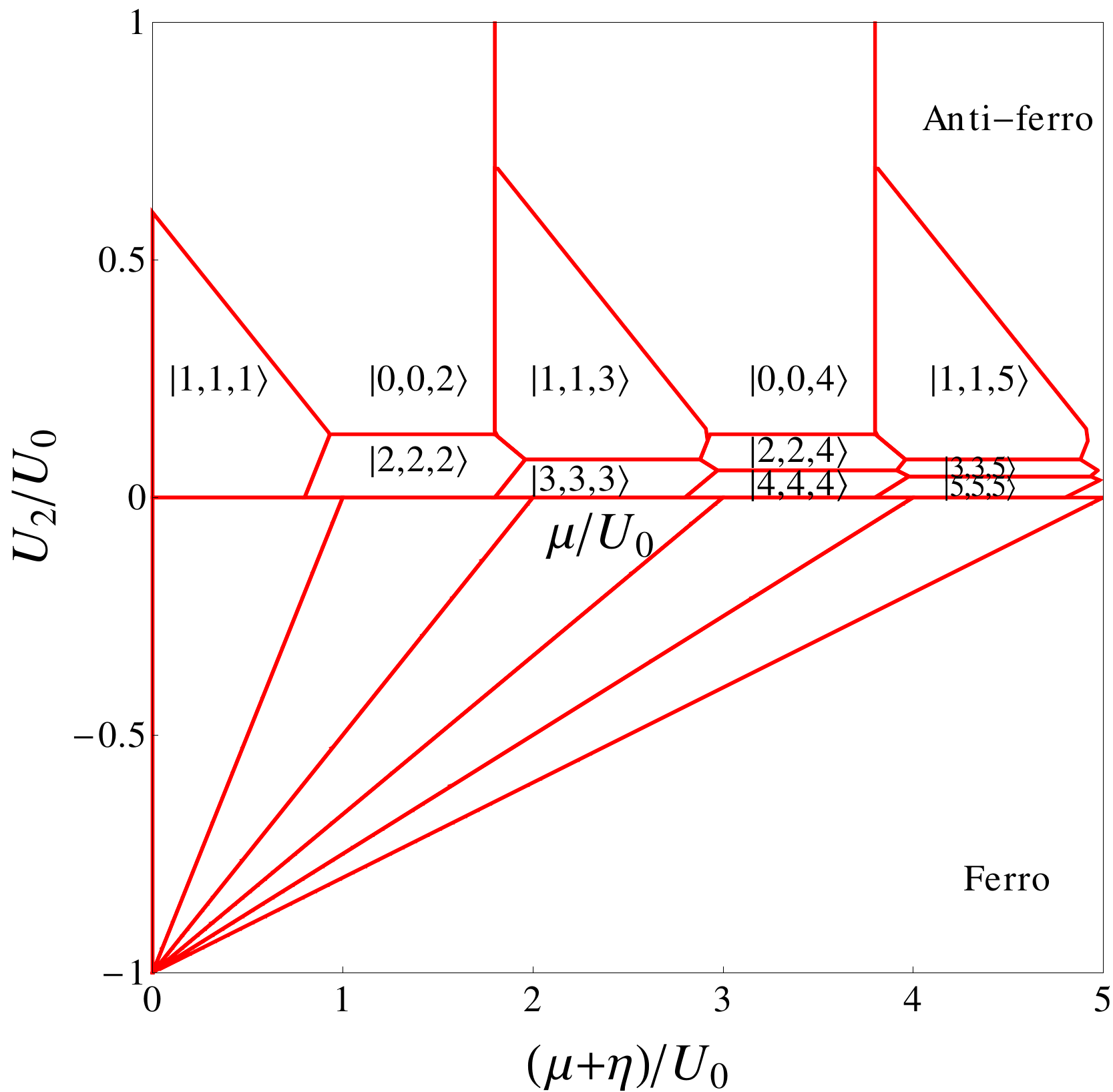}
\caption{\label{fig:Phase-diagram-of}Phase diagram of spinor $F$ = 1 Bose-Hubbard model for a magnetized system
with $\eta=0.2\, U_{0}$ with no hopping at zero temperature. The $x$-axis in the anti-ferromagnetic
case $\left(U_{2}>0\right)$ is the chemical potential, whereas in the ferromagnetic case $\left(U_{2}<0\right)$
the chemical potential is shifted according to (\ref{eq:2-68}).}
\end{figure}

\section{Atomic limit\label{sec:System-Properties-With}}

In the atomic limit the Bose-Hubbard Hamiltonian (\ref{eq:9-1}) reduces to a sum of single-site Hamiltonians
(\ref{eq:13-1}). Since the operators $\hat{\mathbf{S}}_{i}^{2}$, $\hat{S}_{iz}$ and $\hat{n}_{i}$
commute with each other, their eigenvalue problems are solved by the same eigenvectors: $\hat{\mathbf{S}}_{i}^{2}\left|\mathit{S}_{i},m_{i},n_{i}\right\rangle =\mathit{S}_{i}(\mathit{S}_{i}+1)\left|S_{i},m_{i},n_{i}\right\rangle ,$
$\hat{S}_{iz}\left|S_{i},m_{i},n_{i}\right\rangle =m_{i}\left|S_{i},m_{i},n_{i}\right\rangle $ and $\hat{n}_{i}\left|S_{i},m_{i},n_{i}\right\rangle =n_{i}\left|S_{i},m_{i},n_{i}\right\rangle $,
where $S_{i}+n_{i}=$ even \cite{key-42-1,key-29,key-4}. Thus, the eigenvalue problem of the local Hamiltonian
(\ref{eq:13-1}) is given by 
\begin{eqnarray}
\hat{H}_{i}^{\left(0\right)}\left|S_{i},m_{i},n_{i}\right\rangle  & =E_{S_{i},m_{i},n_{i}}^{\left(0\right)}\left|S_{i},m_{i},n_{i}\right\rangle ,
\end{eqnarray}
 where the respective energy eigenvalues are defined as 
\begin{eqnarray}
E_{S_{i},m_{i},n_{i}}^{\left(0\right)}= & \frac{U_{0}}{2}n_{i}(n_{i}-1)+\frac{U_{2}}{2}\bigl[S_{i}(\mathit{S}_{i}+1)-2n_{i}\bigr] & -\mu n_{i}-\eta m_{i}.\label{eq:3}
\end{eqnarray}

Now we go beyond Refs.~\cite{key-60-1,key-16,key-134,key-10-1,key-190,key-15} by considering a system
with an external magnetic field $\eta>0$ at zero temperature $T=0$ and no hopping $J=0$. In that case,
the degeneracy is lifted and the ground state of the Hamiltonian (\ref{eq:13-1}) depends on the concrete
values of the spin-independent interaction $U_{0}$, the spin-dependent interaction $U_{2}$, the chemical
potential $\mu$, and the external magnetic field $\eta$. In addition, the lowest energy state for given
$n_{i}$ and $S_{i}$ is denoted by $\left|S_{i},S_{i},n_{i}\right\rangle $ both for ferromagnetic and
anti-ferromagnetic interactions.

For the following discussion it turns out to be important to determine the degeneracy when two states
have the same energy with equal particle number but different total spin \cite{key-15,key-16}. In order
to define these degeneracy points we put 
\begin{equation}
E_{S_{i},S_{i},n_{i}}^{(0)}=E_{S_{i}+2,S_{i}+2,n_{i}}^{(0)},\label{eq:3-113}
\end{equation}
and, substituting (\ref{eq:3}) into (\ref{eq:3-113}), we get
\begin{eqnarray}
U_{2}^{\mathrm{crit}}=\eta/ & \left(S_{i}+\frac{3}{2}\right).\label{eq:54}
\end{eqnarray}
After having determined how the critical spin-dependent interaction $U_{2}^{\mathrm{crit}}$ depends
on the external magnetic field $\eta$, the calculation of the respective ground state yields the following
results.

\textbf{For ferromagnetic} \textbf{interaction}, i.e. $U_{2}<0$, there is no difference between the
ground state with and without magnetization because all spins are aligned. Thus, the ground state is
given by $\left|n_{i},n_{i},n_{i}\right\rangle $. In addition, the particle number $n_{i}$ is then
defined from the condition
\begin{eqnarray}
E_{n_{i}-1,n_{i}-1,n_{i}-1}^{\left(0\right)}<E_{n_{i},n_{i},n_{i}}^{\left(0\right)} & <E_{n_{i}+1,n_{i}+1,n_{i}+1}^{\left(0\right)},\label{eq:20-1}
\end{eqnarray}
where the ground-state energy $E_{n_{i},n_{i},n_{i}}^{\left(0\right)}$ is compared with the respective
ground-state energies $E_{n_{i}-1,n_{i}-1,n_{i}-1}^{\left(0\right)}$ and $E_{n_{i}+1,n_{i}+1,n_{i}+1}^{\left(0\right)}$
of the neighboring Mott states. Inserting (\ref{eq:3}) into (\ref{eq:20-1}) yields 
\begin{equation}
\Bigl(1+\frac{U_{2}}{U_{0}}\Bigr)(n_{i}-1)<\frac{\mu+\eta}{U_{0}}<n_{i}\Bigl(1+\frac{U_{2}}{U_{0}}\Bigr).\label{eq:2-64}
\end{equation}
By redefining the chemical potential according to 
\begin{eqnarray}
\mu+\eta & \rightarrow\mu,\label{eq:2-68}
\end{eqnarray}
 Eq.~(\ref{eq:2-64}) reduces to 

\begin{equation}
(n_{i}-1)\Bigl(1+\frac{U_{2}}{U_{0}}\Bigr)<\frac{\mu}{U_{0}}<n_{i}\Bigl(1+\frac{U_{2}}{U_{0}}\Bigr),\label{eq:2-64-1}
\end{equation}
which coincides with the unmagnetized result \cite{key-60-1,key-16,key-134,key-10-1,key-190}.

\textbf{For an} \textbf{anti-ferromagnetic interaction}, i.e. $U_{2}>0$, the situation becomes more
complicated. To this end it turns out that we have to consider in total the following four cases for
the ground state as the neighboring ground states change with varying $U_{2}$ and $\eta$.\emph{}\\
The first case\emph{ }is

\begin{eqnarray}
E_{S_{i}-1,S_{i}-1,n_{i}-1}^{\left(0\right)}<E_{S_{i},S_{i},n_{i}}^{\left(0\right)} & <E_{S_{i}+1,S_{i}+1,n_{i}+1}^{\left(0\right)},
\end{eqnarray}
which yields with (\ref{eq:3})

\begin{eqnarray}
n_{i}-1+\left(S_{i}-1\right)\frac{U_{2}}{U_{0}}-\frac{\eta}{U_{0}}<\frac{\mu}{U_{0}} & <n_{i}+S_{i}\frac{U_{2}}{U_{0}}-\frac{\eta}{U_{0}},\label{eq:2-71}
\end{eqnarray}
whereas the second case

\begin{eqnarray}
E_{S_{i}-1,S_{i}-1,n_{i}-1}^{\left(0\right)}<E_{S_{i},S_{i},n_{i}}^{\left(0\right)} & <E_{S_{i}-1,S_{i}-1,n_{i}+1}^{\left(0\right)},
\end{eqnarray}
 becomes
\begin{eqnarray}
n_{i}-1+\left(S_{i}-1\right)\frac{U_{2}}{U_{0}}-\frac{\eta}{U_{0}}<\frac{\mu}{U_{0}} & <n_{i}-\left(S_{i}+1\right)\frac{U_{2}}{U_{0}}+\frac{\eta}{U_{0}}.\label{eq:2-73}
\end{eqnarray}
The third case is

\begin{eqnarray}
E_{S_{i}+1,S_{i}+1,n_{i}-1}^{\left(0\right)}<E_{S_{i},S_{i},n_{i}}^{\left(0\right)} & <E_{S_{i}+1,S_{i}+1,n_{i}+1}^{\left(0\right)},
\end{eqnarray}
which reduces to 
\begin{eqnarray}
1-n_{i}+\left(S_{i}+2\right)\frac{U_{2}}{U_{0}}-\frac{\eta}{U_{0}}<\frac{\mu}{U_{0}} & <n_{i}+S_{i}\frac{U_{2}}{U_{0}}-\frac{\eta}{U_{0}}\label{eq:27}
\end{eqnarray}
and the fourth case

\begin{eqnarray}
E_{S_{i},S_{i},n_{i}-2}^{\left(0\right)}<E_{S_{i},S_{i},n_{i}}^{\left(0\right)} & <E_{S_{i},S_{i},n_{i}+2}^{\left(0\right)}
\end{eqnarray}
yields with (\ref{eq:3}) 
\begin{eqnarray}
\frac{1}{2}\left(2n_{i}-3-2\frac{U_{2}}{U_{0}}\right)<\frac{\mu}{U_{0}} & <\frac{1}{2}\left(1+2n_{i}-2\frac{U_{2}}{U_{0}}\right).
\end{eqnarray}

\Fref{fig:Phase-diagram-of} shows the resulting zero hopping phase diagram of the spin $F=1$
Bose-Hubbard model for a magnetized system at zero temperature for a fixed external magnetic field $\eta$.
Note that in the anti-ferromagnetic case $\left(U_{2}>0\right)$ the $x$-axis is the chemical potential
$\mu$, whereas in the ferromagnetic case $\left(U_{2}<0\right)$ it is shifted by the external magnetic
field $\eta$ according to (\ref{eq:2-68}) for illustrative purposes.

In the case of anti-ferromagnetic interaction with $0<U_{2}/U_{0}<0.5+\eta/U_{0}$ only the first three
cases can occur. At first, we remark that the right boundary of the even lobes occurs for a fixed chemical
potential $\mu=3.8\, U_{0}$ when $U_{2}\geq U_{2\,\rm{even}}^{(3)}=2\eta/3$, where the ground
state for the even lobes is $\left|0,0,n\right\rangle $ which is known as the spin-singlet insulator
\cite{key-29}. When $U_{2}\leq2\eta/3$ both  the spin $S$ and the magnetic quantum number $m$ of the odd
and the even lobes increase step by step by 2. For instance, the ground state for the fourth lobe successively
changes from $\left|0,0,4\right\rangle $ to $\left|4,4,4\right\rangle $ due to the respective critical
values of   $U_{2\,\rm{even}}^{(2)}=2\eta/3$ and
$U_{2\,\rm{even}}^{(3)}=2\eta/7$ where the ground state changes from $\left|0,0,4\right\rangle $
via $\left|2,2,4\right\rangle $ to $\left|4,4,4\right\rangle $ according  to the second case (\ref{eq:2-73})
as discussed above. Another one is the critical value $U_{2\rm{odd}}^{(2)}=2\eta/5$, where the ground
state changes from $\left|1,1,n_{i}\right\rangle $ to $\left|3,3,n_{i}\right\rangle $ which satisfies Eq.~(\ref{eq:27})
for odd lobes. The critical value $U_{2\rm{even}}^{(2)}=2\eta/9$ is finally a value for which the
ground state for the odd lobes becomes $\left|5,5,n_{i}\right\rangle $ which satisfies the first case (\ref{eq:2-71}). 

On the other hand, for $U_{2}/U_{0}\geq0.5+\eta/U_{0}$, the odd lobes vanish while the even lobes
continue. Furthermore, the boundaries for the even lobes occur for a fixed chemical potential $\mu=1.8\, U_{0}$
and $\mu=3.8\, U_{0}$. The reason is that the external magnetic field can not align the spins, so then
the fourth case occurs. Finally, we remark that the even and odd lobes shrink when $U_{2}=0$ as shown
in \fref{fig:Phase-diagram-of}. 

For ferromagnetic interaction, the even and odd lobes decrease with increasing $\left|U_{2}\right|$
and vanish when $U_{2}/U_{0}<-1$. Therefore, there occurs no difference between the ferromagnetic case
with or without magnetization which coincides with the results of Ref.~\cite{key-10-1,key-190,key-15},
because all spins are aligned in the same direction. 

\section{GRAND-CANONICAL FREE ENERGY}

In this article, we follow Ref.~\cite{key-120} and work out a field-theoretic approach to determine
the quantum phase boundary in terms of a Ginzburg-Landau theory where additional source currents are
added to the spin-1 Bose-Hubbard model in order to break the global $U(1)$ symmetry. Furthermore,  a
strong-coupling perturbation theory will be developed by taking into account diagrammatic rules. To this
end we determine a diagrammatic expansion of the grand-canonical free energy in fi{}rst order of the
hopping parameter and in fourth order of the symmetry-breaking currents.

\subsection{Perturbation Theory}

We start with generalizing the usual field-theoretic approach for describing classical phase transitions
\cite{key-8-1,key-9} to the realm of quantum phase transitions. Thus, we add to the Bose-Hubbard Hamiltonian
a term which couples artificial source currents $j_{i\alpha}(\tau),j_{i\alpha}^{*}(\tau)$ to the operators
$\hat{a}_{i\alpha}^{\dagger}$ and $\hat{a}_{i\alpha}$ in order to artificially break the underlying
$U(1)$ symmetry:
\begin{eqnarray}
\hat{H}_{\rm{BH}}(\tau)= & \hat{H}_{\rm{BH}}+\sum_{i}\sum_{\alpha}\left[j_{i\alpha}^{*}(\tau)\hat{a}_{i\alpha}(\tau)+j_{i\alpha}(\tau)\hat{a}_{i\alpha}^{\dagger}(\tau)\right].\label{eq:3-9-1-1}
\end{eqnarray}
This suggests the decomposition
\begin{eqnarray}
\hat{H}_{\rm{BH}}(\tau)= & \hat{H}^{\left(0\right)}+\hat{H}^{\left(1\right)}(\tau),
\end{eqnarray}
where the non-local term is given by 
\begin{eqnarray}
\hat{H}^{\left(1\right)}(\tau)=  -\sum_{ij}\sum_{\alpha}J_{ij}\,\hat{a}_{i\alpha}^{\dagger}\hat{a}_{j\alpha}+\sum_{i}\sum_{\alpha}\left[j_{i\alpha}^{*}(\tau)\hat{a}_{i\alpha}(\tau)
\right. +\left.j_{i\alpha}(\tau)\hat{a}_{i\alpha}^{\dagger}(\tau)\right].\label{eq:20}
\end{eqnarray}
In the following it turns out to be advantageous to introduce the generalized  hopping matrix element 
\begin{eqnarray}
J_{ij} =
\left\{
\begin{array}{ll}
J,  & \mbox{if }  i,j \,\mbox{are next neigbors} \\
0, & \mbox{otherwise } .
\end{array} \label{eq:33}
\right.
\end{eqnarray}

As these artificial currents depend on the imaginary-time variable, we need the time-dependent perturbation
theory to define the perturbative expression for the grand-canonical free energy. To this end we use
the imaginary-time Dirac interaction picture, which is related to the Schrödinger picture with the following
operators: 
\begin{eqnarray}
\hat{O}_{I} & =e^{\tau\hat{H}^{(0)}}\hat{O}e^{-\tau\hat{H}^{(0)}},
\end{eqnarray}
where we use $\hbar=1$ from now on. In order to get the time-evolution operator, we need to solve the
following equation 
\begin{eqnarray}
\frac{\partial}{\partial\tau} & \hat{U}_{I}(\tau,\tau_{0})=-\hat{H}_{I}^{(1)}(\tau)\hat{U}_{I}(\tau,\tau_{0}),\label{eq:3-11-2}
\end{eqnarray}
by using the initial condition
\begin{eqnarray}
\hat{U}_{I}(\tau_{0},\tau_{0}) & =1.\label{eq:3-15}
\end{eqnarray}
An iterative solution yields the the time-evolution operator in the form

\begin{eqnarray}
\hat{U}_{I}(\tau,\tau_{0})= & \hat{T}\exp\left[-\int_{\tau_{0}}^{\tau}d\tau^{\prime}\hat{H}_{I}(\tau^{\prime})\right],
\end{eqnarray}
where $\hat{T}$ is the imaginary-time ordering operator. Therefore, we obtain the generalized grand-canonical
partition function 
\begin{eqnarray}
\mathcal{Z} & =\mathcal{Z}^{(0)}\left\langle \hat{U}_{I}(\beta,0)\right\rangle ^{(0)}.\label{eq:4-155}
\end{eqnarray}
with the unperturbed partition function 
\begin{equation}
\mathcal{Z}^{(0)}=\mathrm{Tr}\biggl[\exp\left\{ -\beta\hat{H}^{(0)}\right\} \biggr]\,,\label{eq:3-10-1-1}
\end{equation}
with $\left\langle \bullet\right\rangle ^{(0)}=\mathrm{Tr}\biggl[\bullet\:\exp\left\{ -\beta\hat{H}^{(0)}\right\} \biggr]/\mathcal{Z}^{(0)}$
defining the thermal average definition with respect to the unperturbed system. Thus, the grand-canonical
partition function, which is a functional of the artificial currents $j_{i\alpha}(\tau)$ and $j_{i\alpha}^{*}(\tau)$,
can be rewritten as 
\begin{eqnarray}
\hspace{-2.7cm}\mathcal{Z}= \mathcal{Z}^{(0)}\Biggl[1+\sum_{n=1}^{\infty}(-1)^{n}\frac{1}{n!}\int_{0}^{\beta}d\tau_{1}\int_{0}^{\beta}d\tau_{2}\cdots\int_{0}^{\beta}d\tau_{n}
\left\langle \hat{T}\left[\hat{H}_{I}(\tau_{1})\hat{H}_{I}(\tau_{2})\cdots\hat{H}_{I}(\tau_{n})\right]\right\rangle ^{(0)}\Biggr].\label{eq:3-27}
\end{eqnarray}
From this follows the grand-canonical free energy via 
\begin{eqnarray}
\mathcal{F}\left[j,j^{*}\right] & =-\frac{1}{\beta}\ln\mathcal{Z}\left[j,j^{*}\right].
\end{eqnarray}

The respective perturbative contributions for $\mathcal{F}$ contain different orders of the hopping
matrix element $J$ and the currents $j$ and $j^{*}$ appearing in (\ref{eq:20}). As we work out a
Ginzburg-Landau theory, we restrict ourselves to the fourth order in the currents. Furthermore, we focus
on the leading non-trivial order in the hopping $J$ which is of first order. Therefore, the free energy
functional can be expressed in terms of imaginary-time integrals over sums of products of thermal Green
functions. The thermal averages in Eq.~(\ref{eq:3-27}) can be expressed in terms of $n$-particle Green
functions of the unperturbed system 
\begin{eqnarray}
\hspace{-2cm}G_{n}^{(0)}(i_{1}^{\prime}\alpha_{1}^{\prime},\tau_{1}^{\prime};\ldots;i_{n}^{\prime}\alpha_{n}^{\prime},\tau_{n}^{\prime}|i_{1}\alpha_{1},\tau_{1};\ldots;i_{n}\alpha_{n},\tau_{n})\label{eq:3-36-1}\\
\hspace{2cm}=\left\langle \hat{T}\left[\hat{a}_{i_{1}^{\prime}\alpha_{1}^{\prime}}^{\dagger}(\tau_{1}^{\prime})\hat{a}_{i_{1}\alpha_{1}}(\tau_{1})\ldots\hat{a}_{i_{n}^{\prime}\alpha_{n}^{\prime}}^{\dagger}(\tau_{n}^{\prime})\hat{a}_{i_{n}\alpha_{n}}(\tau_{n})\right]\right\rangle ^{(0)}.\nonumber 
\end{eqnarray}

\subsection{Cumulant Expansion}

In order to calculate the correlation functions in many-body theory, we usually use the Wick theorem
which allows to decompose the $n$-point correlation function (\ref{eq:3-36-1}) into sums of products
of one-point correlation functions \cite{key-11,key-12}. However, this theorem is not valid for the
considered system here because the unperturbed Bose-Hubbard Hamiltonian (\ref{eq:13-1}) contains terms
which are of fourth order in the creation and annihilation operators. Therefore, instead, we use the
linked cluster theorem \cite{key-13,key-43}, which states that the sum of all connected Green functions
is defined by the logarithm of the partition function. Thus, in order to get these Green functions we
perform functional derivatives with respect to the currents. We note that, according to $\hat{H}^{\left(0\right)}=\sum_{i}\hat{H}_{i}^{\left(0\right)}$,
the unperturbed Hamiltonian (\ref{eq:13-1}) decomposes into a sum over local contributions. For this
reason, the generating functional decomposes into a sum over local terms as

\begin{equation}
C_{0}^{(0)}\left[j,j^{*}\right]=\sum_{i}{}_{i}C_{0}^{(0)}\left[j,j^{*}\right]\label{eq:3-41-2-1-1}
\end{equation}
 with 
\begin{eqnarray}
\hspace{-2.5cm}_{i}C_{0}^{(0)}\left[j,j^{*}\right]=  \ln\left\langle \hat{T}\exp\left\{ -\int_{0}^{\beta}d\tau\sum_{i}\sum_{\alpha}\biggl[j_{i\alpha}(\tau)\hat{a}_{i\alpha}^{\dagger}(\tau)\right.\right.
  \left.+j_{i\alpha}^{*}(\tau)\hat{a}_{i\alpha}(\tau)\biggr]\Biggr\}\right\rangle ^{(0)}.\label{eq:3-41-1}
\end{eqnarray}
In order to obtain higher order cumulants, we calculate the functional derivatives with respect to the
symmetry breaking currents $j_{i\alpha}(\tau)$ and $j_{i\alpha}^{*}(\tau)$:
\begin{eqnarray}
\hspace{-2.5cm}C_{n}^{(0)}(\tau_{1}^{\prime},i_{1}^{\prime}\alpha_{1}^{\prime};\ldots;\tau_{n}^{\prime},i_{n}^{\prime}\alpha_{n}^{\prime}|\tau_{1},i_{1}\alpha_{1};\ldots;\tau_{n},i_{n}
\alpha_{n})\quad\label{eq:3-41-2}\nonumber\\
=\frac{\delta^{2n}C_{0}^{(0)}\left[j,j^{*}\right]}{\delta j_{i_{1}^{\prime}\alpha^{\prime}}(\tau^{\prime})\delta j_{i_{1}\alpha_{1}}^{*}(\tau)\ldots\delta j_{i_{n}^{\prime}
\alpha_{n}^{\prime}}(\tau_{n}^{\prime})\delta j_{i_{n}\alpha_{n}}^{*}(\tau_{n})}\Biggl|_{j=j^{*}=0}. 
\end{eqnarray}
From (\ref{eq:3-41-2-1-1}) and (\ref{eq:3-41-2}) we read off that the cumulants are local quantities
i.e., the $n$-th order cumulant is given by
\begin{eqnarray}
\hspace{-2.5cm}C_{n}^{(0)}(\tau_{1}^{\prime},i_{1}^{\prime}\alpha_{1}^{\prime};\ldots;\tau_{n}^{\prime},i_{n}^{\prime}\alpha_{n}^{\prime}|\tau_{1},i_{1}\alpha_{1};\ldots;\tau_{n},i_{n}\alpha_{n})\nonumber \\
\hspace{-2.5cm}=\:_{i_{1}}C_{n}^{(0)}(\tau_{1}^{\prime},\alpha_{1}^{\prime};\ldots;\tau_{n}^{\prime},\alpha_{n}^{\prime}|\tau_{1},\alpha_{1};\ldots;\tau_{n},\alpha_{n})
\delta_{i_{1},i_{2}}\cdots\delta_{i_{n-1},i_{n}}\delta_{i_{n},i_{1}^{\prime}}\:\delta_{i_{1}^{\prime},i_{2}^{\prime}}\cdots\delta_{i_{n-1}^{\prime},i_{n}^{\prime}}.\label{eq:3-43}
\end{eqnarray}
It is important to know that the cumulants represent the keystone for constructing the Green functions.
In order to see this, we calculate the unperturbed one- and the two-point Green functions with the above
formulas and obtain

\begin{eqnarray}
G_{1}^{(0)}(i_{1}\alpha_{1},\tau_{1}|i_{2}\alpha_{2},\tau_{2}) & =\delta_{i_{1},i_{2}}\:_{i_{1}}C_{1}^{(0)}(\tau_{1},\alpha_{1}|\tau_{2},\alpha_{2}),\label{eq:3-60}
\end{eqnarray}
and
\begin{eqnarray}
\hspace{-2.5cm}G_{2}^{(0)}(i_{1}\alpha_{1},\tau_{1};i_{2}\alpha_{2},\tau_{2}|i_{3}\alpha_{3},\tau_{3};i_{4}\alpha_{4},\tau_{4})
\nonumber\\
=\hspace{0.2cm}\delta_{i_{1},i_{3}}\delta_{i_{2},i_{4}}\delta_{i_{3},i_{4}}\:{}_{i_{1}}C_{1}^{(0)}(\tau_{1},\alpha_{1};\tau_{2},\alpha_{2}|\tau_{3},\alpha_{3};\tau_{4},\alpha_{4})\qquad\nonumber \\
\hspace{0.5cm}+\delta_{i_{1},i_{3}}\delta_{i_{2},i_{4}}\:{}_{i_{1}}C_{1}^{(0)}(\tau_{1},\alpha_{1}|\tau_{3},\alpha_{3})\:{}_{i_{2}}C_{1}^{(0)}(\tau_{2},\alpha_{2}|\tau_{4},\alpha_{4})\nonumber \\
\hspace{0.5cm}+\delta_{i_{1},i_{4}}\delta_{i_{2},i_{3}}\,{}_{i_{1}}C_{1}^{(0)}(\tau_{1},\alpha_{1}|\tau_{4},\alpha_{4})\:{}_{i_{2}}C_{1}^{(0)}(\tau_{2},\alpha_{2}|\tau_{3},\alpha_{3}). 
\label{eq:3-61}
\end{eqnarray}
In order to determine the respective cumulants from combining (\ref{eq:3-36-1})--(\ref{eq:3-61}), we
use for each lattice site the property 
\begin{eqnarray}
\hspace{-2.5cm}\hat{a}_{\alpha}^{\dagger}\left|S,m,n\right\rangle =M_{\alpha,S,m,n}\left|\mathit{S}+1,m+\alpha,n+1\right\rangle 
+N_{\alpha,S,m,n}\left|S-1,m+\alpha,n+1\right\rangle ,\label{eq:3-31-1}
\end{eqnarray}
\begin{eqnarray}
\hspace{-2.5cm}\hat{a}_{\alpha}\left|S,m,n\right\rangle =O_{\alpha,S,m,n}\left|S+1,m-\alpha,n-1\right\rangle 
+P_{\alpha,S,m,n}\left|S-1,m-\alpha,n-1\right\rangle ,\label{eq:3-32-1}
\end{eqnarray}
where $M_{\alpha,S,m,n}$, $N_{\alpha,S,m,n}$, $O_{\alpha,S,m,n}$ and $P_{\alpha,S,m,n}$ are recursively
defined matrix elements of the creation and annihilation operators \cite{key-4,key-10}, see also 
\ref{sec:Matrix-Elements}.

\subsection{Diagrammatic Representation}

We list now the diagrammatic rules which yield a much simpler calculation for the perturbative contributions
of the grand-canonical free energy with the cumulant decomposition of Green functions as follows \cite{key-23-1,key-10}:
\begin{enumerate}
\item At a lattice site a $n$-point cumulant is represented by a vertex with $n$ entering and $n$ leaving
lines.
\item Each line is labelled with both an imaginary-time and a spin index.
\item The currents  $j_{i\alpha}^{*}(\tau)$ $\left(j_{i\alpha}(\tau)\right)$  are described by entering (leaving)
lines.
\item Each line, which connects two vertices, is associated with a factor of the hopping matrix element $J$.
\item For a connected Green function of a given order draw all inequivalent connected diagrams.
\item Sum over all site and spin indices and integrate over all time variables.\label{rule6}
\end{enumerate}
Using this cumulant decomposition the grand-canonical free energy functional  is given by a
diagrammatic expansion up to first order in the hopping parameter and the fourth order in the symmetry-breaking
currents:

\begin{eqnarray} 
\label{eqax}
\hspace{-2.5cm}\mathcal{F}[j,j^{*}]=\mathcal{F}^{(0)}+ \, \raisebox{-6mm}{\includegraphics[width=25mm]{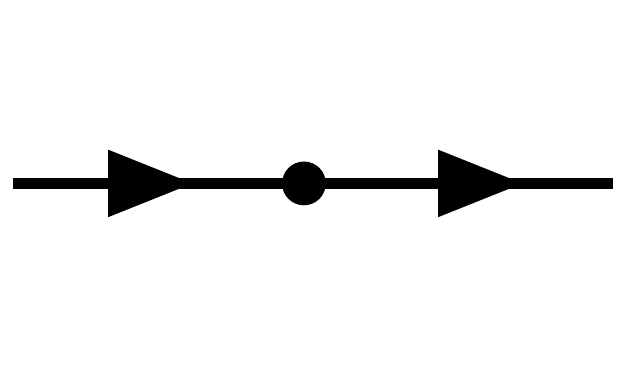}}\,+\, \raisebox{-4mm}{\includegraphics[width=30mm]{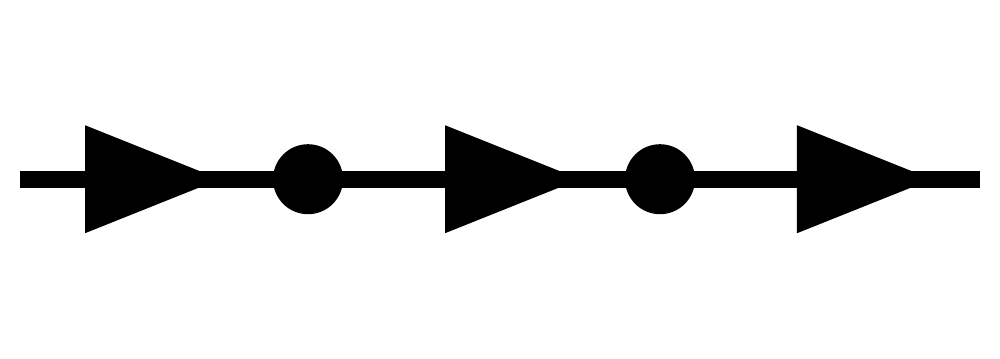}} \,+\frac{1}{4}\,
\raisebox{-3mm}{\includegraphics[width=25mm]{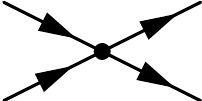}}\,\nonumber\\
+\frac{1}{2}\left(\, \raisebox{-3mm}{\includegraphics[width=25mm]{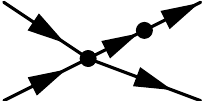}}+\, \raisebox{-5mm}{\includegraphics[width=24mm]{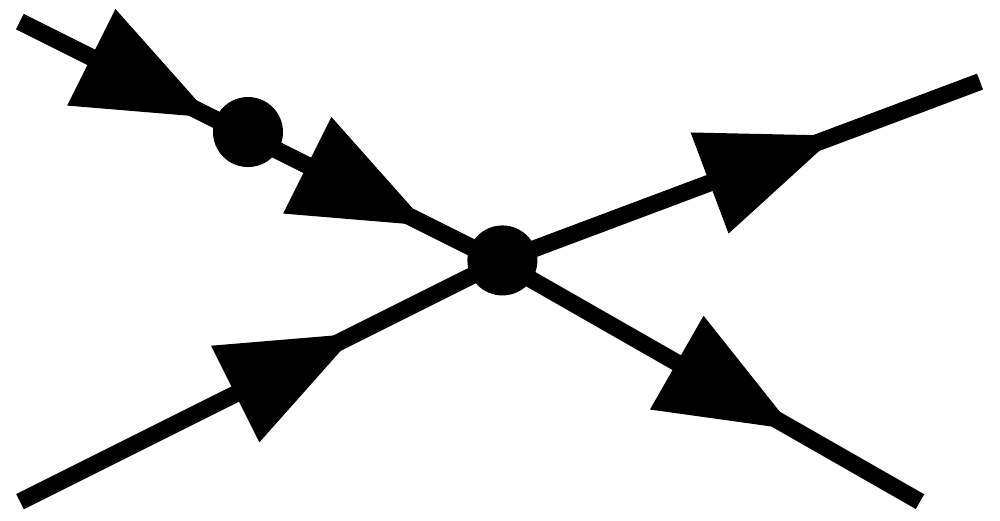}}\right). 
\end{eqnarray}
We remark that all imaginary time, spin and vertex indices can be dropped in order to indicate that
all variables have been integrated out as is demanded by rule \ref{rule6} and the pre-factors show the symmetry
factors of the respective diagrams. Converting the Feynman diagrams into explicit expressions, the grand-canonical
free energy (\ref{eqax}) reads

\begin{eqnarray}
\hspace{-2cm}\mathcal{F}\left[j,j^{*}\right]= \mathcal{F}_{0}-\frac{1}{\beta}\sum_{i}\sum_{\alpha_{1},\alpha_{2}}\int_{0}^{\beta}d\tau_{1}
\int_{0}^{\beta}d\tau_{2}\Biggl\{\: a_{2}^{(0)}(i\alpha_{1},\tau_{1}|i\alpha_{2},\tau_{2})j_{i\alpha_{1}}(\tau_{1})j_{i\alpha_{2}}^{*}(\tau_{2})\nonumber\\
\hspace{-2cm}+\sum_{j}J_{ij}a_{2}^{(1)}(i\alpha_{1},\tau_{1}|j\alpha_{2},\tau_{2})j_{i\alpha_{1}}(\tau_{1})j_{j\alpha_{2}}^{*}(\tau_{2})
  +\frac{1}{4}\sum_{\alpha_{3},\alpha_{4}}\int_{0}^{\beta}d\tau_{3}\int_{0}^{\beta}d\tau_{4}\nonumber \\
 \hspace{-2cm}\times \Biggl[j_{i\alpha_{1}}(\tau_{1})j_{i\alpha_{2}}(\tau_{2})j_{i\alpha_{3}}^{*}(\tau_{3})j_{i\alpha_{4}}^{*}(\tau_{4})a_{4}^{(0)}
 (i\alpha_{1},\tau_{1};i\alpha_{2},\tau_{2}|i\alpha_{3},\tau_{3};i\alpha_{4},\tau_{4})  +\frac{1}{2}\sum_{j}J_{ij}\nonumber \\
 \hspace{-2cm}\times \Bigl[a_{4}^{(1)}(i\alpha_{1},\tau_{1};i\alpha_{2},\tau_{2}|j\alpha_{3},\tau_{3};i\alpha_{4},\tau_{4})
 j_{i\alpha_{1}}(\tau_{1})j_{i\alpha_{2}}(\tau_{2})j_{j\alpha_{3}}^{*}(\tau_{3})j_{i\alpha_{4}}^{*}(\tau_{4})\nonumber \\
 \hspace{-2cm} +a_{4}^{(1)}(i\alpha_{1},\tau_{1};j\alpha_{2},\tau_{2}|i\alpha_{3},\tau_{3};i\alpha_{4},\tau_{4})j_{i\alpha_{1}}(\tau_{1})
 j_{j\alpha_{2}}(\tau_{2})j_{i\alpha_{3}}^{*}(\tau_{3})j_{i\alpha_{4}}^{*}(\tau_{4})\Bigr]\Biggr]\Biggr\},\label{eq:3-36-1-1}
\end{eqnarray}
where 
\begin{eqnarray}
\mathcal{F}_{0} & =-\frac{1}{\beta}\ln\mathcal{Z}^{(0)}
\end{eqnarray}
 is the grand-canonical free energy of the unperturbed system. Furthermore, we have introduced the abbreviations
\begin{eqnarray} 
\hspace{-2cm}a_2^{(0)}(i\alpha_{1},\tau_1|i\alpha_{2},\tau_2)=\tau_1,\alpha_1\, \raisebox{-5mm}{\includegraphics[width=20mm]{a2_1}\put(-30,22){\small {\it i}}}\ \tau_2,\alpha_2
={_iC}_{1}^{(0)}(\tau_1,\alpha_{1}|\tau_2,\alpha_{2}),\label{eqaa} 
\end{eqnarray} 

\begin{eqnarray} 
\hspace{-2cm}a_2^{(1)}(i\alpha_{1},\tau_1|j\alpha_{2},\tau_2)&=\tau_1,\alpha_1\ \raisebox{-3mm}{\includegraphics[width=20mm]{a211}\put(-40,18){\small {\it i}}\put(-20,18){\small {\it j}}}\ 
\tau_2,\alpha_2 \nonumber\\
&=\sum_{\alpha_3}\int_0^{\beta}{\mathrm{d}\tau_3 {_iC}_1^{(0)}(\tau_1,\alpha_{1}|\tau_3,\alpha_{3}){_jC}_1^{(0)}(\tau_3,\alpha_{3}|\tau_2,\alpha_{2})},
\label{eqaaa}\, 
\end{eqnarray} 

\begin{eqnarray} 
\hspace{-2cm}a_4^{(0)}(i\alpha_{1},\tau_1;i\alpha_{2},\tau_2|i\alpha_{3},\tau_3;i\alpha_{4},\tau_4)& = \hspace{1cm}\raisebox{-2mm}{\includegraphics[width=20mm]{a41}\put(-30,17)
{\small {\it i}}\put(-75,30){$\tau_2,\alpha_2$}\put(-75,-5){$\tau_1,\alpha_1$}\put(-5,30){$\tau_3,\alpha_3$}\put(-5,-7){$\tau_4,\alpha_4$}}
\nonumber\\
&={_iC}_2^{(0)}(\tau_1,\alpha_{1};\tau_2,\alpha_{2}|\tau_3,\alpha_{3};\tau_4,\alpha_{4}),\label{eqab}\,
\end{eqnarray} 

\begin{eqnarray} 
\hspace{-2cm}a_4^{(1)}(i\alpha_{1},\tau_1;i\alpha_{2},\tau_2|j\alpha_{3},\tau_3;i\alpha_{4},\tau_4)\label{eq6}=\hspace{0.5cm}\ \raisebox{-2mm}{\includegraphics[width=20mm]{a411}
\put(-35,17){\small {\it i}}\put(-20,25){\small {\it j}}\put(-75,30){$\tau_2,\alpha_2$}\put(-75,-5){$\tau_1,\alpha_1$}\put(-5,30){$\tau_3,\alpha_3$}\put(-5,-7){$\tau_4,\alpha_4$}}\nonumber\\
\hspace{-2cm}=\sum_{\alpha_5}\int_0^{\beta}{\mathrm{d}\tau_5}\,{ _iC}_2^{(0)}(\tau_1,\alpha_{1};\tau_2,\alpha_{2}|\tau_5,\alpha_5;\tau_4,\alpha_{4})\,{_jC}_1^{(0)}
(\tau_5,\alpha_5|\tau_3,\alpha_3).
\end{eqnarray} 
\begin{eqnarray} 
\hspace{-2cm}a_4^{(1)}(i\alpha_{1},\tau_1;j\alpha_{2},\tau_2|i\alpha_{3},\tau_3;i\alpha_{4},\tau_4)\label{eq6xx}=\hspace{0.5cm}\ \raisebox{-2mm}{\includegraphics[width=20mm]{a412.pdf}
\put(-30,20){\small {\it i}}\put(-42,27){\small {\it j}}\put(-75,32){$\tau_2,\alpha_2$}\put(-75,-6){$\tau_1,\alpha_1$}\put(-5,31){$\tau_3,\alpha_3$}\put(-5,-7){$\tau_4,\alpha_4$}}\nonumber\\
\hspace{-2cm}=\sum_{\alpha_5}\int_0^{\beta}{\mathrm{d}\tau_5}\,{ _iC}_2^{(0)}(\tau_1,\alpha_{1};\tau_5,\alpha_{5}|\tau_3,\alpha_3;\tau_4,\alpha_{4})\,{_jC}_1^{(0)}
(\tau_2,\alpha_2|\tau_5,\alpha_5).
\end{eqnarray}

\subsection{Matsubara Transformation}

We can simplify the calculation of these expressions by converting them into frequency space. Thus, we
use the Matsubara transformation where the imaginary-time variable runs from $0$ to $\beta$. The Matsubara
transformation is given by 

\begin{equation}
f(\omega_{m})=\frac{1}{\sqrt{\beta}}\int_{0}^{\beta}d\tau e^{i\omega_{m}\tau}f(\tau),\label{eq:3-50}
\end{equation}
where the Matsubara frequencies are defined according to 
\begin{equation}
\omega_{m}=\frac{2 \pi m}{\beta},\qquad m\in Z.\label{eq:4-196}
\end{equation}
The inverse Matsubara transformation yields 
\begin{equation}
f(\tau)=\frac{1}{\sqrt{\beta}}\sum_{m=-\infty}^{\infty}e^{-i\omega_{m}\tau}f(\omega_{m}).
\end{equation}
 Because of the locality of the cumulants and the conservation of frequency, the coefficient $a_{2}^{(0)}(i\alpha_{1},\omega_{m1}|i\alpha_{2},\omega_{m2})$
in Matsubara space is of the form: 
\begin{eqnarray}
a_{2}^{(0)}(i\alpha_{1},\omega_{m1}|i\alpha_{2},\omega_{m2}) & =a_{2}^{(0)}(i\alpha_{1},\omega_{m1})\delta_{\alpha_{1},\alpha_{2}}\delta_{\omega_{m1},\omega_{m2}}.
\end{eqnarray}
Using Eqs.~(\ref{eq:3-36-1}), (\ref{eq:3-60}), (\ref{eqax}), and (\ref{eq:3-50}), we obtain at first
\begin{eqnarray}
\hspace{-2.5cm}a_{2}^{(0)}(i\alpha_{1},\omega_{m1})=\frac{1}{\mathcal{Z}^{(0)}}\sum_{S_{i},m_{i},n_{i}}e^{-\beta E_{S_{i},m_{i},n_{i}}^{(0)}}
\left[\frac{M_{\alpha_{1},S_{i},m_{i},n_{i}}^{2}}{E_{S_{i}+1,m_{i}+\alpha_{1},n_{i}+1}^{(0)}-E_{S_{i},m_{i},n_{i}}^{(0)}-i\omega_{m1}}\right.\nonumber\\
\hspace{-2.5cm}+\frac{N_{\alpha_{1},S_{i},m_{i},n_{i}}^{2}}{E_{S_{i}-1,m_{i}+\alpha_{1},n_{i}+1}^{(0)}-E_{S_{i},m_{i},n_{i}}^{(0)}-i\omega_{m1}}
-\frac{O_{\alpha_{1},S_{i},m_{i},n_{i}}^{2}}{E_{S_{i},m_{i},n_{i}}^{(0)}-E_{S_{i}+1,m_{i}-\alpha_{1},n_{i}-1}^{(0)}-i\omega_{m1}}\nonumber \\
\hspace{-2.5cm}-\left.\frac{P_{\alpha_{1},S_{i},m_{i},n_{i}}^{2}}{E_{S_{i},m_{i},n_{i}}^{(0)}-E_{S_{i}-1,m_{i}-\alpha_{1},n_{i}-1}^{(0)}-i\omega_{m1}}\right].\label{eq:4-30}
\end{eqnarray}
In view of (\ref{eqaaa}), we use the cumulant multiplicity properties in frequency space and frequency
conservation, which leads to the relation
\begin{eqnarray}
\hspace{-2.5cm}a_{2}^{(1)}(i\alpha_{1},\omega_{m1}|j\alpha_{2},\omega_{m2})= a_{2}^{(0)}(i\alpha_{1},\omega_{m1})a_{2}^{(0)}(j\alpha_{2},\omega_{m2})
\delta_{\omega_{m1},\omega_{m2}}\delta_{\alpha_{1},\alpha_{2}}.
\end{eqnarray}
Similarly, using the conservation of frequency and spin index, we can derive the coefficient of fourth
order in the currents in Matsubara frequency as follows: 
\begin{eqnarray}
\hspace{-2.5cm}a_{4}^{(0)}(i\alpha_{1},\omega_{m1};i\alpha_{2},\omega_{m2}|i\alpha_{3},\omega_{m3};i\alpha_{4},\omega_{m4})=\frac{1}{\beta^{2}}\delta_{\alpha_{1}+\alpha_{2},\alpha_{3}+\alpha_{4}}
\delta_{\omega_{m1}+\omega_{m2},\omega_{m3}+\omega_{m4}}\nonumber \\
\hspace{-2.5cm}\left\{ \int_{0}^{\beta}d\tau_{1}\cdots d\tau_{4}
\left\langle \hat{T}\left[\hat{a}_{i\alpha_{1}}^{\dagger}(\tau_{1})\hat{a}_{i\alpha_{2}}^{\dagger}(\tau_{2})\hat{a}_{i\alpha_{3}}(\tau_{3})\hat{a}_{i\alpha_{4}}(\tau_{4})\right]\right\rangle \right.
 e^{-i(\omega_{m1}\tau_{1}+\omega_{m2}\tau_{2}-\omega_{m3}\tau_{3}-\omega_{m4}\tau_{4})}\nonumber \\
\hspace{-2.5cm} -a_{2}^{(0)}(i\alpha_{1},\omega_{m1}|i\alpha_{3},\omega_{m3}) a_{2}^{(0)}(i\alpha_{2},\omega_{m2}|i\alpha_{4},\omega_{m4})\biggl[\delta_{\alpha_{1},\alpha_{3}}\delta_{\alpha_{2},\alpha_{4}}\delta_{\omega_{m1},\omega_{m3}}
\delta_{\omega_{m2},\omega_{m4}}\nonumber \\
+\delta_{\alpha_{1},\alpha_{4}}\delta_{\alpha_{2},\alpha_{3}}\delta_{\omega_{m1},\omega_{m4}}\delta_{\omega_{m2},\omega_{m3}}
\biggr]\Biggr\}.\label{eq:4-5}
\end{eqnarray}
In   \ref{sec:fourth-order-coefficient} we present several details for the above calculation
because it is complicated and lengthy. The result for $a_{4}^{(0)}$ is displayed in (\ref{eq:4-6}).
The next quantity, which would have to be calculated, is $a_{4}^{(1)}$ according to  in Eqs.~(\ref{eq6}) and (\ref{eq6xx}).
However, it turns out in the next section  that $a_{4}^{(1)}$ will not appear in the effective action, so we do not have
to calculate it explicitly.

We remark that, in order to validate our results, we use the calculated grand-canonical free energy (\ref{eq:3-36-1-1})
to determine the mean-field result. To this end we apply the mean-field approximation to the Bose-Hubbard
Hamiltonian (\ref{eq:8})--(\ref{eq:10}), yielding  with (\ref{eq:33})
\begin{equation}
\hat{H}_{\rm{MF}}=\sum_{i}\left[\hat{H}_{i}^{\left(0\right)}+\hat{H}_{i\rm{MF}}^{(1)}\right],\label{eq:4-12}
\end{equation}
where the localized hopping term reads
\begin{equation}
\hat{H}_{i\rm{MF}}^{(1)}=-zJ\sum_{\alpha}\left(\Psi_{\alpha}\hat{a}_{i\alpha}^{\dagger}+\Psi_{\alpha}^{*}\hat{a}_{i\alpha}-\left|\Psi_{\alpha}\right|^{2}\right).\label{eq:4-13-2}
\end{equation}
By using the formal  identification

\begin{equation}
j_{i\alpha}(\tau)=-zJ\Psi_{\alpha},\label{eq:4-222}
\end{equation}
we obtain from (\ref{eq:3-36-1-1}) an expansion of the mean-field free energy $\mathcal{F}_{\rm{MF}}$
in powers of the order parameter which reads up to fourth order as follows:

\begin{eqnarray}
\hspace{-2.5cm}\mathcal{F}_{\rm{MF}}=\mathcal{F}_{0}-N_{s}\Biggl(\sum_{\alpha}a_{2}^{\rm{MF}}\left(\alpha,0\right)\left|\Psi_{\alpha}\right|^{2}+\sum_{\alpha_{1}}
\sum_{\alpha_{2}}\sum_{\alpha_{3}}\sum_{\alpha_{4}}\Psi_{\alpha_{1}}^{*}\Psi_{\alpha_{2}}^{*}\Psi_{\alpha_{3}}\Psi_{\alpha_{4}}\qquad\qquad\nonumber \\
\times a_{4}^{\rm{MF}}\left(\alpha_{1},0;\alpha_{2},0|\alpha_{3},0;\alpha_{4},0\right)\Biggr),\;\quad\label{eq:4-224}
\end{eqnarray}
where the respective mean-field Landau coefficients are only calculated up to the fourth hopping order:
\begin{equation}
a_{2}^{\rm{MF}}(\alpha,0)=a_{2}^{(0)}(\alpha,0)(zJ){}^{2}-zJ,
\end{equation}
\begin{eqnarray}
\hspace{-2cm}a_{4}^{\rm{MF}}(\alpha_{1},0;\alpha_{2},0|\alpha_{3},0;\alpha_{4},0)=\frac{\beta}{4}
 a_{4}^{(0)}(\alpha_{1},0;\alpha_{2},0|\alpha_{3},0;\alpha_{4},0)(zJ)^{4}.\label{eq:4-13}
\end{eqnarray}
Therefore, the mean-field result (\ref{eq:4-224}) can be determined by using (\ref{eq:4-30}) and (\ref{eq:4-6}).

\section{Ginzburg-Landau effective Action}

In this section, we follow Ref.~\cite{key-3-3,key-120} and deduce the Ginzburg-Landau action for the
spin-1 Bose-Hubbard model. To this end, we use a Legendre transformation to convert the artificially
introduced symmetry-breaking currents $j,j^{*}$ into the order parameter fields. In order to implement
this Legendre transformation in an uncluttered way, the grand-canonical free energy (\ref{eq:3-36-1-1})
can be written in Matsubara space as follows

\begin{eqnarray}
\hspace{-2.5cm}\mathcal{F}\left[j,j^{*}\right]=\mathcal{F}_{0}-\frac{1}{\beta}\sum_{i_{1},i_{2}}\sum_{\alpha_{1},\alpha_{2}}\sum_{\omega_{m1},\omega_{m2}}
\Biggl\{ M_{i_{_{1}}\alpha_{1},i_{2}\alpha_{2}}(\omega_{m1}|\omega_{m2})j_{i_{1}\alpha_{1}}(\omega_{m1})j_{i_{2}\alpha_{2}}^{*}(\omega_{m2})
+\sum_{i_{3},i_{4}}\sum_{\alpha_{3},\alpha_{4}}\nonumber \\
\hspace{-3cm}\sum_{\omega_{m3},\omega_{m4}}N_{i_{1}\alpha_{1},i_{2}\alpha_{2},i_{3}\alpha_{3},i_{4}\alpha_{4}}(\omega_{m1};\omega_{m2}|\omega_{m3};\omega_{m4})
 j_{i_{1}\alpha_{1},}(\omega_{m1})j_{i_{2}\alpha_{2}}(\omega_{m2})j_{i_{3}\alpha_{3}}^{*}(\omega_{m3})j_{i_{4}\alpha_{4}}^{*}(\omega_{m4})\Biggr\},\quad\label{eq:4-13-1}
\end{eqnarray}
where the respective coefficients are given by 

\begin{eqnarray}
\hspace{-2.5cm}M_{i_{1}\alpha_{1},i_{2}\alpha_{2}}(\omega_{m1}|\omega_{m2})=&\delta_{\omega_{m1},\omega_{m2}}\biggl[a_{2}^{(0)}(i_{1}\alpha_{1},\omega_{m1})\delta_{i_{1},i_{2}}
+J_{i_{1}i_{2}}\, a_{2}^{(0)}(i_{1}\alpha_{1},\omega_{m1})\nonumber \\
&\times a_{2}^{(0)}(i_{2}\alpha_{2},\omega_{m2})\biggr]\delta_{\alpha_{1},\alpha_{2}},
\end{eqnarray}
and
\begin{eqnarray}
\hspace{-2.5cm}N_{i_{1}\alpha_{1},i_{2}\alpha_{2},i_{3}\alpha_{3},i_{4}\alpha_{4}}(\omega_{m1};\omega_{m2}|\omega_{m3};\omega_{m4})=\frac{1}{4}
\delta_{\omega_{m1}+\omega_{m2},\omega_{m3}+\omega_{m4}}\delta_{\alpha_{1}+\alpha_{2},\alpha_{3}+\alpha_{4}}\nonumber\\
\hspace{-2.5cm}a_{4}^{(0)}(i_{1}\alpha_{1},\omega_{m1};i_{1}\alpha_{2},\omega_{m2}|i_{1}\alpha_{3},\omega_{m3};i_{1}\alpha_{4},\omega_{m4})
\Biggl\{\delta_{i_{1},i_{2}}\delta_{i_{2},i_{3}}\delta_{i{}_{3},i_{4}}+2\delta_{i_{1},i_{4}}\bigg[J_{i_{1}i_{2}}\nonumber\\
\hspace{-2.5cm}\times a_{2}^{(0)}(i_{2}\alpha_{2},\omega_{m2})\delta_{i_{1},i_{3}}\bigg.
+J_{i_{1}i_{3}}\left.a_{2}^{(0)}(i_{3}\alpha_{3},\omega_{m3})\delta_{i_{1},i_{2}}\right]\Biggr\}.
\label{eq:5-263}
\end{eqnarray}
Now, the order parameter field $\psi_{i\alpha}(\omega_{m})$ is defined as 
\begin{equation}
\Psi_{i\alpha}(\omega_{m})=\left\langle \hat{a}_{i\alpha}(\omega_{m})\right\rangle =\beta\frac{\delta\mathcal{F}}{\delta j_{i\alpha}^{*}(\omega_{m})}.\label{eq:5-218}
\end{equation}
Eq.~(\ref{eq:5-218}) motivates to perform a Legendre transformation, where the currents as the degrees
of freedom are converted to order parameter fields. Using Eq.~(\ref{eq:5-218}) the Ginzburg-Landau
action $\Gamma$ has the following form 
\begin{eqnarray}
\hspace{-2.5cm}\Gamma\left[\Psi_{i\alpha}(\omega_{m}),\Psi_{i\alpha}^{*}(\omega_{m})\right]=\mathcal{F}\left[j,j^{*}\right]-\frac{1}{\beta}\sum_{i}\sum_{\omega_{m}}\sum_{\alpha}
\biggl[\Psi_{i\alpha}(\omega_{m})j_{i\alpha}^{*}(\omega_{m})+\Psi_{i\alpha}^{*}(\omega_{m})j_{i\alpha}(\omega_{m})\biggr],\label{eq:4-20}
\end{eqnarray}
where $\Psi$, $\Psi^{*}$and $j^{*}$, $j$ are conjugate variables which satisfy the Legendre relations
\begin{eqnarray}
j_{i\alpha}(\omega_{m})=-\beta\frac{\delta\Gamma}{\delta\Psi_{i\alpha}^{*}(\omega_{m})},\; j_{i\alpha}^{*}(\omega_{m})=-\beta\frac{\delta\Gamma}{\delta\Psi_{i\alpha}(\omega_{m})}.\label{eq:5-267}
\end{eqnarray}
In order to recover the interesting physical situation, the artificially currents $j^{*}$, $j$ should
vanish. Therefore, we obtain from (\ref{eq:5-267}) the equations of motion as follows 
\begin{equation}
\left.\frac{\delta\Gamma}{\delta\Psi_{i\alpha}^{*}(\omega_{m})}\right|_{\Psi=\Psi_{\rm{eq}}}=0,\qquad\left.\frac{\delta\Gamma}{\delta\Psi_{i\alpha}(\omega_{m})}\right|_{\Psi=\Psi_{\rm{eq}}}=0.\label{eq:4-22}
\end{equation}
Hence, the effective action is stationary with respect to fluctuations around the equilibrium order parameter
field $\Psi_{\mathrm{eq}}$. Additionally, we read off from Eq.~(\ref{eq:4-20}) that the physical grand-canonical
free energy in the case of the vanishing currents $j^{*}$, $j$ is equal to evaluating the effective
action at the equilibrium order parameter field $\mathrm{\Psi_{eq}}$: 

\begin{equation}
\Gamma\left[\Psi=\Psi_{\mathrm{eq}},\Psi^{*}=\Psi_{\mathrm{eq}}\right]=\mathcal{F}\left[j^{*}=0,j=0\right]=\mathcal{F}.\label{eq:4-22-1}
\end{equation}
To determine the explicit form of the effective action as a functional of the order parameter, we have
to calculate the currents as a functionals of the Ginzburg-Landau order parameter field. At first, we
insert (\ref{eq:4-13-1}) in (\ref{eq:5-218}) and find that the order parameter field is given by
\begin{eqnarray}
\hspace{-2.5cm}\Psi_{i\alpha}(\omega_{m})=-\sum_{p}\sum_{\alpha_{1}}\sum_{\omega_{m1}}\biggl[M_{p\alpha,i\alpha_{1}}(\omega_{m1}|\omega_{m})j_{p\alpha}(\omega_{m1})
-2\sum_{i_{2},i_{3}}\sum_{\omega_{m2,}\omega_{m3}}\sum_{\alpha_{2},\alpha_{3}}\nonumber \\
\hspace{-2.5cm} \times N_{p\alpha,i_{2}\alpha_{2},i_{3}\alpha_{3},i_{1}\alpha_{1}}(\omega_{m1};\omega_{m2}|\omega_{m3};\omega_{m})
 j_{p\alpha}(\omega_{m1})j_{i_{2}\alpha_{2}}(\omega_{m2})j_{i_{3}\alpha_{3}}^{*}(\omega_{m3})\biggr].
\label{eq:4-15}
\end{eqnarray}
Afterwards, in order to invert relation (\ref{eq:4-15}) up to first order in the tunneling parameter
$J$, we calculate the inverse matrix of $M_{p\alpha,i\alpha_{1}}(\omega_{m1}|\omega_{m})$, yielding 

\begin{eqnarray}
\hspace{-2.5cm}M_{i_{1}\alpha_{1},i_{2}\alpha_{2}}^{-1}(\omega_{m1}|\omega_{m2})=\frac{\delta_{\alpha_{1},\alpha_{2}}\delta_{\omega_{m1},\omega_{m2}}}{a_{2}^{(0)}(i_{1}
\alpha_{1},\omega_{m1})}\Biggl[\delta_{i_{1},i_{2}}-J_{i_{1}i_{2}}\: a_{2}^{(0)}(i_{2}\alpha_{2},\omega_{m2})\Biggr].
\end{eqnarray}
Multiplying Eq.~(\ref{eq:4-15}) with the inverse matrix $M^{-1}$ then leads to

\begin{eqnarray}
\hspace{-2.5cm}j_{i\alpha}(\omega_{m})=-\sum_{p}\sum_{\alpha_{1}}\sum_{\omega_{m1}}M_{i_{1}\alpha_{1},p\alpha}^{-1}(\omega_{m}|\omega_{m1})\biggl\{\Psi_{p\alpha}(\omega_{m1})
-2\sum_{q,i_{2},i_{3}}\sum_{\omega_{m2,}\omega_{m3}}\sum_{\alpha_{2},\alpha_{3}}\nonumber \\
\hspace{-2.5cm}\times N_{q\alpha_{1},i_{2}\alpha_{2},i_{3}\alpha_{3},p\alpha}(\omega_{m1};\omega_{m2}|\omega_{m3};\omega_{m})t_{q\alpha_{1}}(\omega_{m1})t_{i_{2}
\alpha_{2}}(\omega_{m2})t_{i_{3}\alpha_{3}}^{*}(\omega_{m3})\biggr\},
\label{eq:4-15-1}
\end{eqnarray}
with the abbreviation 
\begin{eqnarray}
t_{i\alpha}(\omega_{m}) & =-\sum_{\alpha_{1}}\sum_{p,\omega_{m1}}M_{p\alpha_{1},i\alpha}^{-1}(\omega_{m1}|\omega_{m})\Psi_{p\alpha}(\omega_{m1}).
\end{eqnarray}
Inserting Eqs.~(\ref{eq:4-13-1}) and (\ref{eq:4-15-1}) into Eq.~(\ref{eq:4-20}) up to the first
order in the tunneling parameter, we finally get for the effective potential

\begin{eqnarray}
\hspace{-2.5cm}\Gamma\left[\Psi_{i\alpha}(\omega_{m}),\Psi_{i\alpha}^{*}(\omega_{m})\right]=\mathcal{F}_{0}+\frac{1}{\beta}\sum_{i}
\Biggl\{\sum_{\alpha}\sum_{\omega_{m}}\:\left[\frac{\left|\Psi_{i\alpha}(\omega_{m})\right|^{2}}{a_{2}^{(0)}(i\alpha,\omega_{m})}
-\sum_{j}J_{ij}\Psi_{i\alpha}(\omega_{m})\Psi_{j\alpha}^{*}(\omega_{m})\right]\label{eq:4-21}\nonumber\\
\hspace{-2.5cm}-\sum_{\alpha_{1},\alpha_{2},\alpha_{3},\alpha_{4}}\;\sum_{\omega_{m1},\omega_{m2},\omega_{m3},\omega_{m4}}
\frac{\Psi{}_{i\alpha_{1}}(\omega_{m1})\Psi_{i\alpha_{2}}
(\omega_{m2})\Psi_{i\alpha_{3}}^{*}(\omega_{m3})\Psi_{i\alpha_{4}}^{*}(\omega_{m4})}{4a_{2}^{(0)}(i\alpha_{1},\omega_{m1})a_{2}^{(0)}(i\alpha_{2},\omega_{m2})a_{2}^{(0)}
(i\alpha_{3},\omega_{m3})a_{2}^{(0)}(i\alpha_{4},\omega_{m4})}\nonumber\\
\hspace{-2.5cm}\times a_{4}^{(0)}(i\alpha_{1},\omega_{m1};i\alpha_{2},\omega_{m2}|i\alpha_{3},\omega_{m3};i\alpha_{4},\omega_{m4})\Biggr\}.
\end{eqnarray}
We note that the coefficient $a_{4}^{(1)}$ from (\ref{eq6}) in the grand-canonical free energy (\ref{eq:4-13-1}),
(\ref{eq:5-263}) is no longer present in the Ginzburg-Landau action (\ref{eq:4-21}). The reason is
that the grand-canonical free energy, which represents a sum over all connected vacuum diagrams, yields
via the Legendre transformation an effective action, which represents a sum over all one-particle irreducible
vacuum diagrams \cite{key-8-1,key-9}. For obtaining physical results,
we insert the effective action Eq.~(\ref{eq:4-21}) into the equations of motion (\ref{eq:4-22}) and yield:
\begin{eqnarray}
\hspace{-2.5cm}0=\left[\frac{1}{a_{2}^{(0)}(i\alpha,\omega_{m})}-\sum_{j}J_{ij}\right]\Psi_{j\alpha}^{\rm{eq}}(\omega_{m})
-\sum_{\alpha_{1},\alpha_{2},\alpha_{3}}\;\sum_{\omega_{m1},\omega_{m2},\omega_{m3}}\nonumber\\
\hspace{-2.5cm}\times\frac{a_{4}^{(0)}
(i\alpha_{1},\omega_{m1};i\alpha_{2},\omega_{m2}|i\alpha_{3},\omega_{m3};i\alpha,\omega_{m})\Psi_{i\alpha_{1}}^{\rm{eq}}
(\omega_{m1})\Psi_{i\alpha_{2}}^{\rm{eq}}(\omega_{m2})\Psi_{i\alpha_{3}}^{\rm{eq}*}
(\omega_{m3})}{2a_{2}^{(0)}(i\alpha_{1},\omega_{m1})a_{2}^{(0)}(i\alpha_{2},\omega_{m2})a_{2}^{(0)}(i\alpha_{3},\omega_{m3})a_{2}^{(0)}(i\alpha,\omega_{m})}\Biggr\}. \label{eq:4-21-1}
\end{eqnarray}
From these equations of motions we will determine in the following both the quantum phase transition
and the possible superfluid phases of the considered system.

\section{Quantum Phase Transition}

In this section, we calculate the phase boundary between the Mott insulator and the superfluid phase
at zero temperature. To do this, we specialize the effective action (\ref{eq:4-21}) for a stationary
equilibrium which is site-independent due to homogeneity:

\begin{eqnarray}
\Psi_{i\alpha}^{\rm{eq}}(\omega_{m})=\Psi_{\alpha}\sqrt{\beta}\,\delta_{m,0}\:\:,\:\:\Psi_{i\alpha}^{\rm{eq}*}(\omega_{m})=\Psi_{\alpha}^{*}\sqrt{\beta}\,\delta_{m,0}.\label{eq:5-236}
\end{eqnarray}
Therefore, the effective action (\ref{eq:4-21}) reduces with (\ref{eq:33}) to the effective potential
\begin{eqnarray}
\hspace{-2.5cm}\Gamma=\mathcal{F}_{0}+N_{s}\Biggl\{\sum_{\alpha}\:\left[\frac{\left|\Psi_{\alpha}\right|^{2}}{a_{2}^{(0)}(\alpha,0)}-zJ\left|\Psi_{\alpha}\right|^{2}\right]
-\sum_{\alpha_{1},\alpha_{2},\alpha_{3},\alpha_{4}}\Psi{}_{\alpha_{1}}\Psi_{\alpha_{2}}\Psi_{\alpha_{3}}^{*}\Psi_{\alpha_{4}}^{*}\nonumber \\
\times\frac{\beta a_{4}^{(0)}(\alpha_{1},0;\alpha_{2},0|\alpha_{3},0;\alpha_{4},0)}{4a_{2}^{(0)}(\alpha_{1},0)a_{2}^{(0)}
(\alpha_{2},0)a_{2}^{(0)}(\alpha_{3},0)a_{2}^{(0)}(\alpha_{4},0)}
\Biggr\},\label{eq:5-237}
\end{eqnarray}
where $N_{s}$ is the total number of lattices sites and $z=2D$ denotes the coordination number of a
$D$ dimensional cubic lattice. Note that we drop the site index since the cumulants $C_{n}^{(0)}$ are
independent of the site indices $i$, $j$ due to the locality of $\hat{H}^{(0)}$.

In order to obtain the quantum phase transition according to the Landau theory, the equilibrium order
parameter should vanish. To this end, we read off from Eqs.~(\ref{eq:4-21-1}) and (\ref{eq:5-236}) or from extremizing (\ref{eq:5-237})
\begin{equation}
0= \frac{1}{a_{2}^{(0)}(\alpha,0)}-zJ_{c,\alpha},
\end{equation}
 which yields with Eq.~(\ref{eq:4-30})
\begin{eqnarray}
\hspace{-2cm}zJ_{c,\alpha}=\Bigg[\frac{M_{\alpha,S,m,n}^{2}}{E_{S,m,n}^{(0)}-E_{S+1,m+\alpha,n+1}^{(0)}}
+\frac{N_{\alpha,S,m,n}^{2}}{E_{S,m,n}^{(0)}-E_{S-1,m+\alpha,n+1}^{(0)}}\nonumber \\
-\frac{O_{\alpha,S,m,n}^{2}}{E_{S,m,n}^{(0)}-E_{S+1,m-\alpha,n-1}^{(0)}}
-\frac{P_{\alpha,S,m,n}^{2}}{E_{S,m,n}^{(0)}-E_{S-1,m-\alpha,n-1}^{(0)}}\Bigg]^{-1}.\qquad\label{eq:5-240}
\end{eqnarray}
In order to obtain the location of the quantum phase transition, we have to take the minimum of Eq.~(\ref{eq:5-240})
with respect to the spin index $\alpha$ \cite{key-10}:
\begin{equation}
zJ_{c}={\rm{min}\atop \alpha}J_{c,\alpha}.\label{eq:4-25-1}
\end{equation}
In the following discussion we distinguish the cases  without and with external magnetic field as well
as a ferromagnetic and anti-ferromagnetic interaction. We note that the predictions of both the effective
action approach and the mean-field theory yield the same approximation for the location of the quantum
phase boundary.

\begin{figure}[t!]
\subfloat[\label{fig:5a-1}$\eta/U_{0}=0.05$.]{\raggedright{}\includegraphics[width=7cm,height=5cm]{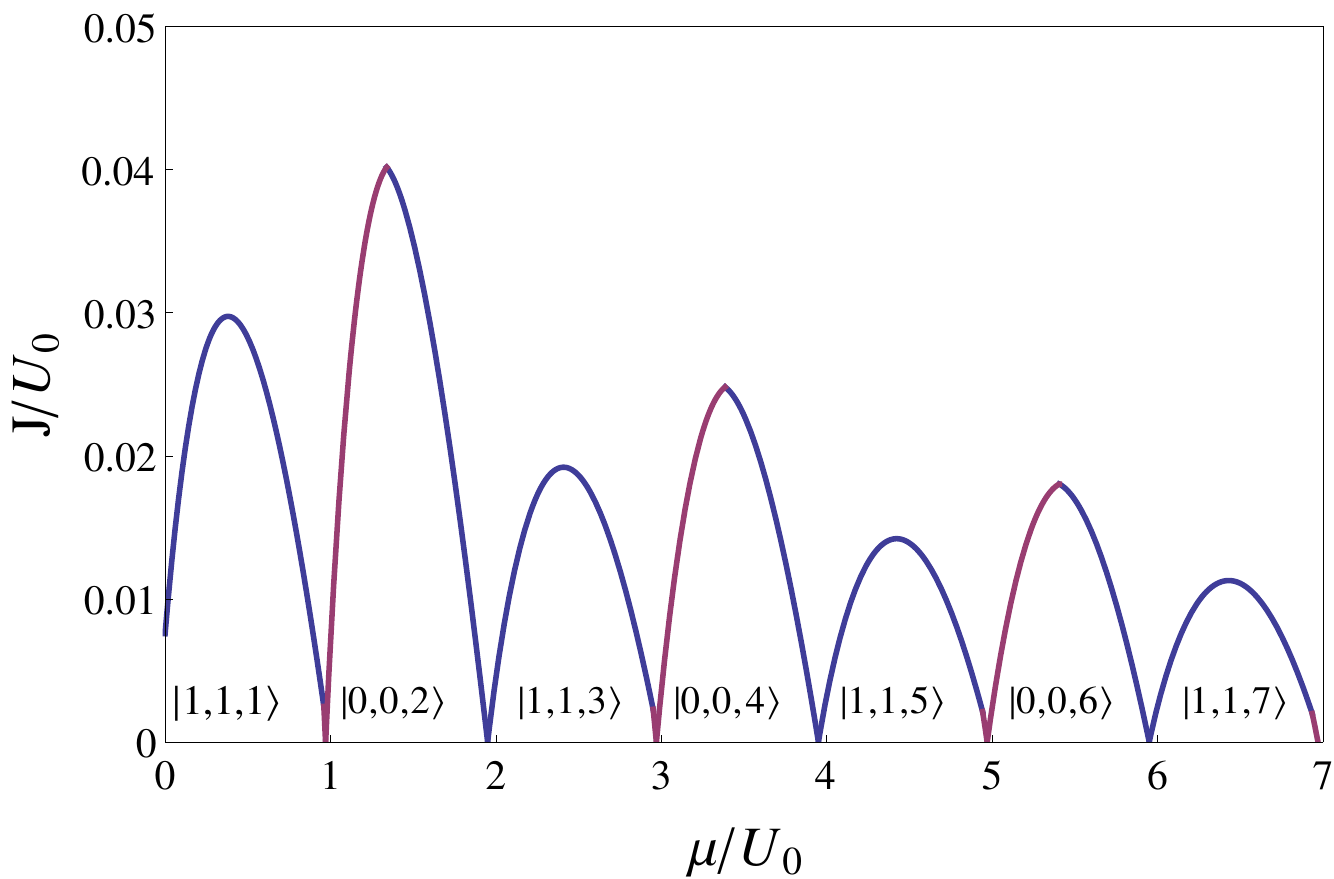}}\hfill{}
\subfloat[\label{fig:5b-1}$\eta/U_{0}=0.07$.]{\raggedright{}~\ \ \includegraphics[width=7cm,height=5cm]{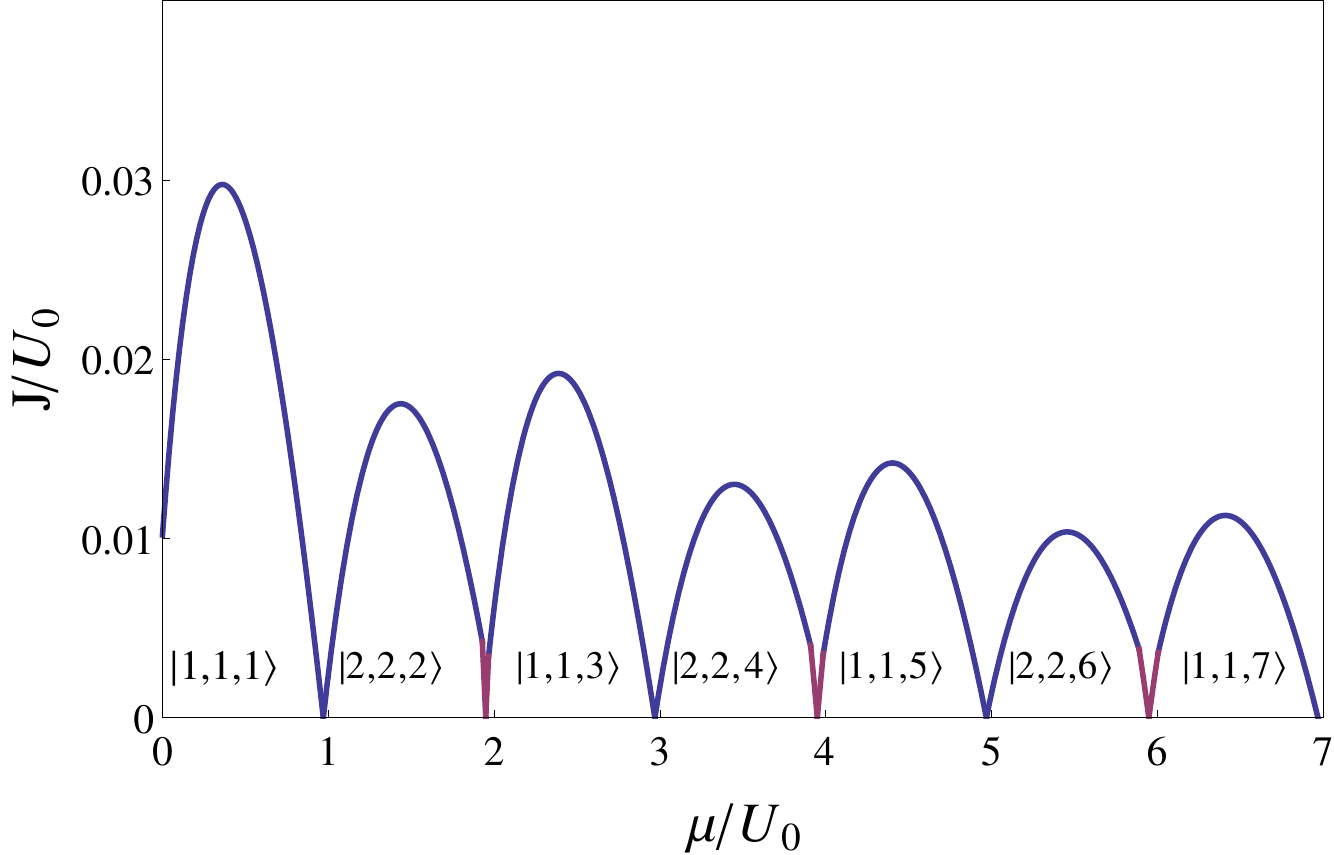}}\hfill{}
\subfloat[\label{fig:5c-1}$\eta/U_{0}=0.125.$]{\begin{raggedright}
\includegraphics[width=7cm,height=5cm]{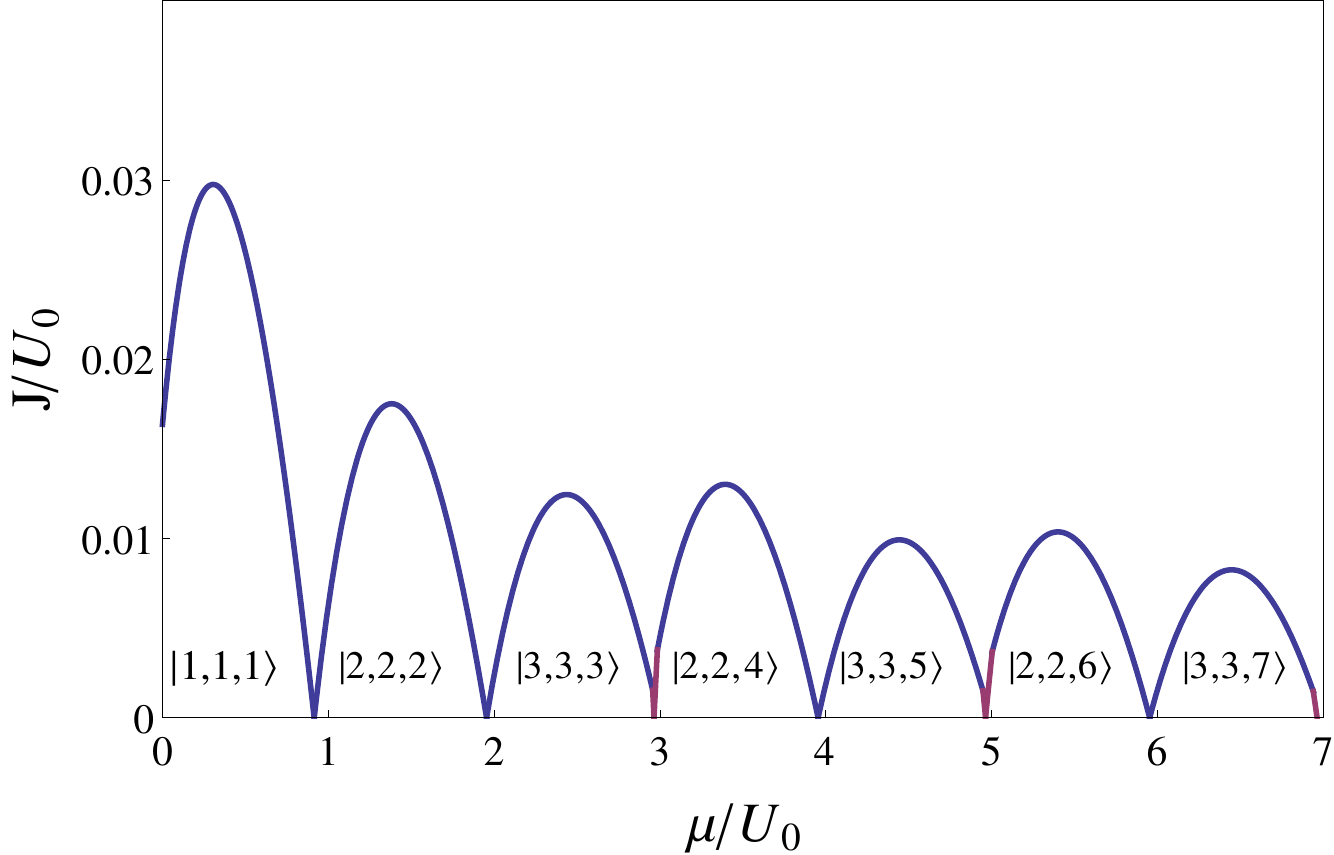}
\par\end{raggedright}
}\hfill{}\subfloat[\label{fig:5d-1}$\eta/U_{0}=0.15$.]{\raggedright{}\includegraphics[width=7cm,height=5cm]{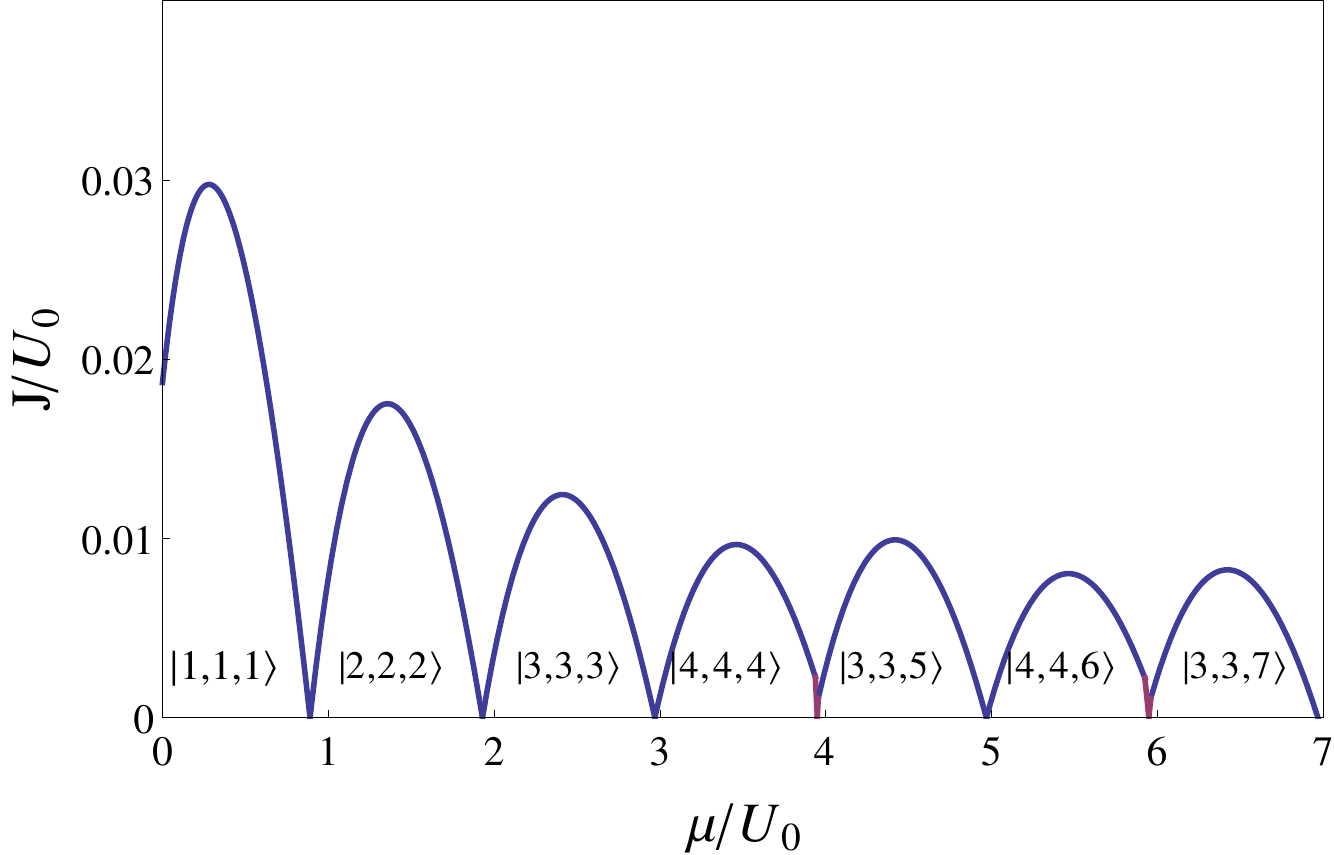}}\hfill{}
\subfloat[\label{fig:5e-1}$\eta/U_{0}=0.2$.]{\raggedright{}\includegraphics[width=7cm,height=5cm]{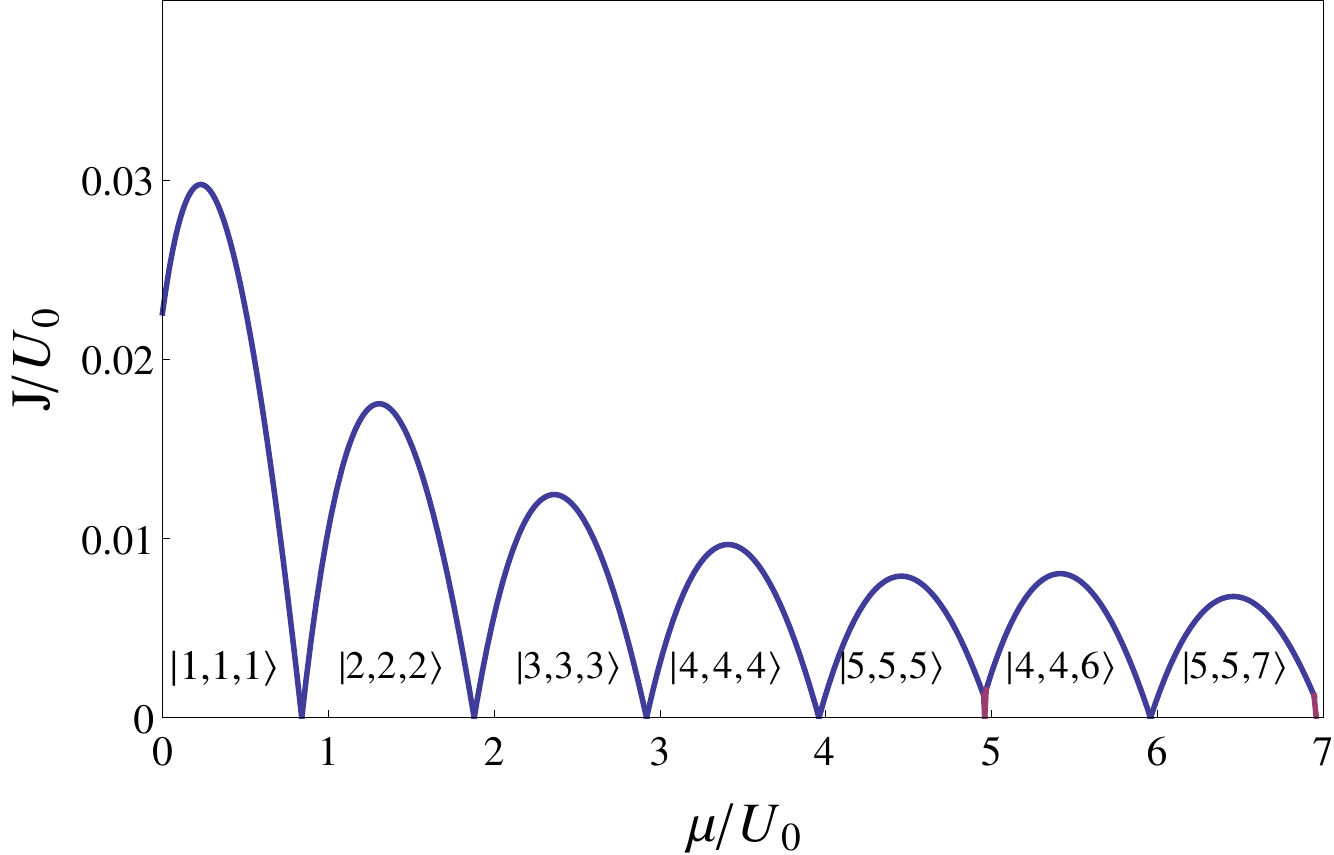}}\hfill{}
\subfloat[\label{fig:5f-1}$\eta/U_{0}=0.3$.]{\raggedright{}\includegraphics[width=7cm,height=5cm]{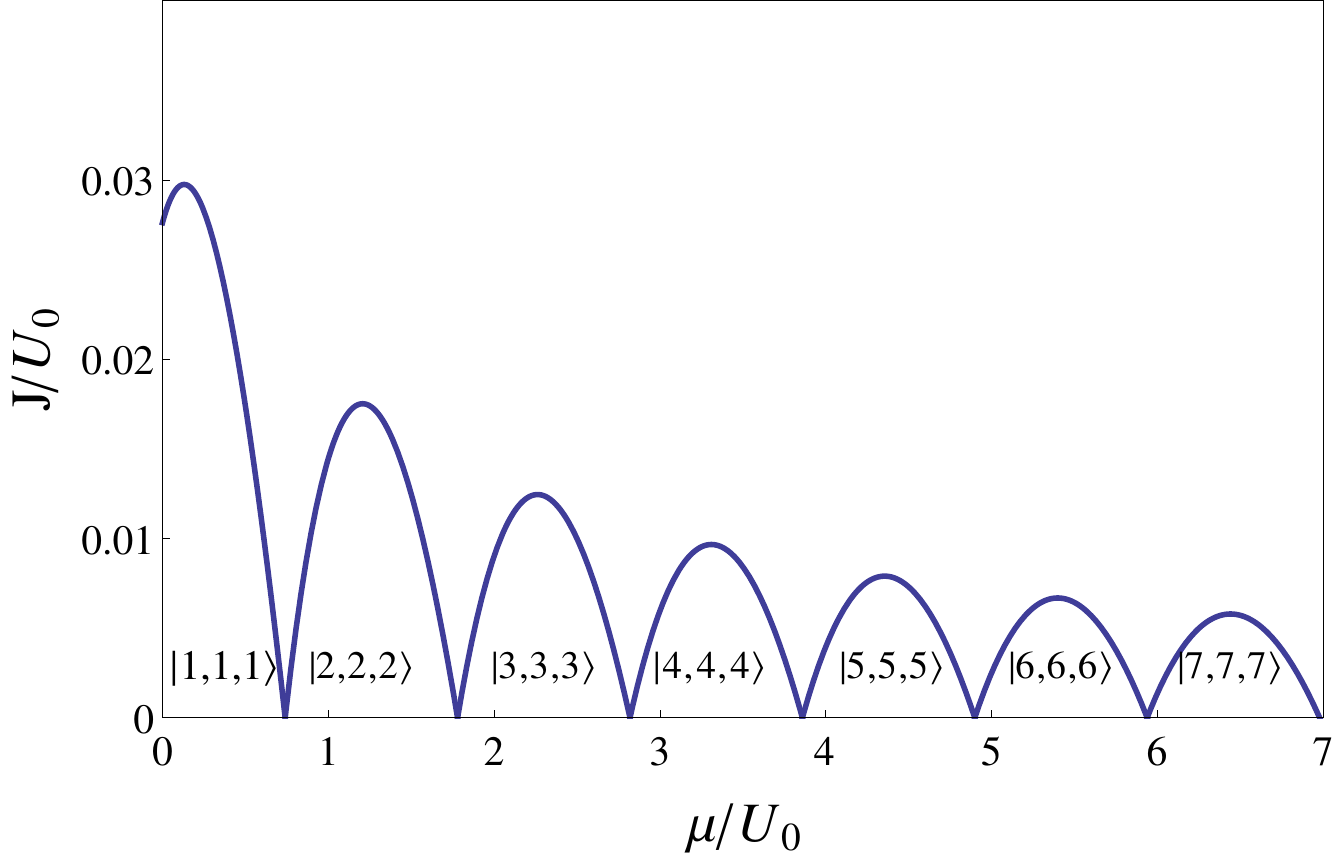}}
\caption{\label{fig:phasetransitionu2fixed}Quantum phase boundary between Mott insulator and superfluid phase
for anti-ferromagnetic interaction with $U_{2}/U_{0}=0.04$. Blue and red line correspond to an instability
of the spin-1 and spin-(-1) component, respectively.}
\end{figure}

\subsection{No Magnetization \label{sub:No-Magnetization}}

For \textbf{ferromagnetic} interactions we find that the Ginzburg-Landau phase boundary (\ref{eq:4-25-1})
with zero external magnetic field is identical to the corresponding same results of Ref.~\cite{key-33}.

On the other side, for \textbf{anti-ferromagnetic interaction} $U_{2}>0$ with $\eta=0$, the minimization
of the energy implies a minimum of the spin value which depends on the number of atoms per site. The
ground state of the nonperturbative Hamiltonian $\hat{H}^{\left(0\right)}$ is $\left|0,0,n\right\rangle $
for even $n$ and $\left|1,m,n\right\rangle $ for odd particle number $n$. In the latter case we have
to determine the value of $m$ to get the minimum of the critical hopping. This means that we have to
find this minimum (\ref{eq:4-25-1}) with respect to both $\alpha$ and $m$ in order to determine the
phase boundary. The result is that the component with $m=0$ forms the superfluid, i.e. $\Psi_{0}\neq0$, so
the SF phase is a polar state with $\Psi_{1}=\Psi_{-1}=0$. We find that the Ginzburg-Landau
phase boundary (\ref{eq:4-25-1}) coincides with the results which were already obtained in Refs.~\cite{key-10,key-4}.

\subsection{With magnetization\label{sub:With-magnetization}}

Afterwards, we study the effect of the external magnetic field $\eta$ on the phase boundary. To this
end we assume without loss of generality that $\eta>0$. 

For a \textbf{ferromagnetic interaction}, there is no change of the quantum phase boundary as the minimization
of the energy implies the maximum of spin value as it is in the case without $\eta$ except the degeneracy
with respect to $m$ is lifted, so the ground state becomes $\left|n,n,n\right\rangle $. Thus, the quantum
phase boundary with $\eta$ is the same as that without it. 

For an \textbf{anti-ferromagnetic interaction}, the situation is more complicated. If $\eta$ is large
compared with $U_{2}$, all spins will be aligned in  $z$-direction, so the ground state will be
a high spin state $\left|n,n,n\right\rangle $ as seen in \fref{fig:5f-1}. In the opposite limit
that $\eta$ is small in comparison with $U_{2}$, the ground state will be $\left|0,0,n\right\rangle $
for even $n$ and $\left|1,1,n\right\rangle $ for odd $n$ as seen in \fref{fig:5a-1}. In between
the ground state can be $\left|S,S,n\right\rangle $ with $0\leq S\leq n$ as discussed in detail in
Section \ref{sec:System-Properties-With}. Using the matrix elements \cite{key-10,key-4} of 
\ref{sec:Matrix-Elements} we show in more detail how the external magnetic field $\eta$ effects the
quantum phase boundary as shown in \fref{fig:phasetransitionu2fixed}. The minimization of (\ref{eq:4-25-1})
with respect to the spin index $\alpha$ yields that either spin-1 or spin-(-1) lead to the quantum phase
boundary, whereas the spin-0 component has no effect \cite{key-10,key-4,key-15,key-60-1}. Furthermore, the size of the
Mott lobes decreases, when the external magnetic field increases, as the increasing Zeeman energy breaks
apart the singlet pairs.
\section{Validity Range of Ginzburg-Landau and Mean Field Theory}

Calculating the condensate density with the Ginzburg-Landau theory within the superfluid phase reveals
that it increases quite fast and that it even diverges between the even and odd lobes \cite{key-3-3}.
This means physically that this theory has a limited range of validity in the superfluid phase. In order
to investigate this delicate issue in more detail, we focus in this section on the scalar Bose-Hubbard
model, which is recovered from our spin-1 theory in the ferromagnetic case, i.e. $\Psi_{1}\neq0,\,\Psi_{-1}=\Psi_{0}=0$,
where we have $\eta=0$ and $S=m=n$ as well as we perform the identification $U_{2}+U_{0}=U$. Thus,
we can specialize the matrix elements according to  \ref{sec:Matrix-Elements}. The Landau coefficients
Eq.~(\ref{eq:4-30}) and Eq.~(\ref{eq:4-6}) reduce at zero temperature to the explicit expressions

\begin{eqnarray}
a_{2}^{(0)}(1,0)=\frac{n+1}{E_{n+1,n+1,n+1}^{(0)}-E_{n,n,n}^{(0)}}-\frac{n}{E_{n,n,n}^{(0)}-E_{n-1,n-1,n-1}^{(0)}},
\end{eqnarray}
and
\begin{eqnarray}
\hspace{-2.5cm}\beta a_{4}^{(0)}(1,0;1,0|1,0;1,0)=  2\left\{ \frac{2n\left(n-1\right)}{(\triangle E_{n-1,n-1,n-1}^{(0)})^{2}\triangle E_{n-2,n-2,n-2}^{(0)}}\right.
+n^{2}\left[-\frac{2}{(\triangle E_{n-1,n-1,n-1}^{(0)})^{3}}\right]\nonumber \\
+\frac{2\left(n+1\right)\left(n+2\right)}{(\triangle E_{n+1,n+1,n+1}^{(0)})^{2}\triangle E_{n+2,n+2,n+2}^{(0)}}
 -\left(n+1\right)^{2}\left[\frac{2}{(\triangle E_{n+1,n+1,n+1}^{(0)})^{3}}\right]\nonumber \\
  \left.-n\left(n+1\right)\left[\frac{2\left(\triangle E_{n+1,n+1,n+1}^{(0)}+\triangle E_{n-1,n-1,n-1}^{(0)}\right)}{\left(\triangle E_{n-1,n-1,n-1}^{(0)}\right)^{2}
 \left(\triangle E_{n+1,n+1,n+1}^{(0)}\right)^{2}}\right]\right\} .
\end{eqnarray}
Using (\ref{eq:4-21-1}) the condensate density becomes
\begin{equation}
\hspace{0.5cm}\left|\Psi_{1}\right|^{2}=\frac{2(a_{2}^{(0)}(1,0))^{3}\left[1-zJa_{2}^{(0)}(1,0)\right]}{\beta a_{4}(1,0;1,0|1,0;1,0)},
\label{eq:5-272}
\end{equation}
and the particle density is given due to (\ref{eq:4-22-1}) by 
\begin{eqnarray}
\left\langle n\right\rangle=\left.-\frac{1}{N_{s}}\frac{\partial\Gamma}{\partial\mu}\right|_{\Psi=\Psi_{\rm{eq}}}.
\label{eq:5-273}
\end{eqnarray}

\begin{figure}[t!]
\centering{}\subfloat[\label{fig:The-condensate-densitya}]
{\includegraphics[width=6cm,height=5cm]{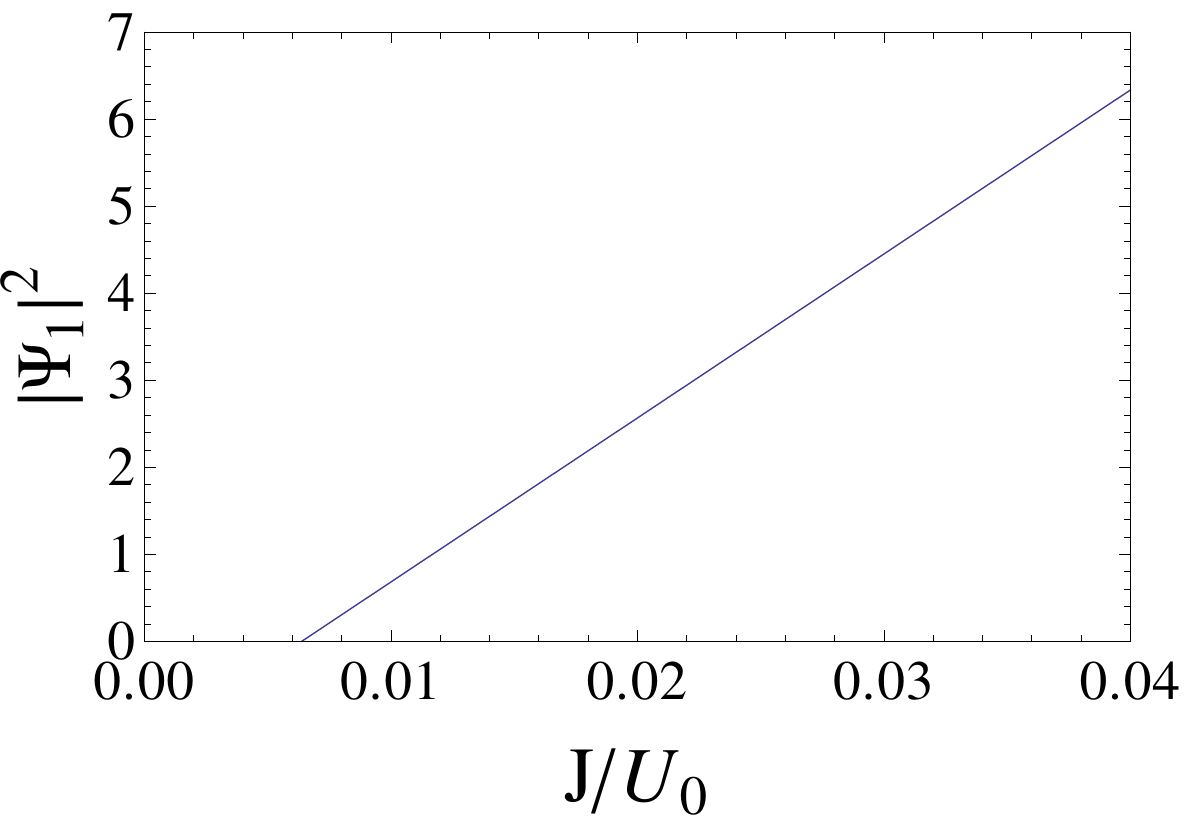}}\quad \quad\quad\quad\quad
\subfloat[\label{fig:The-condensate-densityb}]
{\includegraphics[width=6cm,height=5cm]{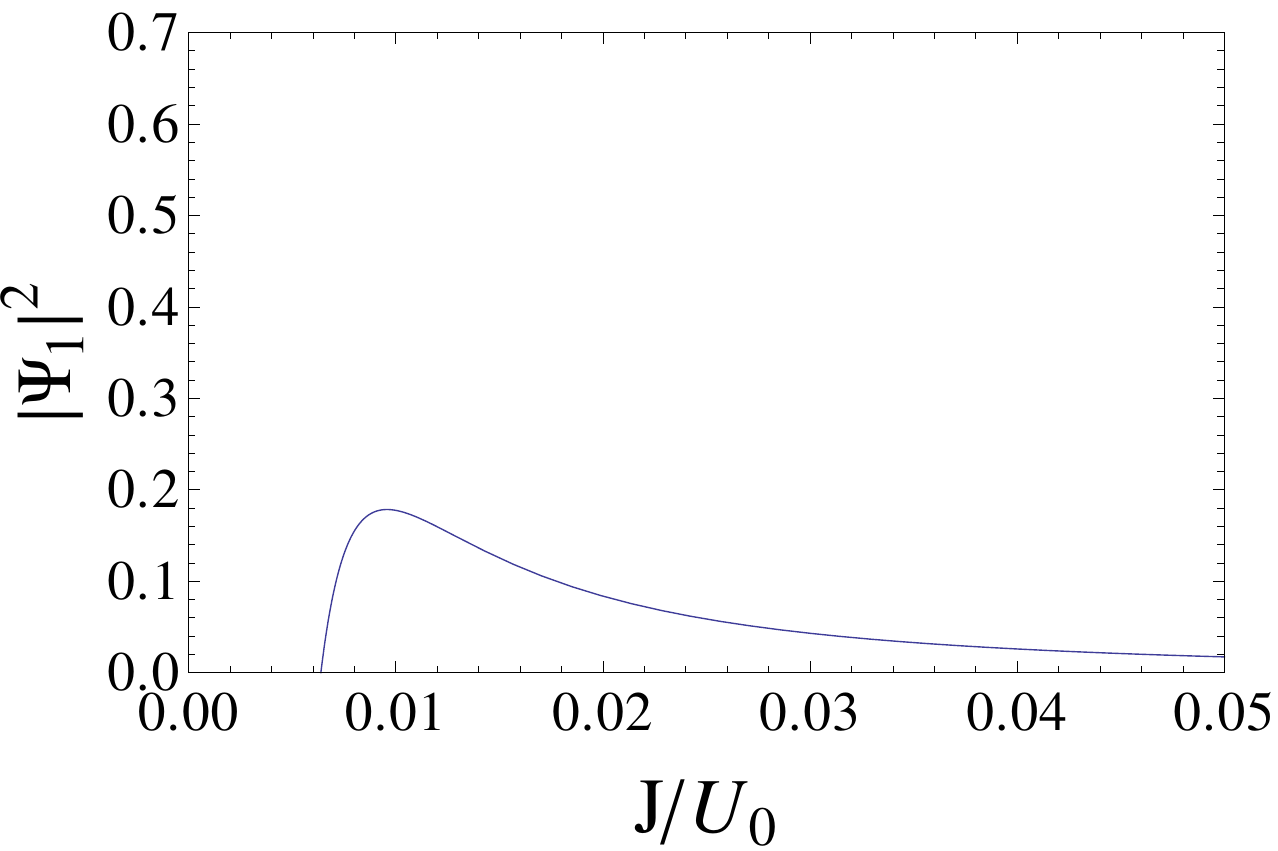}}
\caption{The condensate density as a function of the tunneling parameter $J/U$
in the ferromagnetic case for both (a) the effective action theory   and (b) the mean-field theory
with $\mu/U=0.92$ at zero temperature. }
\end{figure}
\begin{figure}[t!]
\centering{}\subfloat[\label{fig:The-validity-limit-1}]
{\includegraphics[width=6cm,height=5cm]{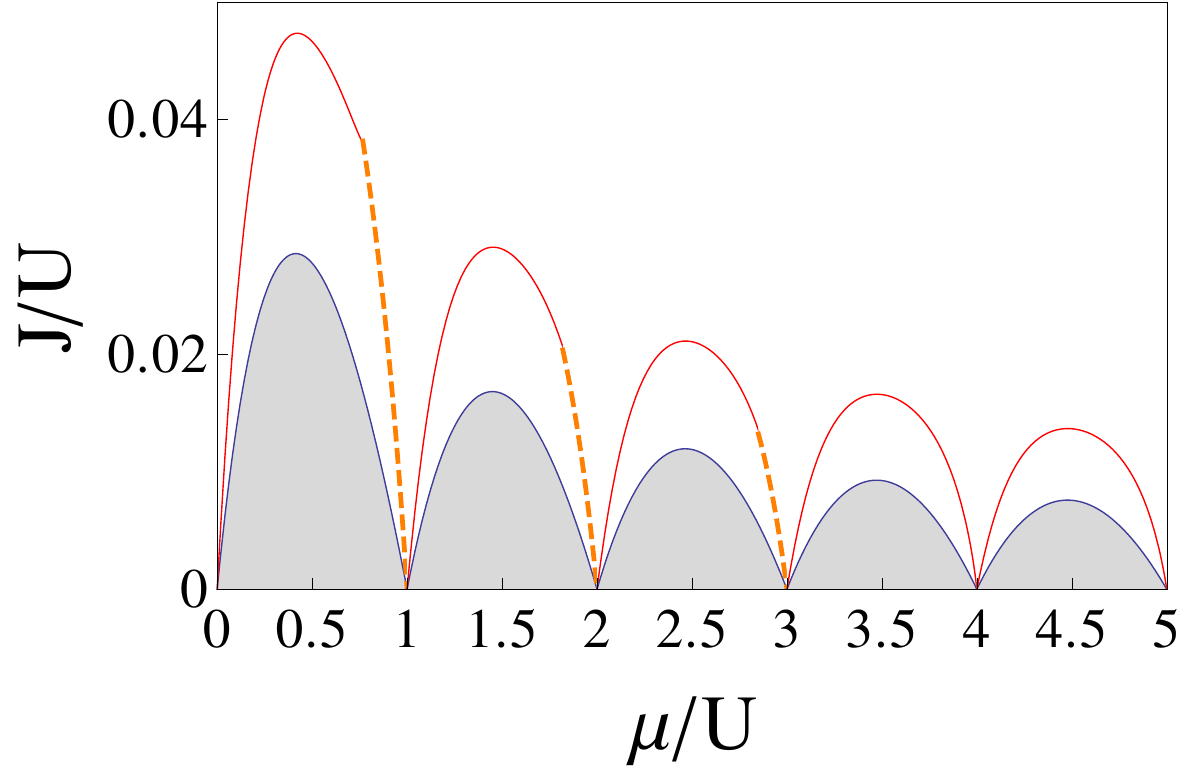}}\quad \quad\quad\quad\quad
\subfloat[\label{fig:The-range-ofmean}]
{\includegraphics[width=6cm,height=5cm]{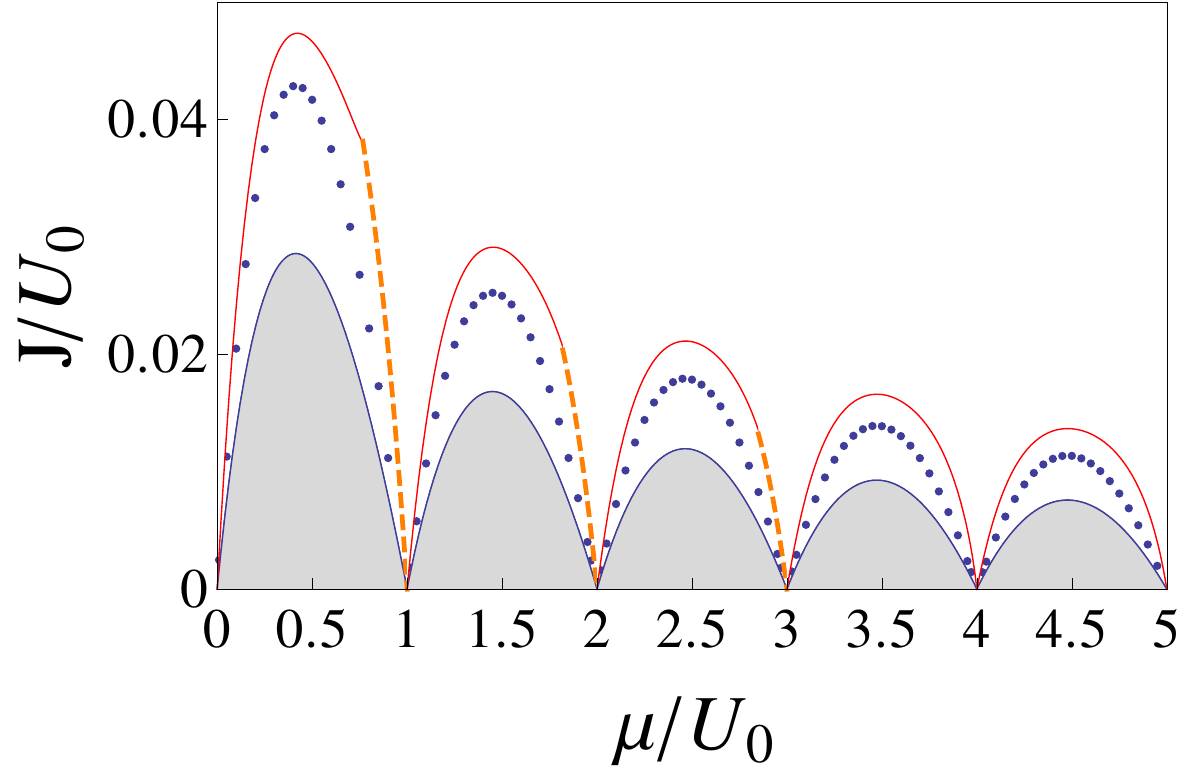}}
\caption{\label{fig:The-validity-limit}Validity range of Ginzburg-Landau theory and mean-field theory for scalar
Bose-Hubbard model in ferromagnetic case at zero temperature. (a) Range of validity of our theory where
the red line depicts the condition that the average particle number equals the condensate density, i.e.
$\left\langle n\right\rangle =\left|\Psi_{1}\right|^{2}$ and the dashed orange line corresponds to the
situation that the condensate density is given by $n+1$. (b) Comparison of the validity ranges of Ginzburg-Landau
theory (orange line) and mean-field theory (blue dots). }
\end{figure}

Calculating the condensate density in the superfluid phase above the first Mott lobe shows, indeed, a
sharp increase, see \fref{fig:The-condensate-densitya} and Ref.~\cite{key-3-3}. Thus, the condensate
density (\ref{eq:5-272}) can not be valid deep in the superfluid phase. In order to determine the range
of validity of the Ginzburg-Landau theory, we remark that, obviously, we can not have more particles
in the condensate than we have in the lattice. This leads to the condition
\begin{eqnarray}
\left|\Psi_{1}\right|^{2} & = & \left\langle n\right\rangle ,\label{eq:5-273-1}
\end{eqnarray}
which is shown in \fref{fig:The-validity-limit-1} as a red line. For Mott lobes with $n\geq4$
this condition is completely sufficient to characterize the range of validity. But we read off from \fref{fig:The-validity-limit-1}
that condition (\ref{eq:5-273-1}) breaks down at the end of the Mott lobes $n=1,2,3$. There we have
to use an additional criterion to obtain a finite range of validity. To this end we complement condition
(\ref{eq:5-273-1}) by the additional ad-hoc restriction that above Mott lobe $n$ the condensate density
can not be larger than $n+1$, yielding the boundary 
\begin{eqnarray}
\left|\Psi_{1}\right|^{2} & = & n+1,\label{eq:5-274}
\end{eqnarray}
which is depicted in \fref{fig:The-validity-limit-1} as a dashed orange line.

By the same way, the condensate density of the mean-field theory is obtained by minimizing the mean-field
energy (\ref{eq:4-224})
\begin{eqnarray}
\left|\Psi_{1}^{\rm{MF}}\right|^{2} & =-\frac{2a_{2}^{\rm{MF}}\left(1,0\right)}{a_{4}^{\rm{MF}}\left(1,0;1,0|1,0;1,0\right)}.
\end{eqnarray}
We remark that, when the chemical potential $\mu$ is fixed, the mean-field condensate density $\left|\Psi_{1}^{\rm{MF}}\right|^{2}$
with spin-1 is not monotonically increasing with the hopping $J$ as shown in \fref{fig:The-condensate-densityb},
see also Ref.~\cite{key-3-3}. Thus, the mean-field prediction for the condensate density is not physical
provided that the hopping is too large. We use this circumstance to our advantage and define also a validity
range for the mean-field theory as follows. For a fixed chemical potential we determine the hopping value
at which the condensate density has its maximal value. Until this hopping value the condensate density
increases with increasing hopping, so that this point defines the validity limit for a fixed $\mu$.
Beyond this hopping value, we can not use the prediction of the mean-field theory because the condensate
density decreases with increasing  hopping parameter as shown in \fref{fig:The-condensate-densityb}.
Thus, we can expect a range of validity until a critical hopping $J$ as shown in \fref{fig:The-range-ofmean}.
Similarly, we could apply the same procedure for the anti-ferromagnetic interaction with and without
magnetization. When we compare the range of validity of Ginzburg-Landau with the corresponding one of
mean-field theory, we find that the Ginzburg-Landau theory has a larger range of validity than that of
mean-field theory as shown in \fref{fig:The-range-ofmean}. Therefore, we discuss now in more
detail the results of the Ginzburg-Landau theory within its validity range.

\section{Superfluid Phases\label{sec:Superfluid-Phases}}

In order to determine the respective superfluid phases, we rewrite the on-site effective potential (\ref{eq:5-237})
according to 
\begin{eqnarray}
\Gamma\left(\Psi_{\alpha},\Psi_{\alpha}^{*}\right)=  \mathcal{F}_{0}+\sum_{\alpha}B_{\alpha}\left|\Psi_{\alpha}\right|^{2}+\sum_{\alpha_{1},\alpha_{2},\alpha_{3},\alpha_{4}}
A_{\alpha_{1}\alpha_{2}\alpha_{3}\alpha_{4}}\Psi_{\alpha_{1}}^{*}\Psi_{\alpha_{2}}^{*}\Psi_{\alpha_{3}}\Psi_{\alpha_{4}},\label{eq:5-1-1}
\end{eqnarray}
with the coefficients

\begin{eqnarray}
B_{\alpha} & =\frac{1}{a_{2}^{(0)}(\alpha,0)}-zJ\label{eq:5-284}
\end{eqnarray}
\begin{eqnarray}
A_{\alpha_{1}\alpha_{2}\alpha_{3}\alpha_{4}} & =-\;\frac{\beta a_{4}^{(0)}(\alpha_{1},0;\alpha_{2},0|\alpha_{3},0;\alpha_{4},0)}{4a_{2}^{(0)}(\alpha_{1},0)a_{2}^{(0)}(\alpha_{2},0)a_{2}^{(0)}(\alpha_{3},0)a_{2}^{(0)}(\alpha_{4},0)},\label{eq:5-285}
\end{eqnarray}
where the symmetries 
\begin{eqnarray}
A_{\alpha_{1}\alpha_{2}\alpha_{3}\alpha_{4}} & =A_{\alpha_{2}\alpha_{1}\alpha_{3}\alpha_{4}}=A_{\alpha_{1}\alpha_{2}\alpha_{4}\alpha_{3}}=A_{\alpha_{2}\alpha_{1}\alpha_{4}\alpha_{3}}\label{eq:5-286}
\end{eqnarray}
 follow from (\ref{eq:4-30}), (\ref{eq:5-285}), and (\ref{eq:4-6}). Using (\ref{eq:5-286}), Eq.~(\ref{eq:5-1-1})
reads explicitly 

\begin{eqnarray}
\hspace{-2.5cm}\Gamma\left(\Psi_{\alpha},\Psi_{\alpha}^{*}\right)=B_{1}\left|\Psi_{1}\right|^{2}+B_{0}\left|\Psi_{0}\right|^{2}+B_{-1}\left|\Psi_{-1}\right|^{2}
+A_{1111}\left|\Psi_{1}\right|^{4}+A_{0000}\left|\Psi_{0}\right|^{4}\nonumber \\
+A_{-1-1-1-1}\left|\Psi_{-1}\right|^{4}+4A_{-100-1}\left|\Psi_{-1}\right|^{2}\left|\Psi_{0}\right|^{2}+4A_{1-11-1}\left|\Psi_{-1}\right|^{2}\left|\Psi_{1}\right|^{2}\nonumber \\
+4A_{1001}\left|\Psi_{1}\right|^{2}\left|\Psi_{0}\right|^{2}+2A_{1-100}\Psi_{1}^{*}\Psi_{-1}^{*}\Psi_{0}\Psi_{0}+2A_{001-1}\Psi_{0}^{*}\Psi_{0}^{*}\Psi_{1}\Psi_{-1}.
\label{eq:5-239}
\end{eqnarray}
As the effective potential (\ref{eq:5-239}) must be extremized with respect to the order parameter $\Psi_{\alpha}$,
we obtain the following self-consistency equations
\begin{eqnarray}
\hspace{-2.5cm}\left(B_{1}+2A_{1111}\left|\Psi_{1}\right|^{2}+4A_{1001}\left|\Psi_{0}\right|^{2}+4A_{1-11-1}\left|\Psi_{-1}\right|^{2}\right)
\Psi_{1}+2A_{1-100}\left|\Psi_{0}\right|^{2}\Psi_{-1}^{*}=0,\label{eq:5-10-1}
\end{eqnarray}
\begin{eqnarray}
\hspace{-2.5cm}\left(B_{-1}+2A_{-1-1-1-1}\left|\Psi_{-1}\right|^{2}+4A_{-100-1}\left|\Psi_{0}\right|^{2}\right.
+\left.4A_{1-11-1}\left|\Psi_{1}\right|^{2}\right)\Psi_{-1}\nonumber\\
\hspace{8cm}+2A_{1-100}\left|\Psi_{0}\right|^{2}\Psi_{1}^{*}=0,\label{eq:5-10-2}
\end{eqnarray}
\begin{eqnarray}
\hspace{-2.5cm}\left(B_{0}+2A_{0000}\left|\Psi_{0}\right|^{2}+4A_{1001}\left|\Psi_{1}\right|^{2}+4A_{-100-1}\left|\Psi_{-1}\right|^{2}\right)
\Psi_{0}+2A_{001-1}\Psi_{1}\Psi_{-1}\Psi_{0}^{*}=0.\label{eq:5-10-3}
\end{eqnarray}
If there is more than one solution, we must take the one which minimizes the effective potential (\ref{eq:5-239}) for
some system parameter. In this way we are able to find the different superfluid phases above both the
even and the odd Mott lobes.

Now we list all possible superfluid phases which could follow from solving Eqs.~(\ref{eq:5-10-1})--(\ref{eq:5-10-3})
with or without magnetization. To this end, we calculate the condensate densities for all these cases:
\begin{figure}[t!]
\subfloat[\label{fig:5a}$\eta/U_{0}=0.05$.]{\raggedright{}\includegraphics[width=7.5cm]{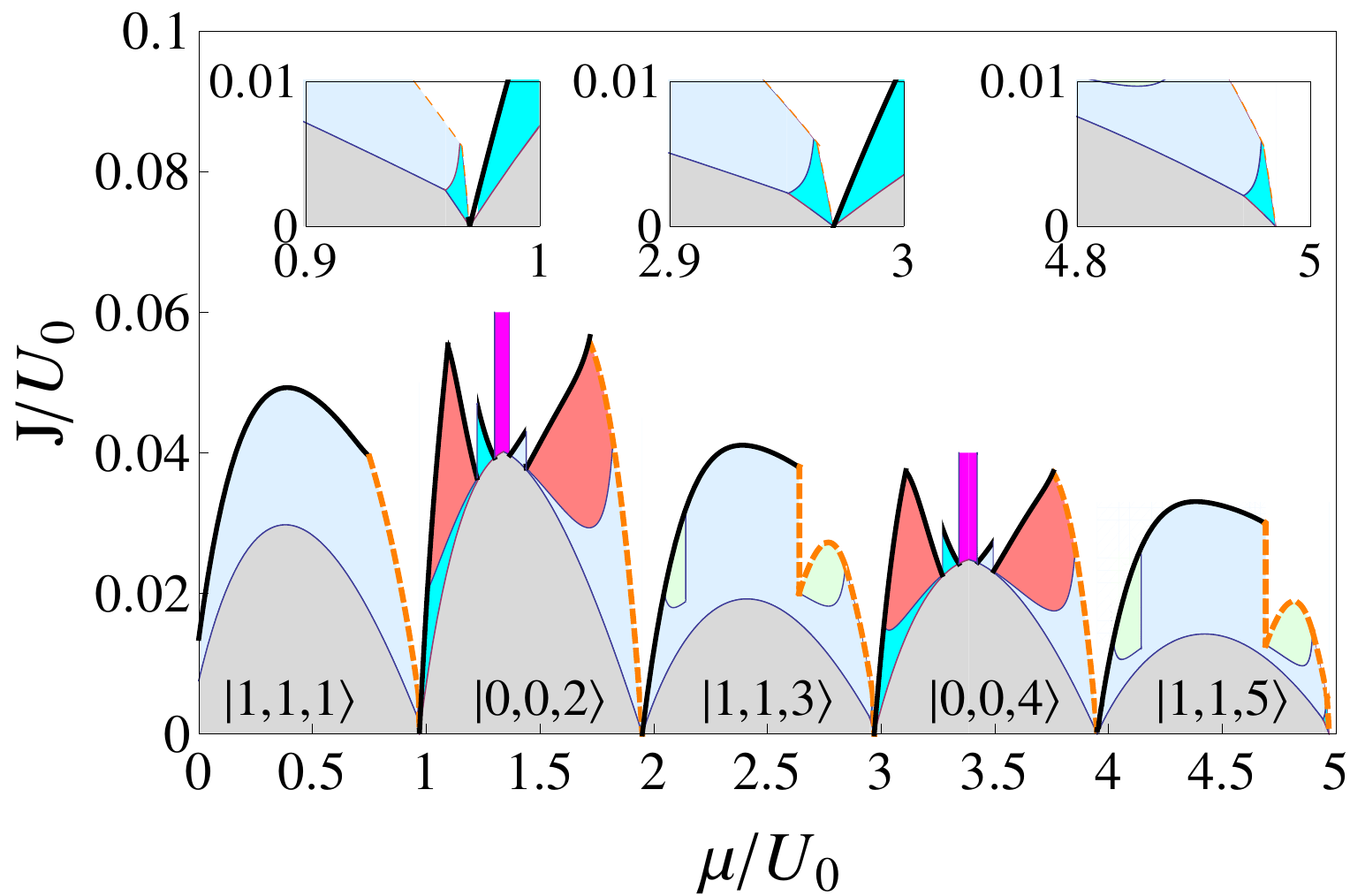}}\hfill{}
\subfloat[\label{fig:5b}$\eta/U_{0}=0.07$.]{\raggedright{}\includegraphics[width=7.5cm]{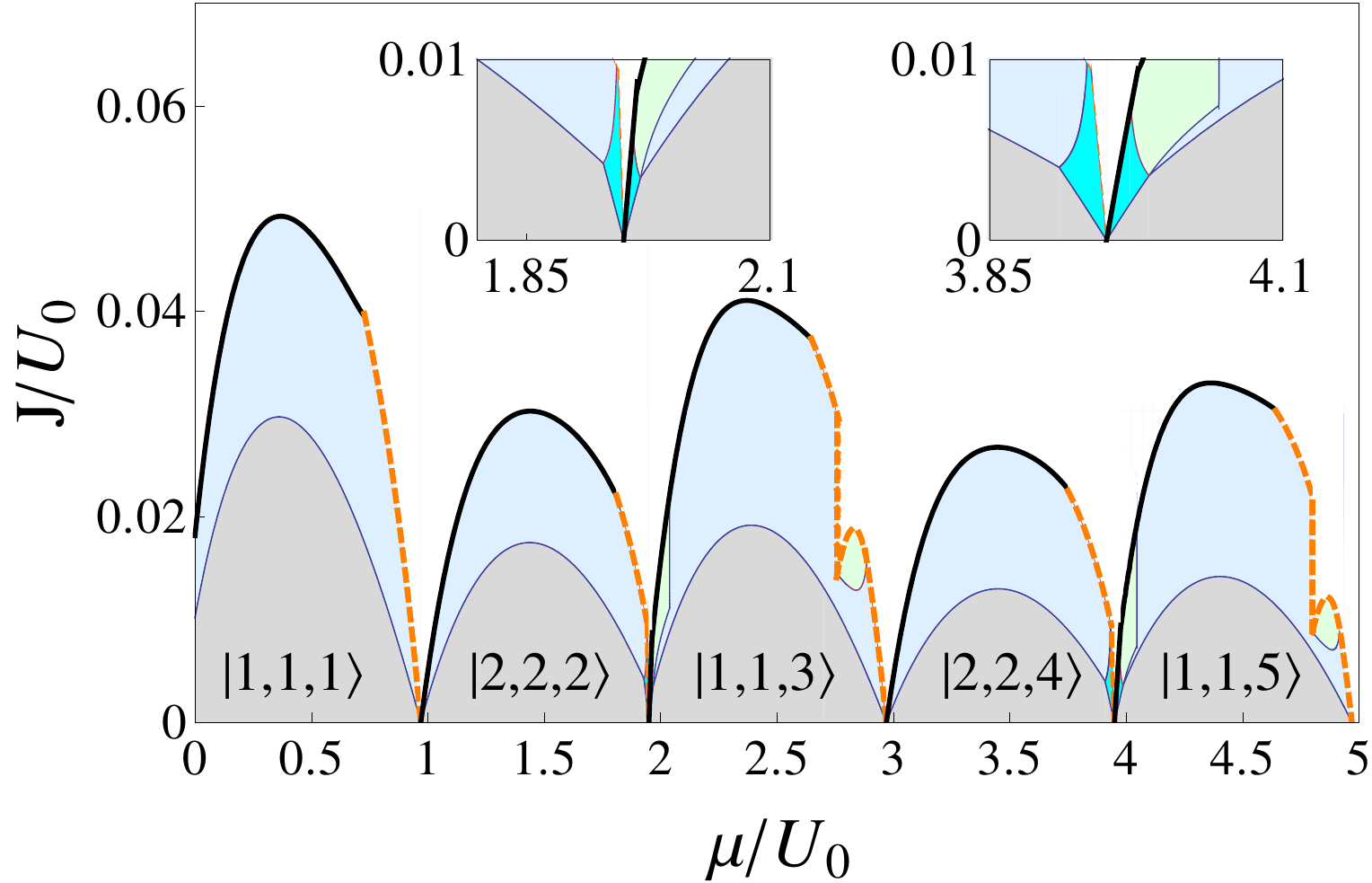}}\hfill{}
\subfloat[\label{fig:5c}$\eta/U_{0}=0.125$.]{\raggedright{}\includegraphics[width=7.5cm]{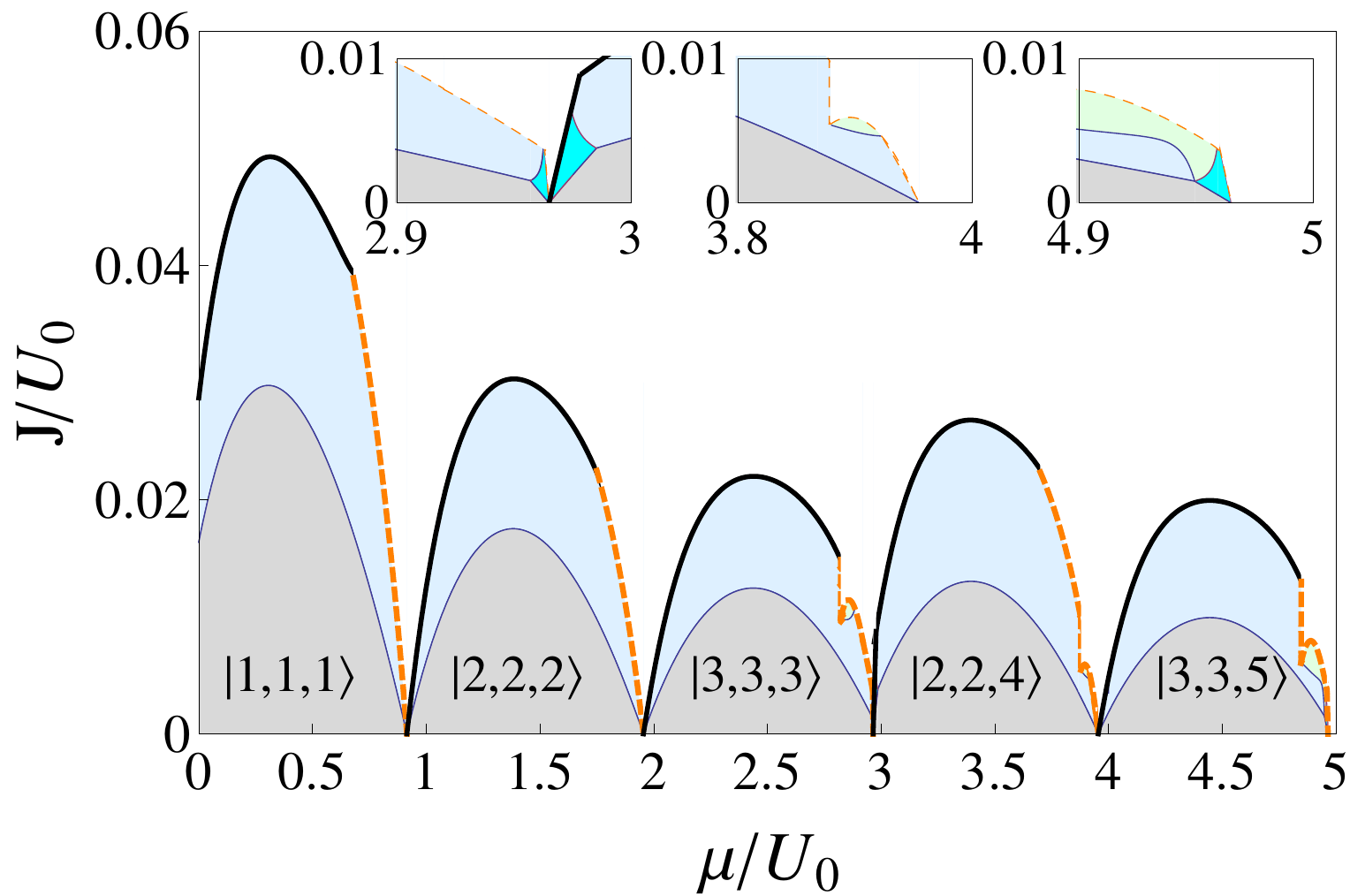}}\hfill{}
\subfloat[\label{fig:5d}$\eta/U_{0}=0.15$.]{\raggedright{}\includegraphics[width=7.5cm]{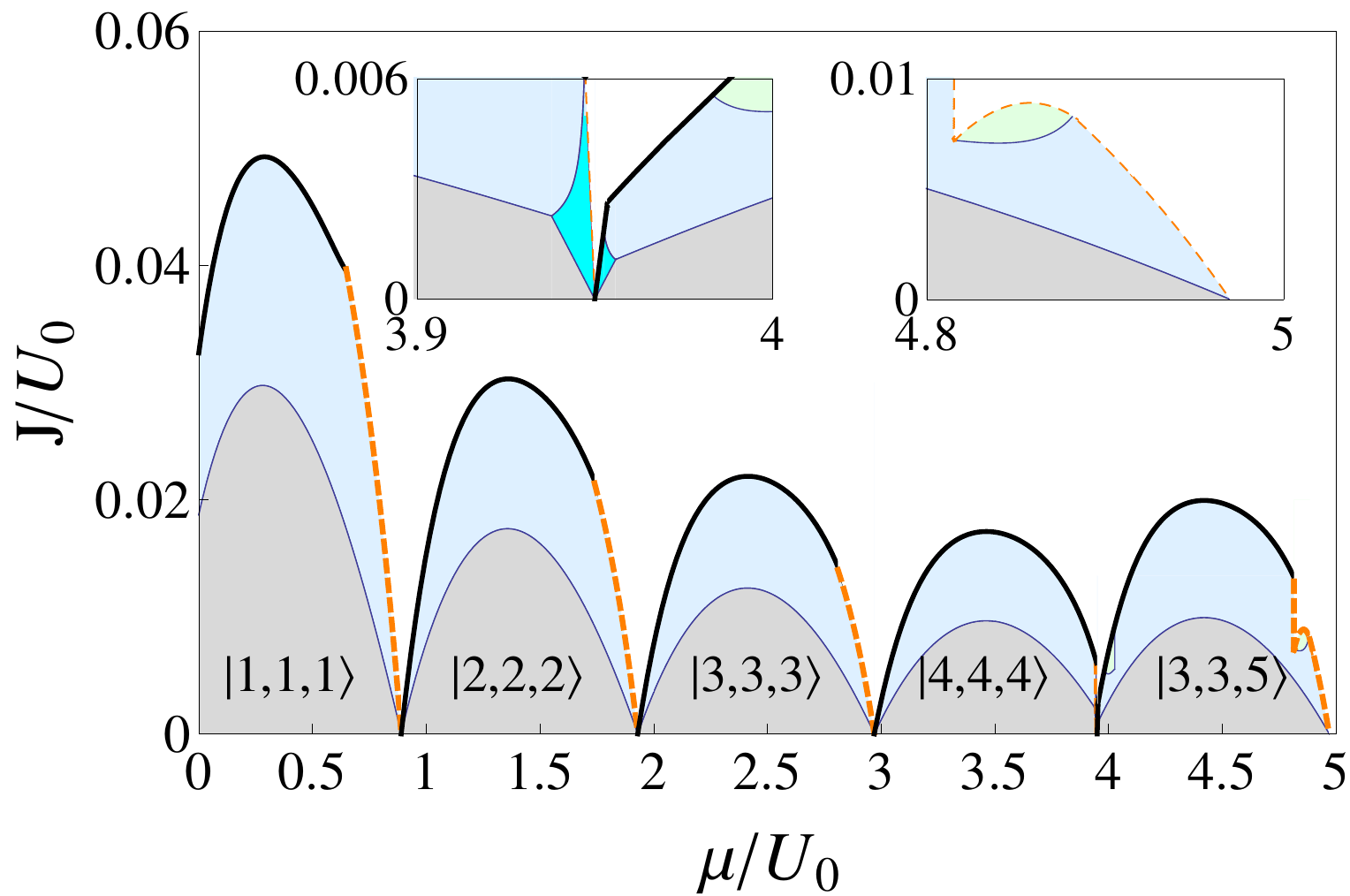}}\hfill{}
\subfloat[\label{fig:5e}$\eta/U_{0}=0.2$.]{\raggedright{}\includegraphics[width=7.5cm]{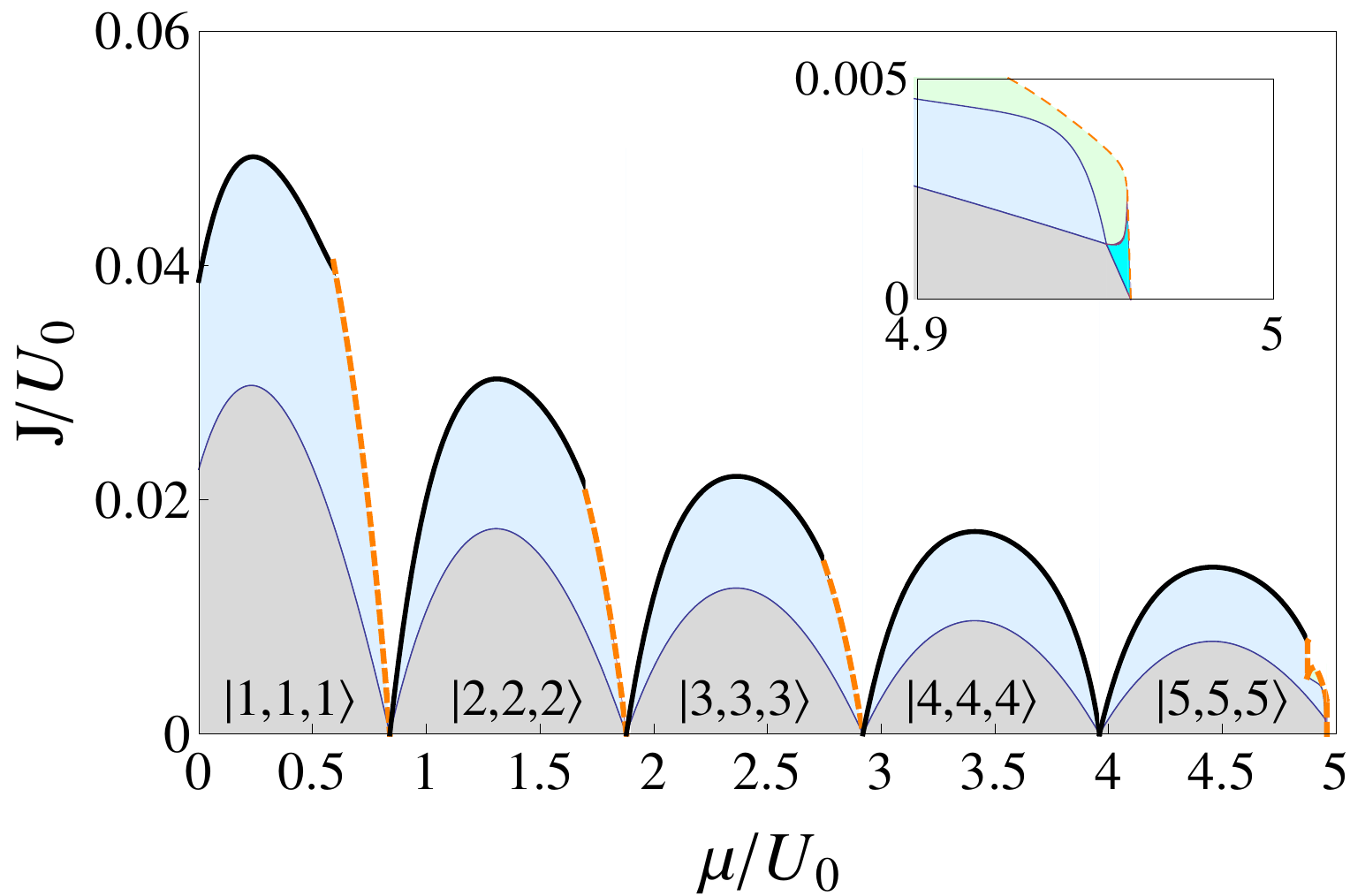}}\hfill{}
\subfloat[\label{fig:5f}$\eta/U_{0}=0.3$.]{\raggedright{}\includegraphics[width=7.5cm]{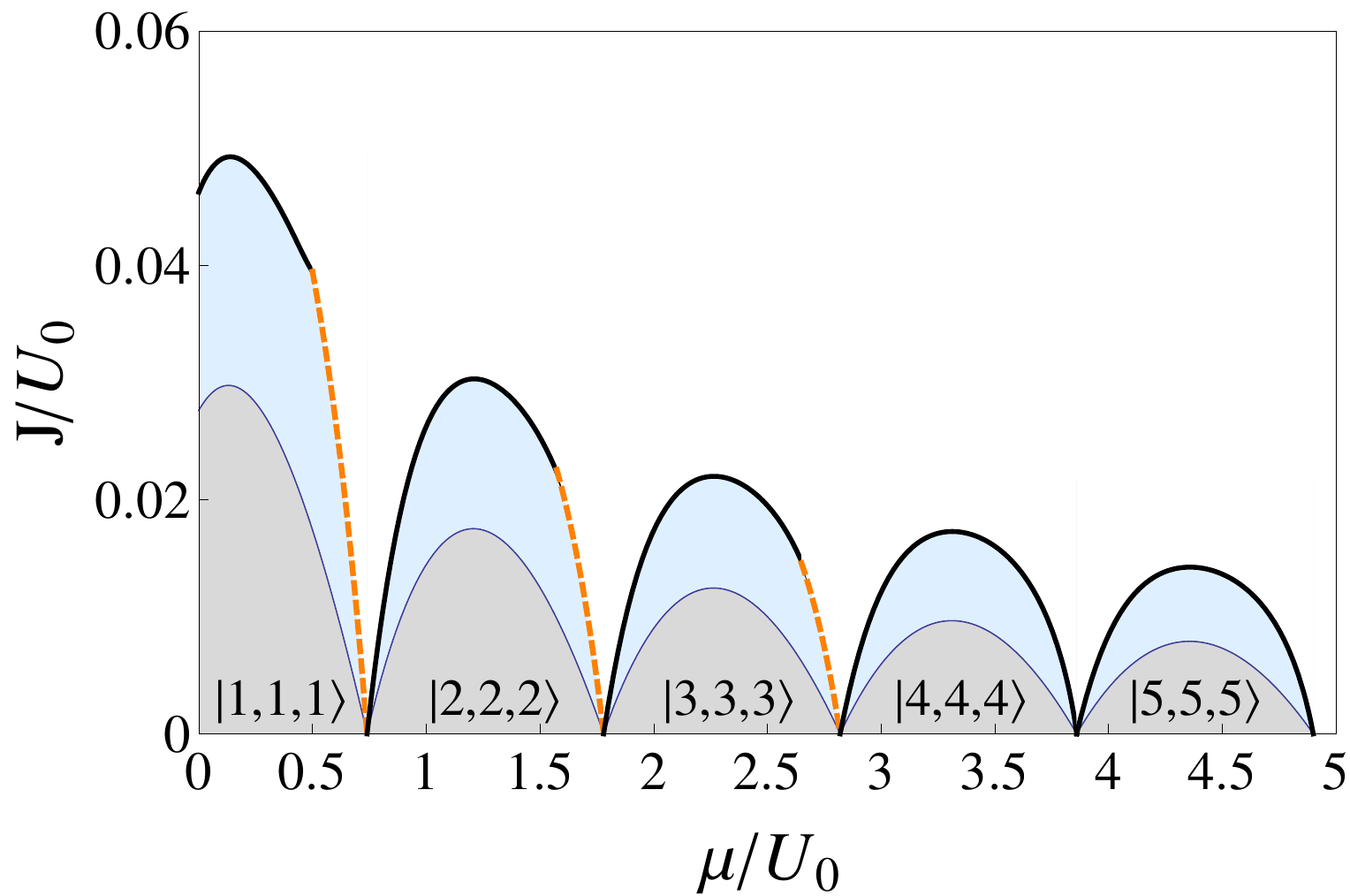}}
\caption{\label{fig:Superfluid-phases1}Superfluid phases with different spin-dependent interaction strengths
for $U_{2}/U_{0}=0.04$. $\Psi_{1}\neq0,\:\Psi_{0}=\Psi_{-1}=0$ (blue); $\Psi_{0}\neq0,\:\Psi_{1}\neq0,\,\Psi_{-1}\neq0$
(red); $\Psi_{-1}\neq0,\:\Psi_{0}=\Psi_{1}=0$ (cyan); $\Psi_{0}\neq0,\:\Psi_{1}=\,\Psi_{-1}=0$ (magenta);
and $\Psi_{1}\neq0,\:\Psi_{-1}\neq0,\,\Psi_{0}=0$ (green), respectively, whereas, the black and dashed
orange lines correspond to the validity ranges (\ref{eq:5-273-1}) and (\ref{eq:5-274}).}
\end{figure}
\begin{enumerate}
\item $\Psi_{1}\neq0,\:\Psi_{-1}=\Psi_{0}=0$ yields with Eq.~(\ref{eq:5-10-1})\label{1it}
\begin{eqnarray}
\left|\Psi_{1}\right|^{2} & =-\frac{B_{1}}{2A_{1111}}.\label{eq:5-291}
\end{eqnarray}

\item $\Psi_{-1}\neq0,\:\Psi_{1}=\Psi_{0}=0$ yields from Eq.~(\ref{eq:5-10-2}) \label{2it}
\begin{eqnarray}
\left|\Psi_{-1}\right|^{2} & =-\frac{B_{-1}}{2A_{-1-1-1-1}}.\label{eq:5-293}
\end{eqnarray}

\item $\Psi_{0}\neq0,\:\Psi_{1}=\Psi_{1}=0$ reduces Eq.~(\ref{eq:5-10-3}) to \label{3it}
\begin{eqnarray}
\left|\Psi_{0}\right|^{2} & =-\frac{B_{0}}{2A_{0000}}.
\end{eqnarray}

\item $\Psi_{1}\neq0,\,\Psi_{-1}\neq0,\,\Psi_{0}=0$ yields from (\ref{eq:5-10-1}) and (\ref{eq:5-10-2})\label{4it}
\begin{eqnarray}
\left|\Psi_{1}\right|^{2} & =\frac{4A_{1-11-1}B_{-1}-2A_{-1-1-1-1}B_{1}}{4A_{1111}A_{-1-1-1-1}-16A_{1-11-1}^{2}},
\end{eqnarray}
\begin{eqnarray}
\left|\Psi_{-1}\right|^{2} & =\frac{4A_{1-11-1}B_{1}-2A_{1111}B_{-1}}{4A_{1111}A_{-1-1-1-1}-16A_{1-11-1}^{2}}.
\end{eqnarray}

\item $\Psi_{1}\neq0,\,\Psi_{0}\neq0,\,\Psi_{-1}=0$ yields from (\ref{eq:5-10-1}) and (\ref{eq:5-10-3})\label{5it}
\begin{eqnarray}
\left|\Psi_{1}\right|^{2} & =\frac{4A_{1001}B_{0}-2A_{0000}B_{1}}{4A_{1111}A_{0000}-16A_{1001}^{2}},
\end{eqnarray}
\begin{eqnarray}
\left|\Psi_{0}\right|^{2} & =\frac{4A_{1001}B_{1}-2A_{1111}B_{0}}{4A_{1111}A_{0000}-16A_{1001}^{2}}.
\end{eqnarray}

\item $\Psi_{-1}\neq0,\,\Psi_{0}\neq0,\,\Psi_{1}=0$ yields from (\ref{eq:5-10-2}) and (\ref{eq:5-10-3})\label{6it}
\begin{eqnarray}
\left|\Psi_{-1}\right|^{2} & =\frac{4A_{-100-1}B_{0}-2A_{0000}B_{-1}}{4A_{-1-1-1-1}A_{0000}-16A_{-100-1}^{2}},
\end{eqnarray}
\begin{eqnarray}
\left|\Psi_{0}\right|^{2} & =\frac{4A_{-100-1}B_{-1}-2A_{-1-1-1-1}B_{0}}{4A_{-1-1-1-1}A_{0000}-16A_{-100-1}^{2}}.
\end{eqnarray}

\item In the general case $\Psi_{1}\neq0,\,\Psi_{-1}\neq0,\,\Psi_{0}\neq0$ it is not possible to solve (\ref{eq:5-10-1})--(\ref{eq:5-10-3})
analytically, so this has to be done numerically. From such  a numerical evaluation we find that the solution
is always  approximately given by either $\Psi_{1}\neq0,\,\Psi_{0}\neq0,\,\Psi_{-1}=0$ with a very small $\Psi_{-1}$
in comparison with $\Psi_{1}$ and $\Psi_{0}$ or by $\Psi_{-1}\neq0,\,\Psi_{0}\neq0,\,\Psi_{1}=0$ when
$\Psi_{1}$ is very small in comparison with $\Psi_{-1}$ and $\Psi_{0}$, which coincides with the above
cases \ref{5it} and \ref{6it}.
\end{enumerate}

\subsection{Without Magnetization}

Now we show for the example of zero temperature that our Ginzburg-Landau theory distinguishes various
ferromagnetic and anti-ferromagnetic superfl{}uid phases for a ferromagnetic and an anti-ferromagnetic
interaction with and without magnetization in its validity range. Without external magnetization the
superfluid phase is a polar (ferromagnetic) state for anti-ferromagnetic (ferromagnetic) interactions,
which is characterized by $\Psi_{1}\neq0,\,\Psi_{-1}=\Psi_{0}=0$ $\left(\Psi_{0}\neq0,\,\Psi_{-1}=\Psi_{1}=0\right)$,
in accordance with previous mean-field results \cite{key-4,key-29}. With magnetization the phase diagram
does not change for the ferromagnetic interaction as the minimization of the energy implies the maximal
spin value as it is in the case without $\eta$ except the degeneracy with respect to $m$ is lifted,
so the ground state becomes $\left|n,n,n\right\rangle .$ For an anti-ferromagnetic interaction the situation
is more complicated with an external magnetic field due to the appearance of different superfluid phases.
Furthermore, we can no longer put $\Psi_{1}=\Psi_{-1}$ as for a non-vanishing $\eta$ as shown in \fref{fig:Superfluid-phases1}.

\subsection{With Magnetization}

In this subsection, we study the predictions of the Ginzburg-Landau theory in view of an effect of the
magnetic field upon the superfluid phases in case of an anti-ferromagnetic interaction, i.e. $U_{2}>0$,
as in $^{23}\rm{Na}$. To this end we show in \fref{fig:Superfluid-phases1} the resulting phase
diagrams before and after the external magnetic field $\eta$ reaches one of the critical values following
from (\ref{eq:54}): 
\begin{eqnarray}
\eta^{\rm{crit}}= & \left(S_{i}+\frac{3}{2}\right)U_{2}.\label{eq:54-1}
\end{eqnarray}

If $\eta$ is small compared to $U_{2}$, spin pairs are produced to get the minimal energy. Therefore,
the ground state becomes $\left|0,0,n\right\rangle $ for an even $n$ and $\left|1,1,n\right\rangle $
for an odd $n$ as shown in \fref{fig:5a}. Thus, the magnetic field is not able to align all spins.
Therefore, both spin-1 and spin-(-1) affect the phase boundary between Mott insulator and superfluid
phases. The phases $\Psi_{1}\neq0,\,\Psi_{-1}\neq0,\,\Psi_{0}=0$; $\Psi_{-1}\neq0,\,\Psi_{1}=\Psi_{0}=0$
and $\Psi_{1}\neq0,\,\Psi_{-1}=\Psi_{0}=0$ appear in the  SF phase for the odd lobes with $n\geq3$ and the
phases $\Psi_{1}\neq0,\,\Psi_{-1}\neq0,\,\Psi_{0}\neq0$; $\Psi_{-1}\neq0,\,\Psi_{1}=\Psi_{0}=0$; $\Psi_{0}\neq0,\:\Psi_{1}=\,\Psi_{-1}=0$
and $\Psi_{1}\neq0,\,\Psi_{-1}=\Psi_{0}=0$ for the even lobes. When $\eta$ is increased above the first
critical value $\eta_{\rm{even}}^{(1)}=0.06\, U_{0}$, both the
spin $S$ and the magnetic quantum number $m$ change from $\left|0,0,n\right\rangle $ to $\left|2,2,n\right\rangle $
for even lobes as shown in \fref{fig:5b}. Correspondingly, the MI phases for the even lobes are
decreased. The phases $\Psi_{-1}\neq0,\,\Psi_{1}=\Psi_{0}=0$ and $\Psi_{1}\neq0,\,\Psi_{-1}=\Psi_{0}=0$
appear in the SF phase for the even lobes and the phase $\Psi_{-1}\neq0,\,\Psi_{1}=\Psi_{0}=0$ is seen
in the SF phase at the beginning of the odd lobes. We note that the phase $\Psi_{1}\neq0,\,\Psi_{-1}\neq0,\,\Psi_{0}\neq0$
no longer appears as a stronger magnetic field leads to a preferred alignment of spins in $z$-direction. 

Beyond the critical value $\eta_{\rm{odd}}^{(1)}=0.1\, U_{0}$ the quantum number $S$ and $m$ for
the odd lobes change from $\left|1,1,n\right\rangle $ to $\left|3,3,n\right\rangle $ as shown in \fref{fig:5c}.
The left phase $\Psi_{1}\neq0,\,\Psi_{-1}\neq0,\,\Psi_{0}=0$ in the odd lobes $n\geq3$ has disappeared
because increasing the magnetic field $\eta$ results in a stronger alignment of the spins, but it is
still not enough to align all the spins. The increase of $\eta$ is enough to align all the spins for
the second lobe and its SF phase is $\Psi_{1}\neq0,\,\Psi_{-1}=\Psi_{0}=0$, but the SF phases for the
fourth lobe are $\Psi_{1}\neq0,\,\Psi_{-1}=\Psi_{0}=0$; $\Psi_{-1}\neq0,\,\Psi_{1}=\Psi_{0}=0$ and
$\Psi_{1}\neq0,\,\Psi_{-1}\neq0,\Psi_{0}=0$. The phase $\Psi_{-1}\neq0,\,\Psi_{1}=\Psi_{0}=0$ appears
now only at the end of the odd lobes. After $\eta_{\rm{even}}^{(2)}=0.14\, U_{0}$ the quantum numbers
$S$ and $m$ for the even lobes change from $\left|2,2,n\right\rangle $ to $\left|4,4,n\right\rangle $
as shown in \fref{fig:5d}. This increase of the magnetic field is not enough to align all the spins
of the fourth lobe, but it is enough to align them for the third lobe. Similarly, the phase $\Psi_{-1}\neq0,\,\Psi_{1}=\Psi_{0}=0$
appears in the SF phase at the contact point between the fourth and the fifth lobe. 

Beyond the critical value $\eta_{\rm{odd}}^{(2)}=0.18\, U_{0}$ the quantum numbers  $S$ and $m$
for the even lobes change from $\left|3,3,n\right\rangle $ to $\left|5,5,n\right\rangle $ as shown
in \fref{fig:5e}. This increase in the magnetic field is not enough to align all the spins for the
fifth lobe, but it is enough to align them for the fourth lobe. Similarly, the phase $\Psi_{-1}\neq0,\,\Psi_{1}=\Psi_{0}=0$
appears in the SF phase at the end of the fifth lobe. As happened in the third
lobe, the left phase $\Psi_{1}\neq0,\,\Psi_{-1}\neq0,\,\Psi_{0}=0$ appears once in the fifth odd lobe.
If $\eta$ increases to 0.3 $U_{0}$ after the critical value $\eta_{\mathrm{even}}^{(3)}=0.22\, U_{0}$
and $\eta_{\mathrm{odd}}^{(3)}=0.26\, U_{0}$, $S$ and $m$ change from $\left|4,4,n\right\rangle $
to $\left|6,6,n\right\rangle $ for the even lobes and from $\left|5,5,n\right\rangle $ to $\left|7,7,n\right\rangle $
for the odd lobes as shown in \fref{fig:5e}, So all seven lobes have $S=m=n$. Therefore, we have
now a full spin alignment in the shown quantum  phase diagram, where only the phase $\Psi_{1}\neq0,\,\Psi_{-1}=\Psi_{0}=0$
exists in the SF phase.
\section{Order of Phase Transition}

In this section, we study which kind of order occurs for the quantum phase transition from the Mott insulator
to the superfluid phase and for the transitions between the respective superfluid phases of spin-1 bosons
in a cubic optical lattice under the effect of the external magnetic field at zero temperature.

\subsection{Quantum Phase Transition}

\begin{figure}
\subfloat[$\mu=2.963\, U_{0}$]{\centering{}\includegraphics[width=6cm,height=5cm]{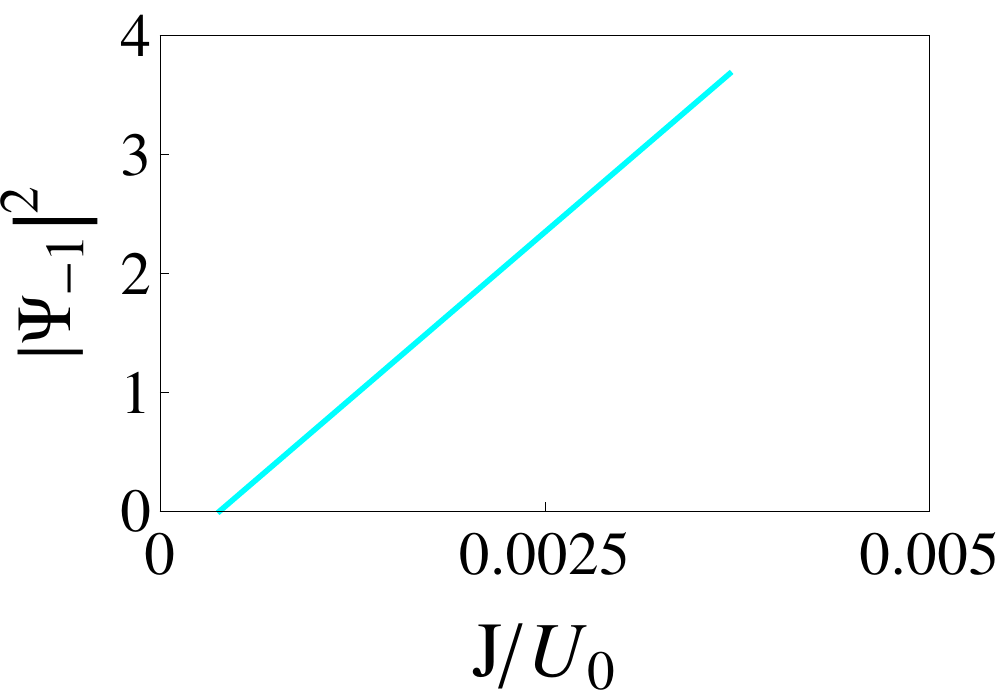}}\hfill
\subfloat[$\mu=2\, U_{0}$]{\centering{}\includegraphics[width=6cm,height=5cm]{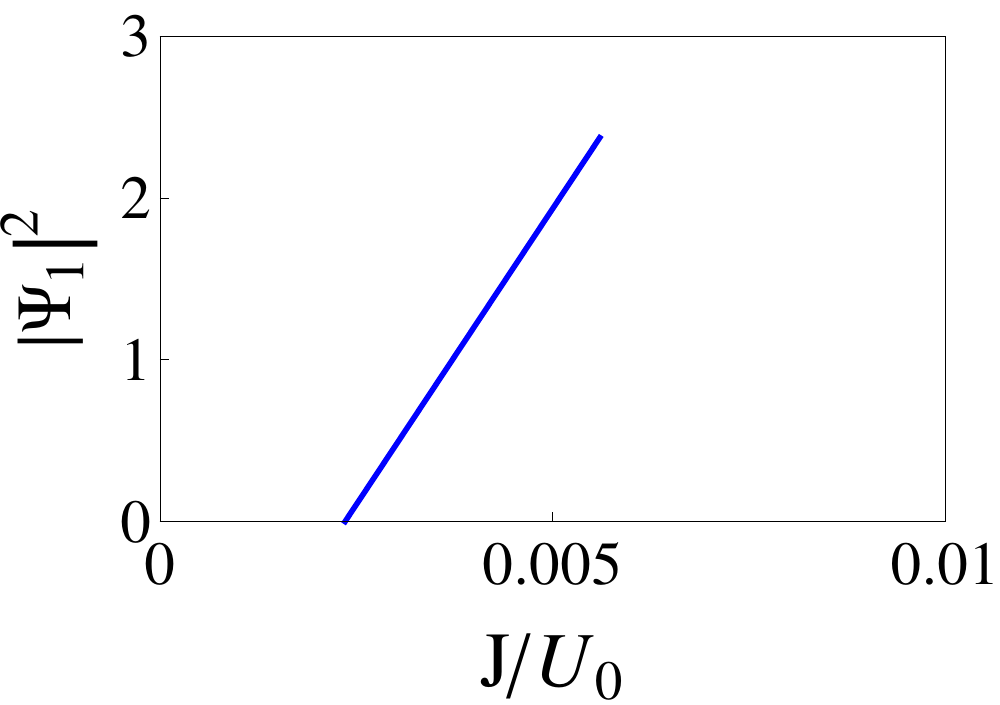}}
\caption{\label{fig:The-condensate-density-1}Condensate density for the phases $\Psi_{-1}\neq0,\:\Psi_{0}=\Psi_{1}=0$
and $\Psi_{1}\neq0,\:\Psi_{-1}=\Psi_{0}=0$ as a function of the tunneling parameter $J/U_{0}$ of spin-1
Bose-Hubbard model in the anti-ferromagnetic case with $\eta=0.125\, U_{0}$ and $U_{2}=0.04\, U_{0}$
at zero temperature. }
\end{figure}

It is well-known that for spinless bosons the superfluid-Mott insulator phase transition of the Bose-Hubbard
model in three dimensions is of second order \cite{key-23-1}. In order to determine the kind of the
order of the quantum phase transition for the spin-1 Bose-Hubbard model, we focus at first on the transition
from the Mott insulator to a superfluid phase at a fixed chemical potential $\mu$ around a external
magnetic field $\eta$ and spin-dependent interaction $U_{2}$ as shown in \fref{fig:The-condensate-density-1}.
There the condensate density is shown as a function of the hopping parameter $J$ at fixed chemical and
external magnetic field values for the third lobe. In this figure, we note that the condensate density
for the two phases $\Psi_{-1}\neq0,\:\Psi_{0}=\Psi_{1}=0$ and $\Psi_{1}\neq0,\:\Psi_{-1}=\Psi_{0}=0$
increases linearly with $J$ until the validity range of Ginzburg-Landau theory is reached and the critical
$J$ values, which are calculated from (\ref{eq:5-293}) and (\ref{eq:5-291}), are 0.00038843 $U_{0}$
and 0.00235782 $U_{0}$, respectively. Therefore, the corresponding superfluid-Mott insulator phase transitions
are of second order.

\subsection{Transitions Between Superfluid Phases\label{sub:Transitions-between-Superfluid}}

\begin{figure}
\subfloat[\label{fig:the condensate density a}]{\includegraphics[width=6cm,height=5cm]{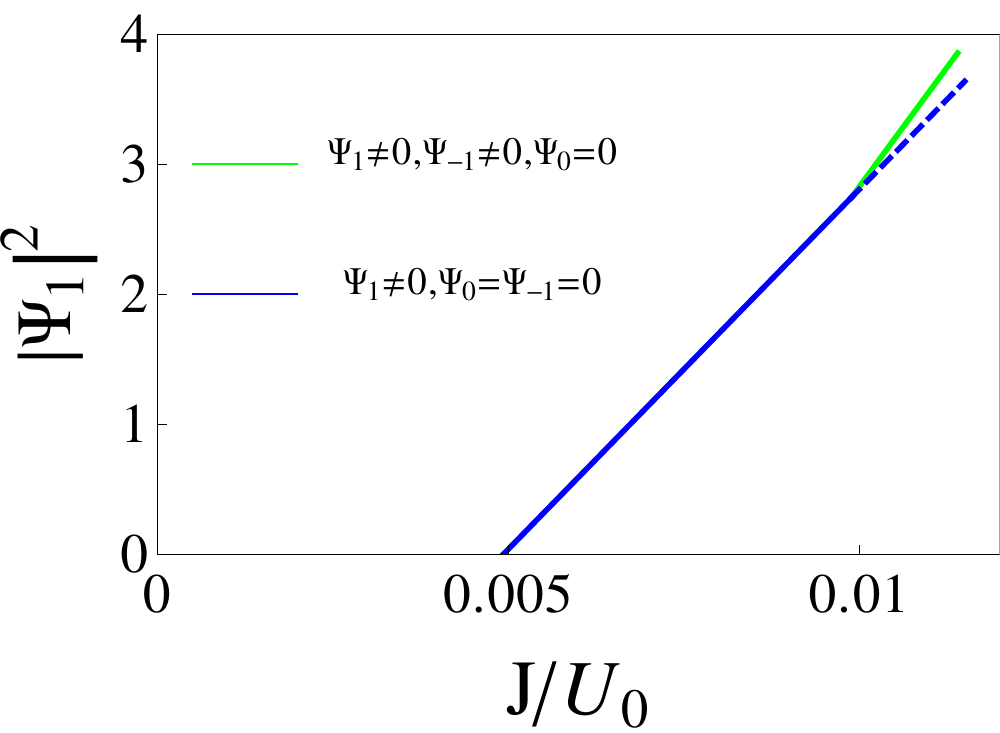}

}\hfill{}\subfloat[\label{fig:the condensate density b}]{\includegraphics[width=6cm,height=5cm]{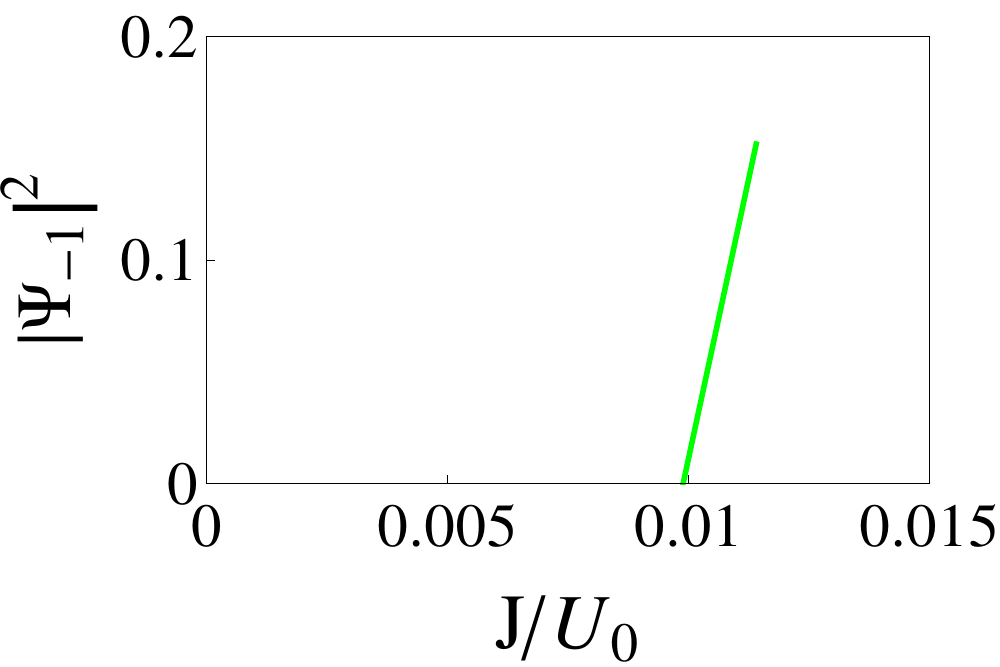}

}\caption{\label{fig:The-condensate-density-2}Condensate density for two spin components as a function of the
tunneling parameter $J/U_{0}$ of the spin-1 Bose-Hubbard model in the anti-ferromagnetic case with $\eta=0.125\, U_{0}$,
$U_{2}=0.04\, U_{0}$ and $\mu=2.864\, U_{0}$ at zero temperature. Solid (dashed) lines correspond to
solutions of minimal (not minimal) energy, compare with \fref{fig:Plot-of-the-1}.}
\end{figure}

The effect of the external magnetic field on spin-1 bosons with anti-ferromagnetic interaction leads
to the appearance of different phases in the superfluid phase as discussed in Sec.~\ref{sec:Superfluid-Phases}.
In order to define the order of the transitions between the phases in the superfluid region, we focus
on the example of the transition from the $\Psi_{1}\neq0,\:\Psi_{0}=\Psi_{-1}=0$ to the $\Psi_{1}\neq0,\:\Psi_{-1}\neq0,\,\Psi_{0}=0$
phase at a fixed chemical potential $\mu$ around an external magnetic field $\eta$ and spin-dependent
interaction $U_{2}$ for the third lobe.  We find that the condensate density $\left|\Psi_{1}\right|^{2}$ continuously increases from the phase  $\Psi_{1}\neq0,\:\Psi_{0}=\Psi_{-1}=0$
to the phase  $\Psi_{1}\neq0,\:\Psi_{-1}\neq0,\,\Psi_{0}=0$  as shown in \fref{fig:the condensate density a} and 
\fref{fig:the condensate density b} shows that the condensate density $\left|\Psi_{-1}\right|^{2}$
continuously increases with increasing the hopping parameter $J$. Furthermore, the latter  condensate density
starts at the critical hopping point $J=0.0099\, U_{0}$ which marks the boundary between the phases
$\Psi_{1}\neq0,\:\Psi_{0}=\Psi_{-1}=0$ and $\Psi_{1}\neq0,\:\Psi_{-1}\neq0,\,\Psi_{0}=0$. In addition,
this point is the same point where the two solutions for $\Psi_{1}$ intersect in  \fref{fig:Plot-of-the-1}.
Therefore, the transition between these two phases is of second order. Similarly, the transition from
$\Psi_{1}\neq0,\:\Psi_{0}=\Psi_{-1}=0$ to $\Psi_{1}\neq0,\:\Psi_{-1}\neq0,\,\Psi_{0}\neq0$ turns out
to be second order as shown in \fref{fig:(a)-Condensate-density}. 

\begin{center}
\begin{figure}[t]
\begin{centering}
\includegraphics[width=6cm,height=5cm]{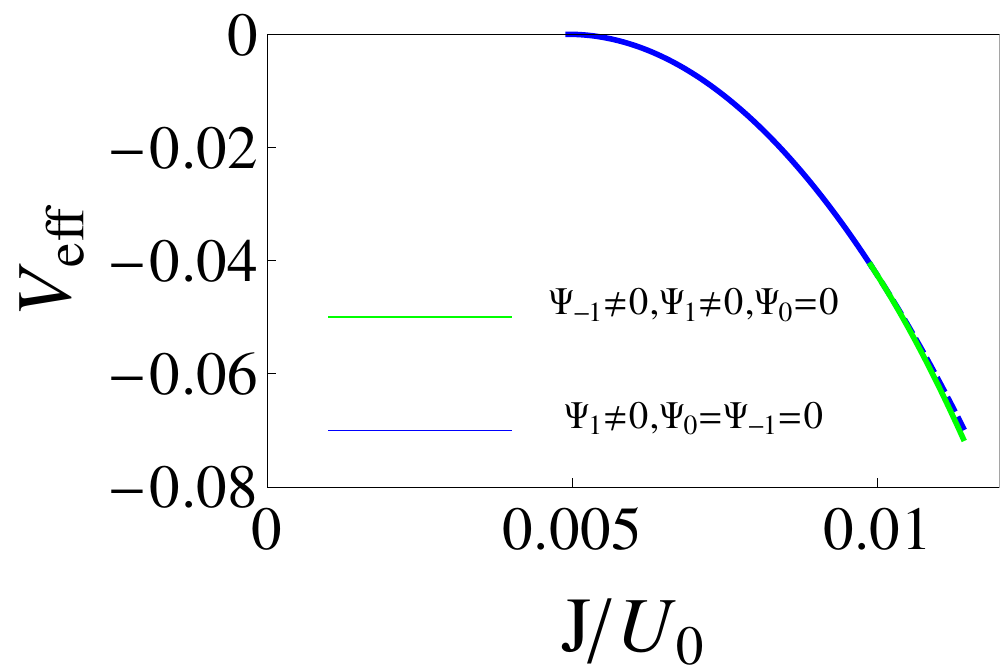}
\par\end{centering}

\caption{\label{fig:Plot-of-the-1}Effective potential for the phases $\Psi_{1}\neq0,\,\Psi_{-1}=0,\,\Psi_{0}=0$
and $\Psi_{1}\neq0,\:\Psi_{-1}=\Psi_{0}=0$ as a function of the tunneling parameter $J/U_{0}$ of spin-1
Bose-Hubbard model in the anti-ferromagnetic case with $\eta=0.125\, U_{0}$, $U_{2}=0.04\, U_{0}$ and
$\mu=2.864\, U_{0}$ at zero temperature. }
\end{figure}
\begin{figure}[t]
\centering{}\subfloat[\label{fig:The-condensate-density-2-1}]{

\includegraphics[width=6cm,height=5cm]{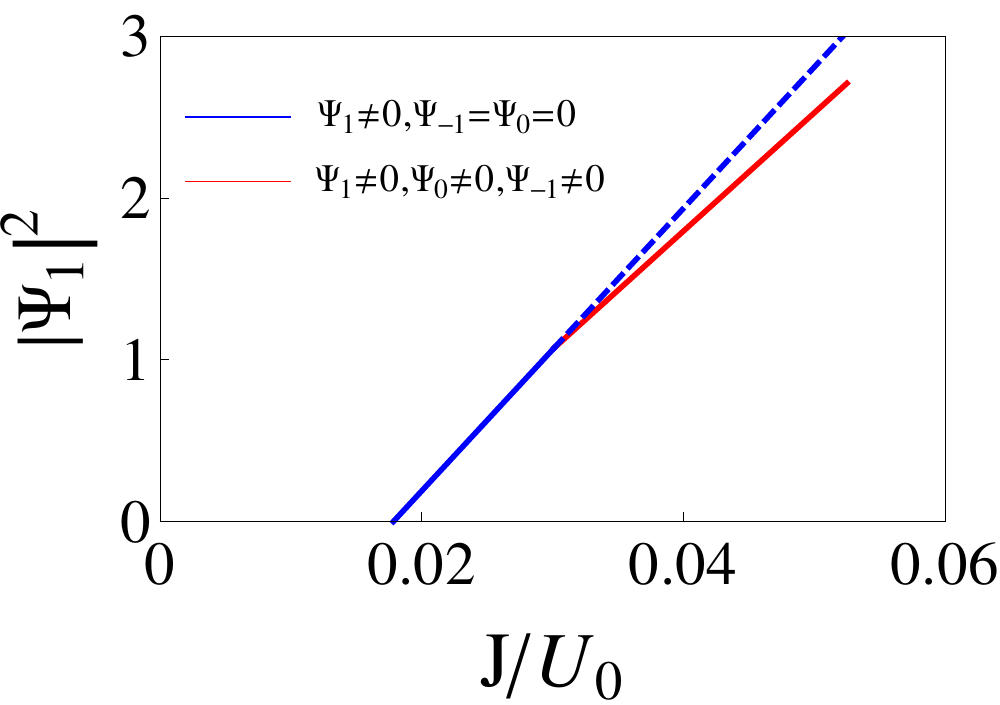}}\hfill
\subfloat[\label{fig:Plot-of-the-1-1}]{

\includegraphics[width=6cm,height=5cm]{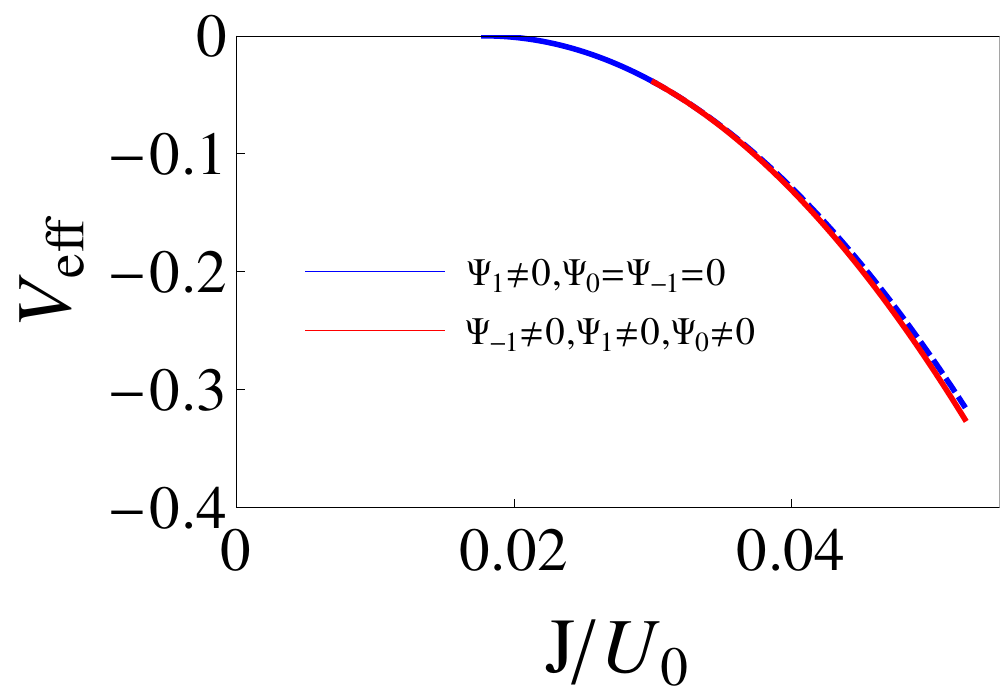}

}\caption{\label{fig:(a)-Condensate-density}(a) Condensate density and (b) effective potential for the phases
$\Psi_{1}\neq0,\:\Psi_{-1}=\Psi_{0}=0$ and $\Psi_{1}\neq0,\:\Psi_{-1}\neq0,\,\Psi_{0}\neq0$ as a function
of the tunneling parameter $J/U_{0}$ of spin-1 Bose-Hubbard model in the anti-ferromagnetic case with
$\eta=0.05\, U_{0}$, $U_{2}=0.04\, U_{0}$ and $\mu=1.756\, U_{0}$ at zero temperature. }
\end{figure}

\par\end{center}

\begin{figure}[t]
\centering{}
\subfloat[\label{fig:The-condensate-density-3}]{\includegraphics[width=6cm,height=5cm]{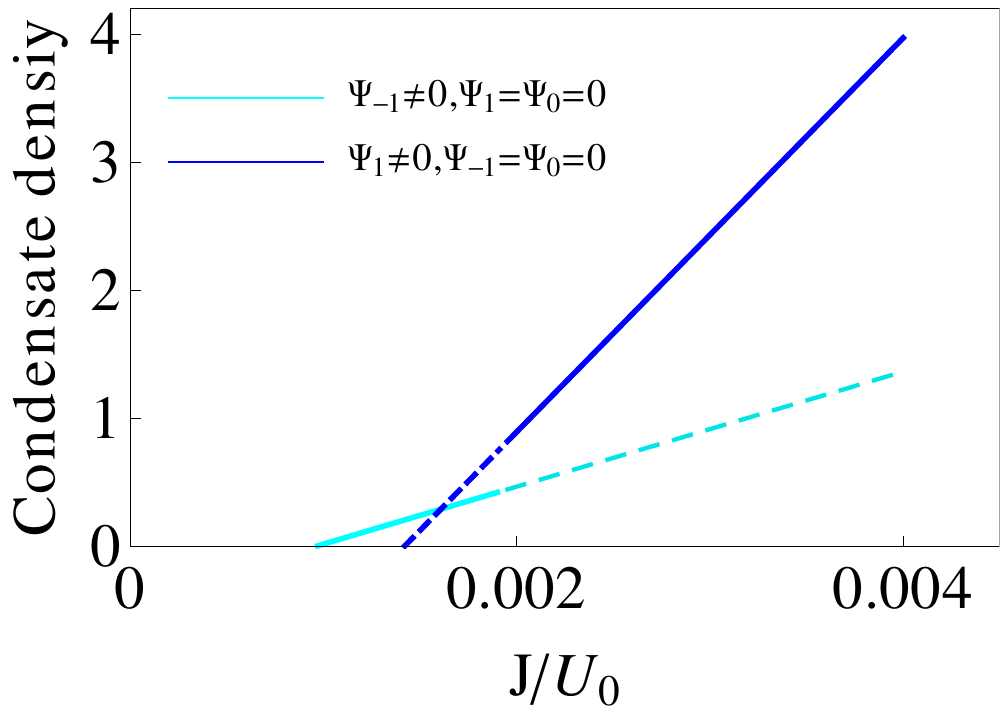}}\hfill
\subfloat[\label{fig:Plot-of-the}]{\includegraphics[width=6.5cm,height=5.2cm]{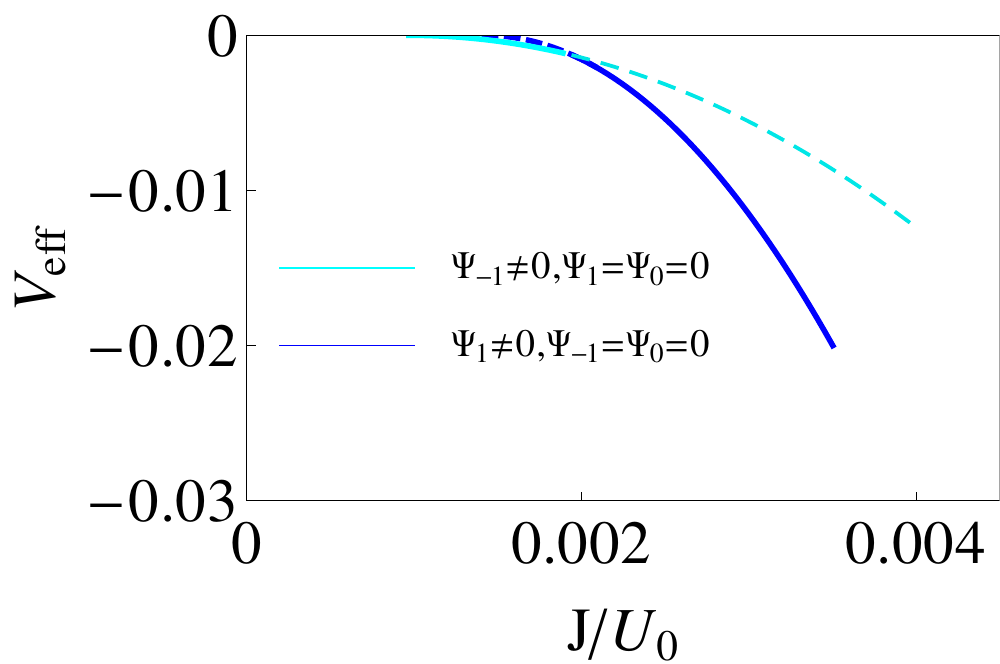}}
\caption{(a) Condensate density and (b) effective potential for the phases $\Psi_{-1}\neq0,\:\Psi_{0}=\Psi_{1}=0$
and $\Psi_{1}\neq0,\:\Psi_{-1}=\Psi_{0}=0$ as a function of the tunneling parameter $J/U_{0}$ of spin-1
Bose-Hubbard model in the anti-ferromagnetic case with $\eta=0.125\, U_{0}$ and $U_{2}=0.04\, U_{0}$
and $\mu=2.96\, U_{0}$ at zero temperature. }
\end{figure}

A different situation occurs when we study the transition from the phase $\Psi_{1}\neq0,\:\Psi_{0}=\Psi_{-1}=0$
to  the phase $\Psi_{-1}\neq0,\:\Psi_{1}=\Psi_{0}=0$ or vice versa. To this end, we focus on the transition from
the phase  $\Psi_{-1}\neq0,\:\Psi_{0}=\Psi_{1}=0$ to the phase $\Psi_{1}\neq0,\:\Psi_{-1}=\Psi_{0}=0$ at a
fixed $\mu=2.96\, U_{0}$ around $\eta=0.125\, U_{0}$ and $U_{2}=0.04\, U_{0}$ for the third lobe as
seen in \fref{fig:The-condensate-density-3}. We note that the phase $\Psi_{1}\neq0,\:\Psi_{-1}=\Psi_{0}=0$
jumps at the intersection point, which is 0.00195199 $U_{0}$ according to \fref{fig:Plot-of-the}.
Furthermore, the dashed line in \fref{fig:Plot-of-the} indicates that we can not take this phase 
because it does not provide a minimal energy. Therefore, the transition between the phases 
$\Psi_{-1}\neq0,\:\Psi_{0}=\Psi_{1}=0$ and $\Psi_{1}\neq0,\:\Psi_{-1}=\Psi_{0}=0$  is of first order.

\section{Conclusion}

In conclusion, we have worked out a Ginzburg-Landau theory for spin-1 bosons in a cubic optical lattice
within its range of validity and investigated  at zero temperature the resulting different superfluid
phases for an anti-ferromagnetic interaction in the presence of an external magnetic field. Inspecting
the energies of the respective phases in the vicinity of their boundaries even  allows to determine the order
of the quantum phase transition. With this we find that the quantum phase transition from the Mott insulator
to the superfluid phase is of second order for spin-1 bosons in a cubic optical lattice under the effect
of the magnetic field at zero temperature. Thus, our finding disagrees with Kimura et al. \cite{key-61-1},
where a first-order SF-MI phase transition was found at some  part of the phase boundary by using the
Gutzwiller variational approach. Furthermore, depending on the particle number, the spin-dependent interaction
and the value of the magnetic field we find new superfluid phases with a macroscopic occupation of the
two spin states $\pm1$ or even of all three spin states $0,\pm1$. This is different from the mean-field
approximation, which only predicts two superfluid phases with spins aligned or opposite to the field
direction \cite{key-15,key-60-1}. Finally, we find that the transition between the SF phases $\Psi_{1}\neq0,\:\Psi_{0}=\Psi_{-1}=0$
to $\Psi_{1}\neq0,\:\Psi_{-1}\neq0,\,\Psi_{0}=0$ is of second order and the transition between the SF
phases $\Psi_{-1}\neq0,\:\Psi_{0}=\Psi_{1}=0$ and $\Psi_{1}\neq0,\:\Psi_{-1}=\Psi_{0}=0$ is of first
order at a fixed values of chemical potential, an external magnetic field, and spin-dependent interaction.
It is interesting to observe that both a first- and second-order phase transition  can occur above the same
Mott lobe in the superfluid phase. 

In this paper we have restricted ourselves to apply the Ginzburg-Landau theory for studying the emergence
of different magnetic Mott insulator and superfluid phases. However, we note that this theory would also
allow, in principle, to investigate the collective excitations of all these different phases. In Ref.~\cite{key-160}
already the corresponding spin-0 case was treated, where particle- and hole excitations characterize
the Mott insulator phase, whereas the superfluid phase yields both a Goldstone and a Higgs mode \cite{key-162,key-163-1,key-200}.
In principle, even nonequilibrium problems could be investigated within the realm of our Ginzburg-Landau
theory. For instance, it would be challenging to investigate how the quench dynamics 
differs when we sweep through  a first-order or a second-order phase transition in the superfluid phase.

Certainly, it would be interesting to study in detail also how all these results would change for more
general spinor Bose gas systems. One example is provided by the competition between the linear Zeeman
effect, considered here, and its quadratic counterpart (see, for instance, Refs. \cite{key-161,key-164}),
another one would be substituting the nonfrustrated cubic by a frustrated triangular optical lattice
\cite{key-53-2}. Finally, one can expect even more complex magnetic Mott insulator and superfluid phases
for spin-2 or spin-3 bosons, which could be realized, for instance, with $^{87}$Rb \cite{key-42-3}
and $^{52}$Cr  atoms \cite{key-163}. 

\section{Acknowledgments}
We thank Mathias Ohliger for many useful discussions at an early stage of this work. Furthermore, we
acknowledge financial support from the Egyptian Government as well as from the German Research Foundation
(DFG) via the Collaborative Research Center SFB/TR49 Condensed Matter Systems with Variable Many-Body
Interactions.

\appendix

\section{Matrix Elements\label{sec:Matrix-Elements}}

The matrix elements $M$, $N$, $O$ and $P$ from Eqs.~(\ref{eq:3-31-1}), (\ref{eq:3-32-1}) represent
the mathematical backbone for analyzing spin-1 bosons in a lattice. Initially, they were calculated individually
in a stepwise procedure in Refs.~\cite{key-60-1,key-4}. In this appendix, however, we follow Ref.~\cite{key-10}
and determine these matrix elements by a recursive procedure. In particular at finite temperature, when
many of these matrix elements have to be evaluated in (\ref{eq:4-30}) and (\ref{eq:4-6}), this recursive
approach turns out to be more efficient than the original stepwise procedure.

We start with characterizing the ground state of the on-site Hamiltonian (\ref{eq:13-1}) via \cite{key-114}

\begin{eqnarray}
\left|S,S,n\right\rangle = & \frac{1}{\sqrt{f(n,S)}}\hat{a}_{1}^{\dagger S}\left(\hat{\Theta}^{\dagger}\right)^{\left(n-S\right)/2}\left|0,0,0\right\rangle ,\label{eq:a7}
\end{eqnarray}
where the normalization factor is given by 
\begin{eqnarray}
f(n,S) & =S!\left(\frac{n-S}{2}\right)!2^{\left(n-S\right)/2}\frac{(n+S+1)!!}{(2S+1)!!},\label{eq:a4}
\end{eqnarray}
and $\hat{\Theta}^{\dagger}=\hat{a}_{0}^{\dagger2}-2\hat{a}_{1}^{\dagger}\hat{a}_{-1}^{\dagger}$ represents
the creation operator of a spin singlet pair. By applying the ladder operators $\hat{\mathbf{\mathit{S}}}_{+}=\sqrt{2}(\hat{a}_{1}^{\dagger}\hat{a}_{0}+\hat{a}_{0}^{\dagger}\hat{a}_{-1})$
and $\hat{\mathbf{\mathit{S}}}_{-}=\sqrt{2}(\hat{a}_{0}^{\dagger}\hat{a}_{1}+\hat{a}_{-1}^{\dagger}\hat{a}_{0})$
on the ground state $\left|S,S,n\right\rangle $, we get the excited states $\left|S,m,n\right\rangle $
with $m<S$.

Now, we turn to calculate the matrix elements $M$, $N$, $O$ and $P$ in Eqs.~(\ref{eq:3-31-1}),
(\ref{eq:3-32-1}). The first substantial consideration declares that no state $\left|S,m,n\right\rangle $
with $m>S$ does exist, so we have 
\begin{equation}
N_{1,S,S,n}=N_{0,S,S,n}=P_{0,S,S,n}=P_{-1,S,S,n}=0,
\end{equation}
and
\begin{equation}
\hat{a}_{1}^{\dagger}\left|S,S,n\right\rangle =M_{1,S,S,n}\left|S+1,S+1,n+1\right\rangle .\label{eq:a5}
\end{equation}
On the other hand we conclude from (\ref{eq:a7}) and (\ref{eq:a4})

\begin{eqnarray}
\hat{a}_{1}^{\dagger}\left|S,S,n\right\rangle = & \sqrt{\frac{(S+1)(n+S+3)}{2S+3}}\left|S+1,S+1,n+1\right\rangle ,\label{eq:a6}
\end{eqnarray}
so, comparing (\ref{eq:a5}) and (\ref{eq:a6}) yields
\begin{eqnarray}
M_{1,S,S,n} & =\sqrt{\frac{(S+1)(n+S+3)}{2S+3}}.
\end{eqnarray}
In this manner, we put our hands on the first matrix element with \emph{$m=S$}. In order to calculate
recursively $M_{1,S,m,n}$ with \emph{$m<S$}, we apply $\hat{\mathbf{\mathit{S}}}_{+}$on Eq.~(\ref{eq:3-31-1})
and obtain
\begin{equation}
M_{1,S,m,n}=\sqrt{\frac{S(S+1)-m(m+1)}{(S+1)(S+2)-(m+1)(m+2)}}M_{1,S,m+1,n}.
\end{equation}
This recursion relation is useful to calculate $M_{1,S,m-1,n}$ from $M_{1,S,m,n}$. We can use the same
procedure to calculate the matrix element with $\alpha=0$ 

\begin{eqnarray}
M_{0,S,m,n}=\sqrt{\frac{S(S+1)-m(m+1)}{(S+1)(S+2)-m(m+1)}}M_{0,S,m+1,n}\nonumber \\
+\sqrt{\frac{2}{(S+1)(S+2)-m(m+1)}}M_{1,S,m,n}.
\end{eqnarray}
and also with $\alpha=-1$ 
\begin{eqnarray}
M_{-1,S,m,n}=\sqrt{\frac{S(S+1)-m(m+1)}{(S+1)(S+2)-m(m-1)}}M_{-1,S,m+1,n}\nonumber \\
+\sqrt{\frac{2}{(S+1)(S+2)-m(m-1)}}M_{0,S,m,n}.
\end{eqnarray}
Specializing \emph{$m=S$ }yields finally 
\begin{equation}
M_{-1,S,S,n}=\sqrt{\frac{n+S+3}{(2S+3)(2S+1)}}.
\end{equation}
Now we come to the evaluation of the Matrix elements $N_{\alpha,S,m,n}$. The particle number $n=\left\langle S,S,n\mid\hat{\mathbf{\mathit{n}}}\mid S,S,n\right\rangle $
can be written as
\begin{eqnarray}
n= \sum_{\alpha}\left\langle S,S,n\mid\hat{a}_{\alpha}\hat{a}_{\alpha}^{+}\mid S,S,n\right\rangle -3,
\end{eqnarray}
thus, we obtain with (\ref{eq:3-31-1}) and (\ref{eq:3-32-1}) 
\begin{eqnarray}
N_{-1,S,S,n}  =  -\sqrt{3+n-{\sum_{\alpha}}M_{\alpha,S,S,n}^{2}}.\label{eq:a14}
\end{eqnarray}
Matrix elements $N_{\alpha,S,m,n}$ with $m<S$ can be derived as above, yielding 
\begin{equation}
N_{-1,S,m,n}=\sqrt{\frac{S(S+1)-m(m-1)}{S(S-1)-(m-1)(m-2)}}N_{-1,S,m-1,n},
\end{equation}
 
\begin{eqnarray}
N_{0,S,m,n}= & \sqrt{\frac{S(S+1)-m(m-1)}{S(S-1)-m(m-1)}}N_{0,S,m-1,n}\nonumber \\
 & +\sqrt{\frac{2}{S(S-1)-m(m-1)}}N_{-1,S,m,n},
\end{eqnarray}
and also
\begin{eqnarray}
N_{1,S,m,n}= & \sqrt{\frac{S(S+1)-m(m-1)}{S(S-1)-m(m+1)}}N_{1,S,m-1,n}\nonumber \\
 & +\sqrt{\frac{2}{S(S+1)-m(m-1)}}N_{0,S,m,n}.
\end{eqnarray}
Thus, with this all matrix elements of the creation operators $\hat{a}_{\alpha}^{+}$ in (\ref{eq:3-31-1})
can be calculated. Therefore, we turn now to the calculation of the matrix element of the annihilation
operators $\hat{a}_{\alpha}$ in (\ref{eq:3-32-1}) by the identical method. In order to calculate $O_{-1,S,m,n}$,
we apply the operator $\hat{a}_{-1}$ to (\ref{eq:a7}), yielding 

\begin{equation}
O_{-1,S,S,n}=-\sqrt{\frac{(n-S)(S+1)}{2S+3}}.
\end{equation}
Applying $\hat{\mathbf{\mathit{S}}}_{+}$ on (\ref{eq:3-32-1}), we get
\begin{equation}
O_{-1,S,m,n}=\sqrt{\frac{S(S+1)-m(m+1)}{(S+1)(S+2)-(m+1)(m+2)}}O_{-1,S,m+1,n}.
\end{equation}
Similarly, we obtain the recursion relations
\begin{eqnarray}
O_{0,S,m,n}=\sqrt{\frac{S(S+1)-m(m+1)}{(S+1)(S+2)-m(m+1)}}O_{0,S,m+1,n}\nonumber \\
-\sqrt{\frac{2}{(S+1)(S+2)-m(m+1)}}O_{-1,S,m,n},
\end{eqnarray}
and 
\begin{eqnarray}
O_{1,S,m,n}=\sqrt{\frac{S(S+1)-m(m+1)}{(S+1)(S+2)-m(m-1)}}O_{1,S,m+1,n}\nonumber \\
-\sqrt{\frac{2}{(S+1)(S+2)-m(m-1)}}O_{0,S,m,n}.
\end{eqnarray}
In order to determine $P_{1,S,S,n}$, the particle number $n={\sum_{\alpha}}\bigl\langle S,S,n\mid\hat{a}_{\alpha}^{+}\hat{a}_{\alpha}\mid S,S,n\bigr\rangle$
reduces with (\ref{eq:3-32-1}) to 
\begin{eqnarray}
P_{1,S,S,n} & = & \sqrt{n-{\sum_{\alpha}}O_{\alpha,S,S,n}^{2}}.\label{eq:a15}
\end{eqnarray}
Applying $\hat{\mathbf{\mathit{\mathit{S}}}}_{-}$ on (\ref{eq:3-32-1}) we get
\begin{equation}
P_{1,S,m,n}=\sqrt{\frac{S(S+1)-m(m-1)}{S(S-1)-(m-1)(m-2)}}P_{1,S,m-1,n},
\end{equation}
 
\begin{eqnarray}
P_{0,S,m,n}= & \sqrt{\frac{S(S+1)-m(m-1)}{S(S-1)-m(m-1)}}P_{0,S,m-1,n}\nonumber \\
 & -\sqrt{\frac{2}{S(S-1)-m(m-1)}}P_{1,S,m,n},
\end{eqnarray}
and 
\begin{eqnarray}
P_{-1,S,m,n}= & \sqrt{\frac{S(S+1)-m(m-1)}{S(S-1)-m(m+1)}}P_{-1,S,m-1,n}\nonumber \\
 & -\sqrt{\frac{2}{S(S-1)-m(m+1)}}P_{0,S,m,n}.
\end{eqnarray}
Finally, we derive useful relations between these creation and annihilation matrix elements. Using (\ref{eq:2})
we get 

\begin{eqnarray}
\left\langle S,m,n\mid\hat{a}_{\alpha}^{+}\mid S-1,m-\alpha,n-1\right\rangle  & =M_{\alpha,S-1,m-\alpha,n-1}.
\end{eqnarray}
Taking into account (\ref{eq:3-32-1}) we obtain 

\begin{eqnarray}
\left\langle S,m,n\mid\hat{a}_{\alpha}^{+}\mid S-1,m-\alpha,n-1\right\rangle =P_{\alpha,S,m,n}.
\end{eqnarray}
Thus, we conclude 
\begin{eqnarray}
M_{\alpha,S-1,m-\alpha,n-1}=P_{\alpha,S,m,n}.\label{eq:28}
\end{eqnarray}
In a similar way we also obtain 
\begin{eqnarray}
P_{\alpha,S+1,m+\alpha,n+1}=M_{\alpha,S,m,n},\label{eq:a-74}
\end{eqnarray}
\begin{eqnarray}
N_{\alpha,S+1,m-\alpha,n-1}=O_{\alpha,S,m,n},
\end{eqnarray}
\begin{eqnarray}
O_{\alpha,S-1,m+\alpha,n+1}=N_{\alpha,S,m,n}.\label{eq:a-76}
\end{eqnarray}
Note that we have used the minus sign in (\ref{eq:a14}) and the positive sign in (\ref{eq:a15}) in
order to satisfy the relations (\ref{eq:28})--(\ref{eq:a-76}). With these above equations, any quantum
mechanical expectation value of the particle number operators in the respective hyperfine spin states
could be evaluated.

\section{Fourth-Order Coefficient\label{sec:fourth-order-coefficient}}

In this section, we calculate the fourth-order coefficient (\ref{eq:4-5}), which contains the expectation
values of time-ordered product of four operators. At first, we remark that there are six distinct permutations
leading to different expectation values for the time-ordered product of the annihilation and creation
operators. Each order has four time and four spin variable permutations corresponding to $\tau_{1}\leftrightarrow\tau_{2}$
, $\tau_{3}\leftrightarrow\tau_{4}$ , $\alpha_{1}\leftrightarrow\alpha_{2}$ and $\alpha_{3}\leftrightarrow\alpha_{4}$.
Thus, we have 24 terms for the above expectation value. Fortunately, we need to determine only six different
thermal averages for one specific time-ordering because, due to symmetry reasons, there are some integrals
over different time-variable permutations which yield the same result. Furthermore, as these expectation
values are local, we drop the site indices in the following calculations and calculate the following
expressions:
\begin{eqnarray}
\hspace{-2.5cm}\left\langle \hat{T}\left[\hat{a}_{\alpha_{4}}(\tau_{4})\hat{a}_{\alpha_{3}}(\tau_{3})\hat{a}_{\alpha_{1}}^{\dagger}(\tau_{1})\hat{a}_{\alpha_{2}}^{\dagger}(\tau_{2})\right]\right\rangle ^{(0)} & , & \left\langle \hat{T}\left[\hat{a}_{\alpha_{1}}^{\dagger}(\tau_{1})\hat{a}_{\alpha_{2}}^{\dagger}(\tau_{2})\hat{a}_{\alpha_{3}}(\tau_{3})\hat{a}_{\alpha_{4}}(\tau_{4})\right]\right\rangle ^{(0)},\\
\hspace{-2.5cm}\left\langle \hat{T}\left[\hat{a}_{\alpha_{4}}(\tau_{4})\hat{a}_{\alpha_{1}}^{\dagger}(\tau_{1})\hat{a}_{\alpha_{3}}(\tau_{3})\hat{a}_{\alpha_{2}}^{\dagger}(\tau_{2})\right]\right\rangle ^{(0)} & , & \left\langle \hat{T}\left[\hat{a}_{\alpha_{1}}^{\dagger}(\tau_{1})\hat{a}_{\alpha_{4}}(\tau_{4})\hat{a}_{\alpha_{3}}(\tau_{3})\hat{a}_{\alpha_{2}}^{\dagger}(\tau_{2})\right]\right\rangle ^{(0)},\\
\hspace{-2.5cm}\left\langle \hat{T}\left[\hat{a}_{\alpha_{4}}(\tau_{4})\hat{a}_{\alpha_{1}}^{\dagger}(\tau_{1})\hat{a}_{\alpha_{2}}^{\dagger}(\tau_{2})\hat{a}_{\alpha_{3}}(\tau_{3})\right]\right\rangle ^{(0)} & , & \left\langle \hat{T}\left[\hat{a}_{\alpha_{1}}^{\dagger}(\tau_{1})\hat{a}_{\alpha_{3}}(\tau_{3})\hat{a}_{\alpha_{2}}^{\dagger}(\tau_{2})\hat{a}_{\alpha_{4}}(\tau_{4})\right]\right\rangle ^{(0)}.
\end{eqnarray}
Following the same method as for the second-order expansion coefficient, we perform a Matsubara transformation.
For example we evaluate one expectation value as follows: 
\begin{eqnarray}
\hspace{-2.5cm}\left\langle \hat{T}\bigg[\hat{a}_{\alpha_{4}}(\tau_{4})\hat{a}_{\alpha_{3}}(\tau_{3})\hat{a}_{\alpha_{1}}^{\dagger}(\tau_{1})\hat{a}_{\alpha_{2}}^{\dagger}
(\tau_{2})\bigg]\right\rangle ^{(0)}=\frac{1}{\mathcal{Z}^{(0)}}\sum_{S,m,n}e^{-\beta E_{S,m,n}^{(0)}}\nonumber\\
\hspace{-2.5cm}\times \left\langle S,m,n|\hat{T}\left[\hat{a}_{\alpha_{4}}(\tau_{4})
\hat{a}_{\alpha_{3}}(\tau_{3})\hat{a}_{\alpha_{1}}^{\dagger}(\tau_{1})\hat{a}_{\alpha_{2}}^{\dagger}(\tau_{2})\right]|S,m,n\right\rangle 
=\frac{1}{\mathcal{Z}^{(0)}}\sum_{S,m,n}e^{-\beta E_{S,m,n}^{(0)}}\nonumber\\
\hspace{-2.5cm}\times \left\langle S,m,n|\hat{T}\left[e^{(\tau_{4}-\tau_{2})\hat{H}^{(0)}}
\hat{a}_{\alpha_{4}}e^{(\tau_{3}-\tau_{4})\hat{H}^{(0)}}\hat{a}_{\alpha_{3}}e^{(\tau_{1}-\tau_{3})\hat{H}^{(0)}}\hat{a}_{\alpha_{1}}^{\dagger}e^{(\tau_{2}
-\tau_{1})\hat{H}^{(0)}}\hat{a}_{\alpha_{2}}^{\dagger}\right]|S,m,n\right\rangle ,
\end{eqnarray}
which leads to 

\begin{eqnarray}
\hspace{-2.5cm}\left\langle \hat{T}\left[\hat{a}_{\alpha_{4}}(\tau_{4})\hat{a}_{\alpha_{3}}(\tau_{3})\hat{a}_{\alpha_{1}}^{\dagger}(\tau_{1})\hat{a}_{\alpha_{2}}^{\dagger}(\tau_{2})
\right]\right\rangle ^{(0)}=\frac{\delta_{\alpha_{1}+\alpha_{2},\alpha_{3}+\alpha_{4}}}{\mathcal{Z}^{(0)}}\sum_{S,m,n}e^{-\beta E_{S,m,n}^{(0)}}\Theta(\tau_{4}-\tau_{3})\Theta(\tau_{3}
-\tau_{1})\nonumber \\
\hspace{-2.5cm}\Theta(\tau_{1}-\tau_{2})e^{(\tau_{4}-\tau_{2})E_{S,m,n}^{(0)}}
\Biggl[M_{\alpha_{4},S,m,n}M_{\alpha_{3},S,m,n}M_{\alpha_{1},S,m,n}M_{\alpha_{2},S,m,n}\, e^{(\tau_{2}-\tau_{1})E_{S+1,m+\alpha_{2},n+1}^{(0)}}\nonumber \\
\hspace{-2.5cm}\times e^{(\tau_{1}-\tau_{3})E_{S+2,m+\alpha_{2}+\alpha_{1},n+2}^{(0)}}e^{(\tau_{3}-\tau_{4})E_{S+1,m+\alpha_{4},n+1}^{(0)}}
+M_{\alpha_{4},S,m,n}M_{\alpha_{2},S,m,n}N_{\alpha_{3},S+1,m+\alpha_{4},n+1}\nonumber \\
\hspace{-2.5cm}\times N_{\alpha_{1},S+1,m+\alpha_{2},n+1}\, 
e^{(\tau_{2}-\tau_{1})E_{S+1,m+\alpha_{2},n+1}^{(0)}}e^{(\tau_{1}-\tau_{3})E_{S,m+\alpha_{3}+\alpha_{4},n+2}^{(0)}}
e^{(\tau_{3}-\tau_{4})E_{S+1,m+\alpha_{4},n+1}^{(0)}}\nonumber \\
\hspace{-2.5cm}+M_{\alpha_{4},S,m,n}N_{\alpha_{2},S,m,n}N_{\alpha_{3},S+1,m+\alpha_{4},n+1}M_{\alpha_{1},S-1,m+\alpha_{2},n+1}e^{(\tau_{2}-\tau_{1})E_{S-1,m+\alpha_{2},n+1}^{(0)}}
 e^{(\tau_{1}-\tau_{3})E_{S,m+\alpha_{3}+\alpha_{4},n+2}^{(0)}}\nonumber \\
\hspace{-2.5cm}\times e^{(\tau_{3}-\tau_{4})E_{S+1,m+\alpha_{4},n+1}^{(0)}}+N_{\alpha_{4},S,m,n}M_{\alpha_{2},S,m,n}M_{\alpha_{3},S-1,m+\alpha_{4},n+1}N_{\alpha_{1},S+1,m+\alpha_{2},n+1}\nonumber \\
\hspace{-2.5cm}\times e^{(\tau_{2}-\tau_{1})E_{S+1,m+\alpha_{2},n+1}^{(0)}}e^{(\tau_{1}-\tau_{3})E_{S,m+\alpha_{3}+\alpha_{4},n+2}^{(0)}}e^{(\tau_{3}-\tau_{4})E_{S-1,m+\alpha_{4},n+1}^{(0)}}\nonumber \\
\hspace{-2.5cm}+N_{\alpha_{4},S,m,n}N_{\alpha_{2},S,m,n}M_{\alpha_{3},S-1,m+\alpha_{4},n+1}M_{\alpha_{1},S-1,m+\alpha_{2},n+1}e^{(\tau_{2}-\tau_{1})E_{S-1,m+\alpha_{2},n+1}^{(0)}}
e^{(\tau_{1}-\tau_{3})E_{S,m+\alpha_{3}+\alpha_{4},n+2}^{(0)}}\nonumber \\
\hspace{-2.5cm}\times e^{(\tau_{3}-\tau_{4})E_{S-1,m+\alpha_{4},n+1}^{(0)}}+N_{\alpha_{4},S,m,n}N_{\alpha_{2},S,m,n}N_{\alpha_{3},S-1,m+\alpha_{4},n+1}N_{\alpha_{1},S-1,m+\alpha_{2},n+1}\nonumber\\ 
\hspace{-2.5cm}\times e^{(\tau_{2}-\tau_{1})E_{S-1,m+\alpha_{2},n+1}^{(0)}}e^{(\tau_{1}-\tau_{3})E_{S-2,m+\alpha_{3}+\alpha_{4},n+2}^{(0)}}e^{(\tau_{3}-\tau_{4})E_{S-1,m+\alpha_{4},n+1}^{(0)}}\biggr]
\end{eqnarray}
Using Matsubara transformation (\ref{eq:3-50}) yields the following integral of the form

\begin{eqnarray}
\hspace{-2.5cm}\mathcal{I}=\kappa\int_{0}^{\beta}d\tau_{1}e^{a\tau_{1}}\int_{0}^{\tau_{1}}d\tau_{2}
e^{b\tau_{2}}\int_{0}^{\tau_{2}}d\tau_{3}e^{b\tau_{3}}\int_{0}^{\tau_{3}}d\tau_{4}e^{d\tau_{4}}\nonumber \\
\hspace{-2.5cm}=\kappa\Biggl[\frac{e^{\left(a+b+c+d\right)\beta}-1}{\left(a+b+c+d\right)\left(b+c+d\right)\left(c+d\right)d}-\frac{e^{a\beta}-1}{a\left(b+c+d\right)\left(c+d\right)d}
-\frac{e^{\left(a+b\right)\beta}-1}{b\left(a+b\right)\left(c+d\right)d}\nonumber \\
 \hspace{-2.5cm}-\frac{e^{\left(a+b+c\right)\beta}-1}{\left(a+b+c\right)\left(b+c\right)cd}+\frac{e^{a\beta}-1}{ab\left(c+d\right)d}+\frac{e^{\left(a+b\right)\beta}-1}{\left(a+b\right)bcd}
 -\frac{e^{a\beta}-1}{abcd}+\frac{e^{a\beta}-1}{a\left(b+c\right)cd}\Biggr].\label{eq:4-1-1}
\end{eqnarray}
In the case of $a+b+c+d=0$, we need 
\begin{eqnarray}
\hspace{-3cm}\lim_{a+b+c+d\rightarrow0} & \frac{e^{\left(a+b+c+d\right)\beta}-1}{\left(a+b+c+d\right)\left(b+c+d\right)\left(c+d\right)d}=\frac{\beta}{\left(a+b+c+d\right)\left(b+c+d\right)\left(c+d\right)d}.\label{eq:4-2-1}
\end{eqnarray}
If $b+c=0$, so
\begin{eqnarray}
\hspace{-2.5cm}\lim_{b+c\rightarrow0} & \left(\frac{e^{a\beta}-1}{a\left(b+c\right)cd}-\frac{e^{\left(a+b+c\right)\beta}-1}{\left(a+b+c\right)\left(b+c\right)cd}\right)=\frac{e^{a\beta}-1}{a^{2}cd}-\frac{\beta e^{a\beta}}{acd}.\label{eq:4-3-1}
\end{eqnarray}
Similarly, when $c+d\rightarrow0$ and $a+b\rightarrow0$, we get 
\begin{eqnarray}
\hspace{-2.5cm}\lim_{{c+d\rightarrow0\atop a+b\rightarrow0}}  \left(\frac{e^{\left(a+b+c+d\right)\beta}-1}{\left(a+b+c+d\right)\left(b+c+d\right)\left(c+d\right)d}
-\frac{e^{\left(a+b\right)\beta}-1}{b\left(a+b\right)\left(c+d\right)d}\right).\label{eq:4-4-1}\nonumber\\
  =-\frac{\beta}{b^{2}d}-\frac{a\beta^{2}}{2(b^{2}d)}.
\end{eqnarray}
Making use of the integral (\ref{eq:4-1-1}) and Eqs.~(\ref{eq:4-2-1}), (\ref{eq:4-3-1}) and (\ref{eq:4-4-1}),
we finally  get

\begin{eqnarray}
\hspace{-2.5cm}a_{4}^{(0)}(\alpha_{1},\omega_{m1};\alpha_{2},\omega_{m2}|\alpha_{3},\omega_{m3};\alpha_{4},\omega_{m4})=\frac{1}{\beta}
\frac{1}{\mathcal{Z}^{(0)}}\sum_{S,m,n}e^{-\beta E_{S,m,n}^{(0)}}\delta_{\alpha_{1}+\alpha_{2},\alpha_{3}+\alpha_{4}}\delta_{\omega_{m1}+\omega_{m2},\omega_{m3}+\omega_{m4}}\nonumber\\
\hspace{-2.5cm}\times\left\{ \frac{M_{\alpha_{2},S,m,n}M_{\alpha_{4},S,m,n}M_{\alpha_{3},S+1,m+\alpha_{4},n+1}M_{\alpha_{1},S+1,m+\alpha_{2},n+1}}{\left(\bigtriangleup E_{S+1,m+\alpha_{4},n+1}^{(0)}
+i\omega_{m3}-i\omega_{m1}-i\omega_{m2}\right)\left(\triangle E_{S+2,m+\alpha_{2}+\alpha_{1},n+2}^{(0)}-i\omega_{m1}-i\omega_{m2}\right)}\right.\nonumber\\
\hspace{-2.5cm}\times\frac{1}{\left(\triangle E_{S+1,m+\alpha_{2},n+1}^{(0)}-i\omega_{m2}\right)}+\frac{N_{\alpha_{2},S,m,n}N_{\alpha_{4},S,m,n}N_{\alpha_{3},S-1,m+\alpha_{4},n+1}
N_{\alpha_{1},S-1,m+\alpha_{2},n+1}}{\left(\triangle E_{S-2,m+\alpha_{2}+\alpha_{1},n+2}^{(0)}-i\omega_{m1}-i\omega_{m2}\right)\left(\triangle E_{S-1,m+\alpha_{2},n+1}^{(0)}
-i\omega_{m2}\right)}\nonumber\\
\hspace{-2.5cm}\times\frac{1}{\left(\triangle E_{S-1,m+\alpha_{4},n+1}^{(0)}+i\omega_{m3}-i\omega_{m1}-i\omega_{m2}\right)}
+\frac{O_{\alpha_{1},S,m,n}O_{\alpha_{4},S,m,n}O_{\alpha_{2},S+1,m-\alpha_{1},n-1}O_{\alpha_{3},S+1,m-\alpha_{4},n-1}}
{\left(\triangle E_{S+1,m-\alpha_{1},n-1}^{(0)}+i\omega_{m3}+i\omega_{m4}-i\omega_{m2}\right)}\nonumber\\
\hspace{-2.5cm}\times\frac{1}{\left(\triangle E_{S+2,m-\alpha_{2}-\alpha_{1},n-2}^{(0)}+i\omega_{m3}+i\omega_{m4}\right)\left(\triangle E_{S+1,m-\alpha_{4},n-1}^{(0)}
+i\omega_{m4}\right)}+\frac{1}{\left(\triangle E_{S-1,m-\alpha_{4},n-1}^{(0)}+i\omega_{m4}\right)}\nonumber\\
\hspace{-2.5cm}\times\frac{P_{\alpha_{1},S,m,n}P_{\alpha_{4},S,m,n}P_{\alpha_{2},S+1,m-\alpha_{1},n-1}P_{\alpha_{3},S+1,m-\alpha_{4},n-1}}
{\left(\triangle E_{S-1,m-\alpha_{1},n-1}^{(0)}+i\omega_{m3}+i\omega_{m4}-i\omega_{m2}\right)\left(\triangle E_{S-2,m-\alpha_{2}-\alpha_{1},n-2}^{(0)}
+i\omega_{m3}+i\omega_{m4}\right)}\nonumber\\
\hspace{-2.5cm}+\frac{1}{\left(\triangle E_{S,m+\alpha_{3}+\alpha_{4},n+2}^{(0)}-i\omega_{m1}-i\omega_{m2}\right)}
\left(\frac{M_{\alpha_{4},S,m,n}N_{\alpha_{3},S+1,m+\alpha_{4},n+1}}{\bigtriangleup E_{S+1,m+\alpha_{4},n+1}^{(0)}+i\omega_{m3}-i\omega_{m1}-i\omega_{m2}}\right.\nonumber\\
\hspace{-2.5cm}\left.+\frac{N_{\alpha_{4},S,m,n}M_{\alpha_{3},S-1,m+\alpha_{4},n+1}}{\bigtriangleup E_{S-1,m+\alpha_{4},n+1}^{(0)}+i\omega_{m3}-i\omega_{m1}-i\omega_{m2}}\right)
\left(\frac{M_{\alpha_{2},S,m,n}N_{\alpha_{1},S+1,m+\alpha_{2},n+1}}{\bigtriangleup E_{S+1,m+\alpha_{2},n+1}^{(0)}-i\omega_{m2}}\right.\nonumber\\
\hspace{-2.5cm}\left.+\frac{N_{\alpha_{2},S,m,n}M_{\alpha_{1},S-1,m+\alpha_{2},n+1}}{\bigtriangleup E_{S-1,m+\alpha_{2},n+1}^{(0)}-i\omega_{m2}}\right)+\frac{1}
{\triangle E_{S,m-\alpha_{1}-\alpha_{2},n-2}^{(0)}+i\omega_{m3}+i\omega_{m4}}\nonumber\\
\hspace{-2.5cm}\times\left(\frac{O_{\alpha_{1},S,m,n}P_{\alpha_{2},S+1,m-\alpha_{1},n-1}}{\triangle E_{S+1,m-\alpha_{1},n-1}^{(0)}+i\omega_{m3}+i\omega_{m4}-i\omega_{m2}}
+\frac{P_{\alpha_{1},S,m,n}O_{\alpha_{2},S-1,m-\alpha_{1},n-1}}{\triangle E_{S-1,m-\alpha_{1},n-1}^{(0)}+i\omega_{m3}+i\omega_{m4}-i\omega_{m2}}\right)\nonumber\\
\hspace{-2.5cm}\times\left(\frac{O_{\alpha_{4},S,m,n}P_{\alpha_{3},S+1,m-\alpha_{4},n-1}}{\triangle E_{S+1,m-\alpha_{4},n-1}^{(0)}+i\omega_{m4}}+\frac{P_{\alpha_{4},S,m,n}
O_{\alpha_{3},S-1,m-\alpha_{4},n-1}}{\triangle E_{S-1,m-\alpha_{4},n-1}^{(0)}+i\omega_{m4}}\right)+\frac{1}{\triangle E_{S+2,m+\alpha_{4}-\alpha_{1},n}^{(0)}-i\omega_{m2}+i\omega_{m3}}\nonumber\\
\hspace{-2.5cm}\times\left(\frac{M_{\alpha_{4},S,m,n}O_{\alpha_{1},S+1,m+\alpha_{4},n+1}}{\bigtriangleup E_{S+1,m+\alpha_{4},n+1}^{(0)}+i\omega_{m3}-i\omega_{m1}-i\omega_{m2}}
+\frac{O_{\alpha_{1},S,m,n}M_{\alpha_{4},S+1,m-\alpha_{1},n-1}}{\bigtriangleup E_{S+1,m-\alpha_{1},n-1}^{(0)}+i\omega_{m3}+i\omega_{m4}-i\omega_{m2}}\right)\nonumber\\
\hspace{-2.5cm}\times\left(\frac{M_{\alpha_{2},S,m,n}O_{\alpha_{3},S+1,m+\alpha_{2},n+1}}{\bigtriangleup E_{S+1,m+\alpha_{2},n+1}^{(0)}-i\omega_{m2}}+\frac{O_{\alpha_{3},S,m,n}
M_{\alpha_{2},S+1,m-\alpha_{3},n-1}}{\bigtriangleup E_{S+1,m-\alpha_{3},n-1}^{(0)}+i\omega_{m3}}\right)+\frac{1}{\triangle E_{S-2,m+\alpha_{4}-\alpha_{1},n}^{(0)}-i\omega_{m2}
+i\omega_{m3}}\nonumber\\
\hspace{-2.5cm}\times\left(\frac{N_{\alpha_{4},S,m,n}P_{\alpha_{1},S-1,m+\alpha_{4},n+1}}{\bigtriangleup E_{S-1,m+\alpha_{4},n+1}^{(0)}+i\omega_{m3}-i\omega_{m1}-i\omega_{m2}}
+\frac{P_{\alpha_{1},S,m,n}N_{\alpha_{4},S+1,m-\alpha_{1},n-1}}{\bigtriangleup E_{S-1,m-\alpha_{1},n-1}^{(0)}+i\omega_{m3}+i\omega_{m4}-i\omega_{m2}}\right)\nonumber\\
\hspace{-2.5cm}\times\left(\frac{N_{\alpha_{2},S,m,n}P_{\alpha_{3},S-1,m+\alpha_{2},n+1}}{\bigtriangleup E_{S-1,m+\alpha_{2},n+1}^{(0)}-i\omega_{m2}}+\frac{P_{\alpha_{3},S,m,n}
N_{\alpha_{2},S-1,m-\alpha_{3},n-1}}{\bigtriangleup E_{S-1,m-\alpha_{3},n-1}^{(0)}+i\omega_{m3}}\right)+\delta_{\alpha_{1},\alpha_{4}}\delta_{\omega_{m1},\omega_{m4}}\nonumber\\
\hspace{-2.5cm}\times\left(\frac{M_{\alpha_{4},S,m,n}M_{\alpha_{1},S,m,n}}{\bigtriangleup E_{S+1,m+\alpha_{4},n+1}^{(0)}-i\omega_{m1}}\right.
+\frac{N_{\alpha_{4},S,m,n}N_{\alpha_{1},S,m,n}}{\bigtriangleup E_{S-1,m+\alpha_{4},n+1}^{(0)}-i\omega_{m1}}
+\frac{O_{\alpha_{4},S,m,n}O_{\alpha_{1},S,m,n}}{\bigtriangleup E_{S+1,m-\alpha_{1},n-1}^{(0)}-i\omega_{m4}}\nonumber\\
\hspace{-2.5cm}+\left.\frac{P_{\alpha_{4},S,m,n}P_{\alpha_{1},S,m,n}}{\bigtriangleup E_{S-1,m-\alpha_{1},n-1}^{(0)}-i\omega_{m4}}\right)
\left(\frac{M_{\alpha_{3},S,m,n}M_{\alpha_{2},S,m,n}}{\bigtriangleup E_{S+1,m+\alpha_{2},n+1}^{(0)}-i\omega_{m2}}
\left(\frac{\beta}{2}-\frac{1}{\bigtriangleup E_{S+1,m+\alpha_{2},n+1}^{(0)}-i\omega_{m3}}\right)\right.\nonumber \\
\hspace{-2.5cm}+\frac{N_{\alpha_{3},S,m,n}N_{\alpha_{2},S,m,n}}{\bigtriangleup E_{S-1,m+\alpha_{2},n+1}^{(0)}-i\omega_{m2}}\left(\frac{\beta}{2}
-\frac{1}{\bigtriangleup E_{S-1,m+\alpha_{2},n+1}^{(0)}-i\omega_{m3}}\right)\nonumber \\
\hspace{-2.5cm}+\frac{O_{\alpha_{3},S,m,n}O_{\alpha_{2},S,m,n}}{\bigtriangleup E_{S+1,m-\alpha_{3},n-1}^{(0)}-i\omega_{m3}}\left(\frac{\beta}{2}
-\frac{1}{\bigtriangleup E_{S+1,m-\alpha_{3},n-1}^{(0)}-i\omega_{m2}}\right)\nonumber \\
\hspace{-2.5cm}+\left.\left.\frac{P_{\alpha_{3},S,m,n}P_{\alpha_{2},S,m,n}}{\bigtriangleup E_{S+1,m-\alpha_{3},n-1}^{(0)}-i\omega_{m3}}
\left(\frac{\beta}{2}-\frac{1}{\bigtriangleup E_{S+1,m-\alpha_{3},n-1}^{(0)}-i\omega_{m2}}\right)\right)\right\} _{{\alpha_{1}\leftrightarrow\alpha_{2}\atop {\alpha_{3}
\leftrightarrow\alpha_{4}\atop {\omega_{m1}\leftrightarrow\omega_{m2}\atop \omega_{m3}\leftrightarrow\omega_{m4}}}}}\nonumber \\
\hspace{-2.5cm}-\delta_{\alpha_{1},\alpha_{3}}\delta_{\alpha_{2},\alpha_{4}}\delta_{\omega_{m1},\omega_{m3}}\delta_{\omega_{m2},
\omega_{m4}}a_{2}^{(0)}(\alpha_{1},\omega_{m1}|\alpha_{3},\omega_{m3})a_{2}^{(0)}(\alpha_{2},\omega_{m2}|\alpha_{4},\omega_{m4})\;\;\,\nonumber \\
\hspace{-2.5cm}-\delta_{\alpha_{1},\alpha_{4}}\delta_{\alpha_{2},\alpha_{3}}\delta_{\omega_{m1},\omega_{m4}}\delta_{\omega_{m2},\omega_{m3}}a_{2}^{(0)}(\alpha_{1},
\omega_{m1}|\alpha_{3},\omega_{m4})a_{2}^{(0)}(\alpha_{2},\omega_{m2}|\alpha_{4},\omega_{m3}),
\label{eq:4-6}
\end{eqnarray}
where $\bigtriangleup E_{S^{\prime},m^{\prime},n^{\prime}}^{(0)}=E_{S^{\prime},m^{\prime},n^{\prime}}^{(0)}-E_{S,m,n}^{(0)}$
and $\omega_{m1}\leftrightarrow\omega_{m2}$ and $\alpha_{1}\leftrightarrow\alpha_{2}$ refer to a symmetrization
with respect to the Matsubara frequencies and spin indices.


\section*{References}
\begin{thebibliography}{10}

\bibitem{key-1}M. H. Anderson, J. R. Ensher, M. R. Matthews, C. E. Wieman, and E. A. Cornell, Science,
\textbf{269}, 198 (1995). 

\bibitem{key-3}K. B. Davis, M. O. Mewes, M. R. Andrews, N. J. van Druten, D. S. Durfee, D. M. Kurn,
and W. Ketterle, \emph{Phys. Rev. Lett.} $\mathbf{75}$, 3969 (1995).

\bibitem{key-4-1}C.C. Bradley, C. A. Sackett, J. J. Tollett, and R. G. Hulet,\emph{ }Phys. Rev. Lett.\emph{
}\textbf{75}, 1687 (1995).

\bibitem{key-100}C. C. Bradley, C. A. Sackett, and R. G. Hulet , Phys. Rev. Lett. \textbf{78}, 985 (1997) 

\bibitem{key-50}R. Feynman, Int. J. Theor. Phys. 21, 467 (1982).

\bibitem{key-34}M. Greiner, O. Mandel, T. Esslinger, T. W. H\"ansch, and I. Bloch,\emph{ }Nature \textbf{415},
39 (2002). 

\bibitem{key-35}M. Greiner, O. Mandel, T. W. H\"ansch, and I. Bloch,\emph{ }Nature \textbf{419}, 51
(2002). 

\bibitem{key-33}M. P. A. Fisher, P. B. Weichman, G. Grinstein, and D. S. Fisher, Phys. Rev. B \textbf{40},
546 (1989). 

\bibitem{key-50-1}D. Jaksch, C. Bruder, J. I. Cirac, C. W. Gardiner, and P. Zoller, Phys.\emph{ }Rev.
Lett.\emph{ }\textbf{81}, 3108 (1998).

\bibitem{key-36}S. Sachdev, \emph{Quantum Phase Transitions}, 2nd edition, Cambridge University Press
(2011).

\bibitem{key-22}J. K. Freericks and H. Monien, Phys. Rev. B \textbf{53}, 2691 (1996).

\bibitem{key-18}D. van Oosten, P. Straten and H. T. C. Stoof, Phys. Rev. A \textbf{63}, 053601 (2001). 

\bibitem{key-23-1}F. E. A. dos Santos and A. Pelster, Phys. Rev. A \textbf{79}, 013614 (2009).

\bibitem{key-3-3}B. Bradlyn, F. E. A. dos Santos, and A. Pelster,\emph{ }Phys. Rev. A \textbf{79}, 013615
(2009). 

\bibitem{key-24}N. Teichmann, D. Hinrichs, M. Holthaus, and A. Eckardt, Phys. Rev. B \textbf{79}, 100503
(2009).

\bibitem{key-30}D. Hinrichs, A. Pelster, M. Holthaus, Appl. Phys. B (in press).

\bibitem{key-27}B. Capogrosso-Sansone, N. V. Prokof'ev, and B. V. Svis- tunov,  Phys. Rev. B \textbf{75},
134302 (2007).

\bibitem{key-28}B. Capogrosso-Sansone, S. G. S\"oyler, N. Prokof'ev and B. Svistunov, Phys. Rev.\emph{
}A \textbf{77}, 015602 (2008).

\bibitem{key-31}T. Wang, X.-F. Zhang, S. Eggert, and A. Pelster, Phys. Rev. A \textbf{87}, 063615 (2013).

\bibitem{key-58}G. Modugno, F. Ferlaino, R. Heidemann, G. Roati, M. Inguscio, Phys. Rev. A \textbf{68},
011601(R) (2003).

\bibitem{key-59}A. Albus, F. Illuminati, J. Eisert, Phys. Rev. A \textbf{68}, 023606 (2003).

\bibitem{key-60}H. P. B\"uchler, G. Blatter, Phys. Rev. Lett. \textbf{91}, 130404 (2003).

\bibitem{key-61}M. Lewenstein, L. Santos, M. Baranov, H. Fehrmann, Phys. Rev. Lett. \textbf{92}, 050401
(2004).

\bibitem{key-57}I. Bloch, Nature \textbf{453}, 1016 (2008).

\bibitem{key-51}K.V. Krutitsky, A. Pelster, and R. Graham, New J. Phys.\emph{ }8, 187 (2006). 

\bibitem{key-53-1}G. Roati, C. D'Errico, L. Fallani, M. Fattori, C. Fort, M. Zaccanti, G. Modugno, M.
Modugno, and Massimo Inguscio, Nature \textbf{453}, 895 (2008).

\bibitem{key-52}U. Bissbort, R. Thomale, and W. Hofstetter,\emph{ }Phys. Rev.\emph{ }A \textbf{81},
063643 (2010).

\bibitem{key-37}M. Lewenstein, A. Sanpera, and V. Ahufinger, \emph{Ultracold Atoms in Optical Lattices,
Simulating Quantum Many-Body Systems}, Oxford University Press (2012).

\bibitem{key-5}T.-L. Ho, Phys. Rev. Lett.\emph{ }$\mathbf{81}$, 742 (1998).

\bibitem{key-148}T. Ohmi and K. Machida, \emph{ }J. Phys. Soc. Jpn \textbf{67}, 1822 (1998).

\bibitem{key-145}J. Stenger, S. Inouye, D. M. Stamper-Kurn, H.-J. Miesner, A. P. Chikkatur, and W. Ketterle,
Nature \textbf{396}, 345 (1999).

\bibitem{key-143}D. M. Stamper-Kurn, M. R. Andrews, A. P. Chikkatur, S. Inouye, H. J. Miesner, J. Stenger,
and W. Ketterle, Phys. Rev. Lett\emph{.} \textbf{80}, 2027 (1998).

\bibitem{key-71}M.-S. Chang, C. D. Hamley, M. D. Barrett, J. A. Sauer, K. M. Fortier, W. Zhang, L. You,
and M. S. Chapman, Phys. Rev. Lett. $\mathbf{92}$, 140403 (2004).

\bibitem{key-72}H. Schmaljohann, M. Erhard, J. Kronj\"ager, M. Kottke, S. van Staa, L. Cacciapuoti,
J. J. Arlt, K. Bongs, and K. Sengstock, Phys. Rev. Lett. $\mathbf{92}$, 040402 (2004).

\bibitem{key-73}T. Kuwamoto, T. Araki, T. Eno, and T. Hirano, Phys. Rev. A $\mathbf{69}$, 063604 (2004). 

\bibitem{key-74}J. M. McGuirk, H. J. Lewandowski, D. M. Harber, T. Nikuni, J. E. Williams, and E. A.
Cornell, Phys. Rev. Lett. \textbf{89}, 090402 (2002).

\bibitem{key-75}Q. Gu, K. Bongs, and K. Sengstock, Phys. Rev. A \textbf{70}, 063609 (2004).

\bibitem{key-59-1}C. K. Law, H. Pu, and N. P. Bigelow, Phys. Rev. Lett. \textbf{81}, 5257 (1998).

\bibitem{key-77}H. Pu, C. K. Law, S. Raghavan, J. H. Eberly, and N. P. Bigelow, Phys. Rev. A \textbf{60},
1463 (1999).

\bibitem{key-42-2}A. Widera, F. Gerbier, S. F\"olling, T. Gericke, O. Mandel, and I. Bloch, Phys. Rev.
Lett.\emph{ }\textbf{95}, 190405 (2005).

\bibitem{key-42-3}A. Widera, F. Gerbier, S. F\"olling, T. Gericke, O. Mandel, and I. Bloch, New J.
Phys.\emph{ }\textbf{8}, 152 (2006). 

\bibitem{key-53-2}C. Becker, P. Soltan-Panahi, J. Kronj\"ager, S. D\"orscher, K. Bongs and K. Sengstock,
New J. Phys.\emph{ }\textbf{12}, 065025 (2010). 

\bibitem{key-29}E. Demler and F. Zhou, Phys. Rev. Lett.\emph{ }\textbf{88}, 163001 (2002).

\bibitem{key-4}S. Tsuchiya, S. Kurihara, and T. Kimura, Phys. Rev. A \textbf{70} , 043628 (2004). 

\bibitem{key-15}N. Uesugi, and M. Wadati,\emph{ }J. Phys. Soc. Japan\emph{ }\textbf{72}, 1041 (2003).

\bibitem{key-60-1}A. A. Svidzinsky and S. T. Chui, Phys. Rev. A \textbf{68}, 043612 (2003).

\bibitem{key-120}M. Mobarak and A.  Pelster, Laser Phys. Lett. (in press).

\bibitem{key-42-1}Y. Wu, Phys. Rev. A \textbf{54}, 4534 (1996). 

\bibitem{key-3-1}J. P. Burke, C. H. Green, and J. L. Bohn, Phys. Rev. Lett\emph{.} \textbf{81}, 3355
(1998).

\bibitem{key-119}P. B. Blakie and C. W. Clark, J. Phys. B \textbf{37}, 1391 (2004).

\bibitem{key-16}T. Kimura, S. Tsuchiya, M. Yamashita, and S. Kurihara, J. Phys. Soc. Jpn\emph{ }\textbf{75},
074601 (2006).

\bibitem{key-134}G. G. Batrouni, V. G. Rousseau, and R. T. Scalettar, Phys. Rev. Lett.\emph{ }\textbf{102},\emph{
}140402 (2009).

\bibitem{key-10-1}M. \L{}\k{a}cki, S. Paganelli, V. Ahufinger, A. Sanpera, and J. Zakrzewski, Phys.
Rev. A \textbf{83}, 013605 (2011).

\bibitem{key-190} S. Paganelli, M. \L{}\k{a}cki, V. Ahufinger, J. Zakrzewski, and A. Sanpera, J. Low
Temp. Phys. \textbf{165}, 227 (2011).

\bibitem{key-8-1}H. Kleinert, and V. Schulte-Frohlinde, \emph{Critical Properties of $\Phi^{4}$- Theories},
World Scientific, Singapore (2001).

\bibitem{key-9}J. Zinn-Justin, \emph{Quantum Field Theory and Critical Phenomena}, 4th edition, Claredon
Press, Oxford (2002).

\bibitem{key-11}C. J. Gorter and H. G. B. Casimir, Phys. Z\emph{.} \textbf{35}, 963 (1934).

\bibitem{key-12}L. P. Kadanoff, and G. Baym, \emph{Quantum Statistical Mechanics: Green's Function Methods
in Equilibrium and Non-Equilibr}i\emph{um} \emph{Problems},\emph{ }W. A. Benjamin, New York, (1962).

\bibitem{key-13}W. Metzner, Phys. Rev\emph{. }B \textbf{43}, 8549 (1991).

\bibitem{key-43}M. Ohliger and A. Pelster, World. J. Cond. Matt. Phys. \textbf{3}, 125 (2013). 

\bibitem{key-10}M. Ohliger, \emph{Thermodynamic Properties of Spinor Bosons in Optical Lattices}, Diploma
thesis, Freie Universit\"at Berlin (2008).

\bibitem{key-61-1}T. Kimura, S. Tsuchiya, M. Yamashita, and S. Kurihara, Phys. Rev. Lett.\textbf{ 94},
110403 (2005).

\bibitem{key-160}T. D. Gra\ss{}, F. E. A. dos Santos, and A. Pelster, Phys. Rev. A \textbf{8}4, 013613 (2011). 

\bibitem{key-162}P. T. Ernst, S. Götze, J. S. Krauser, K. Pyka, D.-S. Lühmann, D. Pfannkuche, and K.
Sengstock, Nature Phys.  \textbf{6}, 56 (2010). 

\bibitem{key-163-1}U. Bissbort, S. Götze, Y. Li, J. Heinze, J. S. Krauser, M. Weinberg, C. Becker, K.
Sengstock, and W. Hofstetter, Phys. Rev. Lett.\textbf{ 106}, 205303 (2011).

\bibitem{key-200}M.  Endres, T.  Fukuhara, D.  Pekker, M.  Cheneau, P.  Schau\ss{},	 C.  Gross, E.  Demler,	 S.  Kuhr, and I.  Bloch, Nature  \textbf{487}, 454  (2012). 

\bibitem{key-161}L. Santos, M. Fattori, J. Stuhler, and T. Pfau, Phys. Rev. A \textbf{75}, 053606  (2007). 

\bibitem{key-164}N. T. Phuc, Y. Kawaguchi, and M. Ueda, Phys. Rev. A \textbf{84}, 043645  (2011). 

\bibitem{key-163}A. Griesmaier, J, Werner, S. Hensler, J. Stuhler, and T. Pfau, Phys. Rev. Lett.\textbf{
94}, 160401 (2005). 

\bibitem{key-114}T.-L. Ho and S. K. Yip, Phys. Rev. Lett.\textbf{ 84}, 4031 (2000).
\end{thebibliography}
\end{document}